\documentclass[manuscript, acmlarge,screen]{acmart}

\usepackage{bm}
\usepackage{soul}
\usepackage{array}
\usepackage{subfig}
\usepackage{wrapfig}
\usepackage{amsmath}
\usepackage{listings}
\usepackage{arydshln}
\usepackage[export]{adjustbox}
\usepackage{subfig}
\usepackage{hyperref}
\usepackage{verbatim}
\usepackage[nameinlink]{cleveref}
\usepackage{multirow}
\usepackage{makecell}
\usepackage{adjustbox}
\usepackage{colortbl}
\usepackage{tabularx}
\usepackage{comment}
\usepackage{longtable}

\definecolor{green}{rgb}{0.08, 0.47, 0.16}
\definecolor{darkblue}{HTML}{0C0893} 
\definecolor{brilliantlavender}{rgb}{0.6, 0.4, 0.8}
\definecolor{candypink}{rgb}{0.89, 0.44, 0.48}
\definecolor{lightpurple}{rgb}{0.8, 0.5, 0.98}
\definecolor{green}{rgb}{0.08, 0.47, 0.16}
\definecolor{violet}{rgb}{0.96, 0.5, 0.5}
\definecolor{asparagus}{rgb}{0.53, 0.66, 0.42}
\definecolor{darkpastelpurple}{rgb}{0.59, 0.44, 0.84}
\definecolor{mediumslateblue}{rgb}{0.48, 0.41, 0.93}
\newcolumntype{L}[1]{>{\raggedright\let\newline\\\arraybackslash\hspace{0pt}}m{#1}}
\newcolumntype{C}[1]{>{\centering\let\newline\\\arraybackslash\hspace{0pt}}m{#1}}
\newcolumntype{R}[1]{>{\raggedleft\let\newline\\\arraybackslash\hspace{0pt}}m{#1}}

\sethlcolor{cyan}

\newcommand{\eg}{\textit{e.g.,\ }}
\newcommand{\ie}{\textit{i.e.,\ }}
\newcommand{\etal}{\textit{et al.\ }}

\newcommand\addauthornote[1]{%
  \if@ACM@anonymous\else
    \g@addto@macro\addresses{\@addauthornotemark{#1}}%
  \fi}

\setcopyright{acmlicensed}
\copyrightyear{2025}
\acmYear{2025}
\acmDOI{XXXXXXX.XXXXXXX}

\acmConference[]{}{}{}
\acmISBN{}

\begin{document}

\title{Towards Human-Centered Early Prediction Models for Academic Performance in Real-World Contexts}

\author{Han Zhang}
\email{micohan@cs.washington.edu}
\orcid{0000-0002-1377-1168}\affiliation{%
\institution{University of Washington}
  \country{USA}
}

\author{Yiyi Ren}
\email{yren98@uw.edu}
\orcid{0000-0002-2084-4943}
\affiliation{%
\institution{University of Washington}
  \country{USA}
}

\author{Paula S. Nurius}
\email{nurius@uw.edu}
\orcid{0000-0001-5091-6349}
\affiliation{%
\institution{University of Washington}
  \country{USA}
}

\author{Jennifer Mankoff}
\email{jmankoff@cs.washington.edu}
\orcid{0000-0001-9235-5324}
\affiliation{%
\institution{University of Washington}
  \country{USA}
}

\author{Anind K. Dey}
\email{anind@uw.edu}
\orcid{0000-0002-3004-0770}
\affiliation{%
\institution{University of Washington}
  \country{USA}
}


\renewcommand{\shorttitle}{Towards Human-Centered Early Academic Performance Prediction Models}

\renewcommand{\shortauthors}{Zhang et al.}

\begin{abstract}
Supporting student success requires collaboration among multiple stakeholders. Researchers have explored machine learning models for academic performance prediction; yet key challenges remain in ensuring these models are interpretable, equitable, and actionable within real-world educational support systems. First, many models prioritize predictive accuracy but overlook human-centered machine learning principles, limiting trust among students and reducing their usefulness for educators and institutional decision-makers. Second, most models require at least a month of data before making reliable predictions, delaying opportunities for early intervention. Third, current models primarily rely on sporadically collected, classroom-derived data, missing broader behavioral patterns that could provide more continuous and actionable insights. To address these gaps, we present three modeling approaches---LR, 1D-CNN, and MTL-1D-CNN---to classify students as low or high academic performers. We evaluate them based on \textit{explainability}, \textit{fairness}, and \textit{generalizability} to assess their alignment with key social values. Using behavioral and self-reported data collected within the first week of two Spring terms, we demonstrate that these models can identify at-risk students as early as week one. However, trade-offs across human-centered machine learning principles highlight the complexity of designing predictive models that effectively support multi-stakeholder decision-making and intervention strategies. We discuss these trade-offs and their implications for different stakeholders, outlining how predictive models can be integrated into student support systems. Finally, we examine broader socio-technical challenges in deploying these models and propose future directions for advancing human-centered, collaborative academic prediction systems.
\end{abstract}




\maketitle

\section{Introduction}\label{sec:intro}

Academic performance significantly affects college students' post-graduation opportunities, shaping their career prospects, social well-being, and future potential~\cite{kinkel2006impact,strahan2003effects}. However, stressors such as transitioning to a new environment~\cite{thurber2012homesickness}, insufficient social support~\cite{pluut2015social}, and uncertainties about the future~\cite{hamaideh2011stressors} present challenges that may negatively impact academic performance~\cite{kadapatti2012stressors,wang2015smartgpa}. As a result, understanding students' academic well-being and accurately predicting their academic outcomes---especially in identifying at-risk students---has become a critical area of research within domains such as learning analytics, where the primary focus is on measuring and analyzing student data, often collected from online learning platforms, to enhance learning processes~\cite{silvola2021expectations,ahern2024potential,daud2017predicting,namoun2020predicting,larrabee2019efficacy}. 

However, supporting students is not a task that can be accomplished in isolation---it requires collaborations among multiple stakeholders, including educators, policymakers, technology builders, and students themselves. Recently, researchers in the CSCW and broader HCI communities have examined student discrimination events~\cite{sefidgar2019passively}, academic well-being during COVID-19~\cite{nepal2022covid,zhang2022impact}, and predictive models for student academic performance~\cite{wang2015smartgpa,wang2014studentlife,mendez2021showing}, with the goal of shaping policy and intervention strategies. While these advancements have expanded our understanding of student experiences and shown that academic performance can be predicted with a reasonable degree of accuracy, three key gaps remain in how predictive models are designed and what data can be leveraged to better support multi-stakeholder collaboration and informed interventions.

First, \textbf{many predictive models focus heavily on improving accuracy but overlook how they fit into real-world decision making}. These models are often designed to maximize performance metrics but fail to consider how stakeholders will use them in practice. In educational settings, predictive models need to be more than just technically accurate; they need to generate insights that are clear, actionable, and fair for those making decisions. This means ensuring that predictions are transparent and understandable (\textit{explainability}) so that different stakeholders can interpret, trust, and meaningfully apply them in their workflows~\cite{ehsan2024xai,ehsan2023charting}; equitable across different student groups so they do not reinforce existing educational disparities (\textit{fairness})~\cite{mehrabi2021survey,delmonaco2024you}; and adaptable to different student populations and institutional settings to ensure reliable performance in diverse educational environments  (\textit{generalizability})~\cite{xu2023globem}.

Second, \textbf{existing models require at least a month of data before academic performance predictions can be made, delaying critical interventions}. Early awareness of academic risk is crucial for enabling timely interventions and support systems that benefit both students and other stakeholders~\cite{jayaprakash2014early,costa2017evaluating}. However, most studies rely on data collected mid-term or later to predict end-of-term outcomes, delaying the opportunity for earlier interventions that are essential to preventing students from falling too far behind~\cite{lopez2021early}.  Furthermore, real-world implementation is often hindered by the reliance on single datasets that require waiting until the end of a term to validate their accuracy, making real-time implementation difficult.

Third, \textbf{most existing work focuses on intermittently collected, classroom-derived data, neglecting students' broader daily experiences that affect academic success}. Current predictive models primarily rely on data from online learning systems (OLS), often capturing student interactions with class materials~\cite{lara2014system,lu2018applying,sukhbaatar2019artificial}. While these data provide insights into academic engagement, their sporadic collection can miss critical changes in behavior or performance and often lack actionable intervention strategies. For example, while correlations between forum visit frequency and performance are noted, these insights rarely guide concrete support actions. Moreover, they fail to account for essential daily behaviors and health factors outside the classroom, such as sleep, physical activity, and social interactions, which strongly influence academic outcomes~\cite{okano2019sleep,marques2018does,van2018domains}. 

To address these gaps, this paper examines how predictive models can be designed to better support decision-making. We explore the feasibility of developing predictive models that integrate three key human-centered machine learning (HCML) principles---\textit{explainability}, \textit{fairness}, and \textit{generalizability}---while also enabling earlier predictions and providing insights into academic-related factors beyond formal learning environments. Specifically, we approach this by tackling a binary classification task that predicts whether students' end-of-term GPAs will fall below or above a threshold of 3.2, classifying them as low or high performers. Throughout the paper, we refer to academic performance prediction as this binary classification task. Recognizing the challenges in building models that robustly incorporate even one of these human-centered dimensions in other domains (\eg~\cite{xu2023globem}), we explore three modeling approaches rather than pursuing a single model solution. 

Our dataset comprises daily behavioral data continuously collected from passive sensors, along with self-reports on physical and mental health, captured no later than the first week of the Spring terms in two academic years from a longitudinal study~\cite{xu2022globem,sefidgar2019passively}. We evaluate the three aspects of the HCML principles across the modeling approaches. Our findings underscore the complexities of balancing these three aspects in predictive modeling while providing insights into behavioral patterns associated with academic outcomes. We discuss scenarios where different approaches may benefit various stakeholders and offer insights for early interventions. Our contributions are as follows:

\begin{itemize}
   \item In Section~\ref{sec:method}, we present three academic performance prediction modeling approaches. The first two approaches, LR and 1D-CNN, were trained and tested on data from both 2018 and 2019. The third approach, MTL-1D-CNN, extends the 1D-CNN approach to improve \textit{generalizability} through multi-task learning. This approach trains on two related tasks---predicting end-of-term GPA and prior-term GPA---to enhance cross-year performance when trained on 2018 data and tested on 2019 data. 

    \item In Section~\ref{sec:approach_eval}, we evaluate the effectiveness of our approaches in classifying low and high performers early, followed by assessing their  \textit{explainability}, \textit{fairness}, and \textit{generalizability}. Our results show that the LR and 1D-CNN approaches can accurately predict academic performance comparably to prior studies but provide predictions \textit{at least three weeks earlier}. Each approach demonstrates mixed results across the three HCML principles, revealing inherent trade-offs. Additionally, we summarize academic-related in-class and outside-classroom behavioral patterns and discuss their potential for early interventions.

    \item In Section~\ref{sec:discussion}, we discuss the trade-offs observed across the three approaches, highlighting scenarios where each may benefit specific stakeholders. We outline key insights from our experiments and further discuss their implications for early intervention strategies. Furthermore, we reflect on the limitations of our work and propose future directions to improve human-centered academic predictive models.

\end{itemize}

While further efforts are needed to refine early predictive models for real-world educational settings, our work takes an initial step towards developing models that are not only accurate but also incorporate three HCML principles to better support real-world decision-making. By leveraging passive sensing data collected early in the term, we capture both in-class and out-of-class behavioral patterns, that provide insights into academic performance and facilitate timely, coordinated support strategies. We encourage future research to move beyond accuracy-driven approaches and focus on integrating predictive models into institutional workflows, ensuring they align with ethical and equity considerations, and effectively support multi-stakeholder collaboration.  
\section{Background and Related Work}\label{sec:related_work}

In this section, we review existing literature on academic performance prediction, focusing on the need for early prediction of at-risk students in real-world settings, the promise of passive behavioral data, and the value of adopting a human-centered approach when designing predictive models. We examine prior studies through the lens of three key human-centered principles: fairness, explainability, and generalizability. Our review identifies several gaps in existing literature and sets the stage for our work.

\subsection{Need for Early Prediction of At-risk Students in Real-world Settings}

Predicting student academic performance has been a longstanding focus of research in educational data mining and learning analytics (\eg~\cite{sukhbaatar2019artificial,namoun2020predicting,khan2021student,ojajuni2021predicting,daud2017predicting}), and it has recently gained increasing attention in the CSCW and broader HCI communities (\eg~\cite{sefidgar2019passively,nepal2022covid,pyle2023social,wang2015smartgpa,wang2014studentlife}). A common goal across these studies is to predict end-of-term GPA~\cite{sukhbaatar2019artificial,lu2018applying,chen2020utilizing} (as shown in the \textbf{Task} column of Table~\ref{tab:related_work} in Appendix), which, if predicted early enough, can enable timely interventions to improve student outcomes~\cite{lopez2021early,cassells2018effectiveness}. 

While significant progress has been made, a shared limitation across these efforts is that the earliest predictions occur around a month after the term begins (as indicated in the \textbf{Data} column of Table~\ref{tab:related_work}), limiting the opportunity for timely detection and intervention. Timely identification of at-risk students is crucial for providing the best opportunity for support and behavior correction~\cite{cassells2018effectiveness,thayer2018wise,hlosta2017ouroboros}. Research has consistently demonstrated that students who struggle academically early in the term are more likely to experience cumulative negative effects on their performance if these issues are not addressed promptly~\cite{jayaprakash2014early}. Furthermore, the longer it takes to identify students in need of help, the more difficult it becomes to reverse academic difficulties~\cite{katamei2015intervention}. Early identification allows for the implementation of timely and tailored interventions, such as academic counseling, peer support, and mental health resources, which can help mitigate potential challenges before they escalate~\cite{katamei2015intervention,geiser2007validity}.

In addition to the limitation of delayed early prediction, existing early prediction models also face challenges related to their practical implementation in real-world educational settings. Most prior studies rely on data collected over extended periods, with nearly all requiring end-of-term academic outcomes for model training and evaluation, which inherently limits their ability to be deployed for early interventions. Among all the reviewed work, all but one study, which trained a model on one term's data and applied it to a different term for testing~\cite{chen2020utilizing}, suffer from this limitation.

\subsection{Promise of Passive Behavioral Data for Academic Performance Prediction}

A recent literature review reveals that most studies in this area rely heavily on data derived from online learning systems (OLS), such as Learning Management Systems (LMS) and Massive Open Online Courses (MOOC)~\cite{lopez2021early}. Our review of prior work on academic performance prediction further supports this finding (as seen in the \textbf{Input} column of Table~\ref{tab:related_work}). OLS data typically captures student engagement with course materials, assignment submissions, and exam performance~\cite{chen2020utilizing,mwalumbwe2017using,yaugci2022educational}. While this data provides valuable insights into academic engagement, it is often collected over extended periods, which can lead to missed behavioral or performance changes between data collection points~\cite{fredrickson2000extracting,kawakami2023sensing}. Furthermore, it overlooks crucial daily behaviors and health factors outside the classroom, which can significantly impact academic performance (\eg~\cite{trockel2000health,felez2018activpal,gallagher2006national,wyatt2013comparing}).

Daily habits and behaviors such as sleep, physical activity, substance use, and social interactions have been demonstrated to be strongly associated with student academic performance~\cite{trockel2000health,felez2018activpal,xu2019prediction,cox2007academic,wang2015smartgpa}. For example, one study found that weekday and weekend wake-up times had the most significant relative effects on term GPA~\cite{trockel2000health}. Another study demonstrated that time spent in sedentary breaks during weekdays was positively related to academic achievement~\cite{felez2018activpal}. Additionally, research shows that longer periods of socializing at night, especially as the term progresses, can negatively impact students' term GPA~\cite{wang2015smartgpa}. Beyond daily behaviors, stress and related mental health challenges are growing concerns across colleges and universities, with an increasing number of students experiencing elevated levels of stress and mental health issues~\cite{hunt2010mental,gallagher2006national}. These stressors can significantly undermine academic success, leading to poorer grades, lower GPAs, and higher rates of course withdrawal and dropout~\cite{wyatt2013comparing}.

Incorporating behavioral and health data offers an opportunity for support systems to proactively identify at-risk students and enable earlier, more effective interventions. With the pervasive and unobtrusive collection of behavioral data, passive sensing data holds the promise of providing continuous insights into students' daily habits and behaviors~\cite{kawakami2023sensing}. Within the CSCW and broader HCI communities, passive behavioral data has been increasingly studied and applied in research related to mental well-being (\eg~\cite{sefidgar2019passively,morshed2019prediction,xu2023globem,das2022semantic}), as well as social well-being (\eg~\cite{kawakami2023sensing,das2024sensible,adler2022burnout,das2023algorithmic}). However, the use of such data for predicting student academic performance is still a relatively new area, with only two recognized studies leveraging this approach~\cite{sano2015recognizing,wang2015smartgpa}. One study focused on predicting end-of-term cumulative GPA using 10 weeks of passive sensing data combined with self-reports~\cite{wang2015smartgpa}, while the other classified students in the top 20\% and bottom 20\% of GPA performers based on one month of similar data~\cite{sano2015recognizing}.

\subsection{Human-centered Nature of Academic Performance Prediction Models}

Researchers have increasingly advocated for integrating technical innovations in ML/AI with human needs and social values~\cite{kim2021human,hong2020human,banovic2023being,zhang2023framework,andersen2023introduction,meegahapola2024faircomp,xu2023globem}. This approach, often referred to as \textit{Human-Centered Machine Learning} (HCML), emphasizes the importance of aligning AI systems with the social and ethical contexts in which they operate. Researchers have pointed out that HCML covers a broad scope, including fair and transparent algorithm design, human-in-the-loop decision-making, human-AI collaboration, and assessing the social impact of ML/AI on diverse communities~\cite{chancellor2023toward,amershi2019guidelines}. This approach underscores the need for ML/AI systems to be not only technically robust but also sensitive to the broader socio-technical environments they are deployed in.

Compared to research in other domains, such as mental and social well-being, where HCML has become a guiding principle during model and system development~\cite{kim2021human,yoo2024missed,andersen2023introduction,coleman2023reconsidering}, research in academic support settings has been slower to adopt these principles. While acknowledging the multifaceted nature of HCML, in this review, we focus primarily on examining prior academic performance prediction work in terms of its capability to provide key stakeholders with understandable information and actionable insights for early interventions (\textit{explainability}), ensure equitable deployment across marginalized student groups (\textit{fairness}), and assess the robustness of the models in generalizing across diverse contexts, such as different student populations and academic terms (\textit{generalizability}).

\subsubsection{Explainability}\label{subsubsec:explainability} While there is still no universal consensus on the exact definition of explainability and related terms such as interpretability~\cite{ehsan2024xai,rosenfeld2019explainability}, explainability, or Explainable AI (XAI), it generally refers to the ability of a system to make its decisions or behaviors understandable to humans~\cite{ehsan2023charting,arrieta2020explainable,carvalho2019machine,miller2019explanation}. In educational contexts, explainability is particularly critical, as it provides transparency into how predictive models arrive at certain decisions, thereby guiding ML engineers debug the models, and other stakeholders---such as educators, students, and administrators--- in working together on potential interventions. By clarifying the model's decision-making process, stakeholders can better understand why certain students are identified as at-risk and what steps can be taken to support them. 

One commonly used method for interpreting models is feature importance~\cite{bhatt2020explainable}, which assigns numerical values to each feature based on its influence on the model's predictions. This method helps stakeholders grasp which factors are most significant in determining outcomes, such as students' academic performance (this is the approach we use in this paper). More in-depth descriptions of other explainable or interpretable machine learning methods can be found in works like~\cite{du2019techniques,doshi2017towards,molnar2020interpretable}.

In prior work reviewed in Table~\ref{tab:related_work}, several studies touched upon the \textit{explainability} of their models by discussing the factors that contribute to the prediction of students' academic performance~\cite{bravo2021early,yao2019predicting,wang2015smartgpa,lu2018applying,waheed2023early,sano2015recognizing}. Most identified factors were related to in-class behaviors, including students' interactions with OLS (\eg visits to forums, attempts at questionnaires)~\cite{bravo2021early,lu2018applying,waheed2023early}, as well as total hours of academic activities~\cite{sano2015recognizing,wang2015smartgpa}. Additionally, personality traits (\eg conscientiousness, diligence, and orderliness) have also been identified as factors associated with students' academic performance~\cite{wang2015smartgpa,yao2019predicting}. While these insights are valuable for understanding the contributors to student performance, they often fall short of translating into actionable interventions. For example, knowing that quiz attempts or forum visits correlate with academic performance does not clarify \textit{how} to intervene. Similarly, personality traits are relatively stable over time and are not easily influenced through short-term interventions. Without concrete actions tied to these predictors, educators may struggle to design specific interventions that directly address students' needs, limiting the practical utility of these models in real-world educational settings.

\subsubsection{Fairness}
Exposure to discrimination and social marginalization has long been recognized as contributing to heightened stress levels among students, both directly and through indirect forms such as bearing witness to incidents~\cite{assari2017racial,huynh2017not}. This compounded stress can exacerbate the challenges students face in educational settings, with those experiencing multiple forms of discrimination often showing more pronounced stress responses~\cite{khan2017multifactorial}. The persistent strain from stress-related factors disproportionately impacts socially disadvantaged populations, leading to both health and academic performance disparities~\cite{holt2007multiple,nadal2020injurious,sternthal2011racial}. Such disparities further underscore the importance of addressing equity and fairness, especially in systems that impact students' success.

As ML/AI technologies are increasingly integrated into decision-making processes, especially in highly sensitive areas like education, ensuring fairness is crucial. Researchers in HCI, ML, and CSCW have been highlighting the importance of these technologies being \textit{fair}, meaning non-discriminatory with respect to individuals' \textit{protected traits}, such as gender, ethnicity, or religion~\cite{friedler2016possibility,chancellor2023toward,buyl2024inherent}. Although the field of AI fairness is still developing consensus on both its ontology and methods~\cite{olteanu2019social,friedler2016possibility,suresh2019framework,verma2018fairness}, one well-established source of bias is \textit{algorithmic bias}, where reliance on automated decision-making processes based on ML or statistical methods can amplify biases towards certain subpopulations within the training data~\cite{olteanu2019social}.

Algorithmic bias can arise from multiple sources, including model design choices---such as the architecture, loss function, optimizer, and hyper-parameters used~\cite{hooker2021moving,mehrabi2021survey}. These decisions, along with statistically biased estimators, can result in models that are unintentionally introduce or amplify biases, affecting the fairness of outcomes and decision-making processes~\cite{olteanu2019social}. In the context of predictive models in educational settings, such biases may disproportionately affect marginalized student groups, exacerbating existing inequalities rather than mitigating them. 

To quantify these risks, different fairness definitions and metrics have been proposed(\eg~\cite{xu2022algorithmic}). In this paper, we use three fairness measures that are most applicable to a binary classification setting: demographic parity~\cite{barocas2016big}, equalized odds~\cite{zafar2017fairness}, and equal opportunity~\cite{hardt2016equality}. A more detailed review of these measures can be found in Appendix~\ref{appx:algorithmic_bias}. Despite its significance, fairness in academic performance prediction models remains largely underexplored, with no prior studies we reviewed critically evaluating their models for bias (as shown in the \textbf{Model Fairness} column in Table~\ref{tab:related_work}). Our work seeks to address this gap by incorporating fairness evaluation into the development of predictive models.

\subsubsection{Generalizability} For real-world deployment in educational settings, ensuring that a model can generalize across different contexts---such as varying populations and institutions---is critical. This means training a model on one dataset and ensuring its accuracy remains robust when tested on one or more previously unseen datasets~\cite{xu2023globem}. Generalizability is especially challenging when dealing with longitudinal behavioral data, as behaviors can vary greatly over time (season to season, year to year) and across different locations and individuals. Such variations can alter the data distribution, often leading to a decrease in model accuracy~\cite{xu2023globem,adler2022machine,xuglobem}. 

Despite the importance of generalizability, almost no academic performance prediction studies address this issue directly. In our review, only one such study---by Chen \etal~\cite{chen2020utilizing}---evaluates the generalizability of their model. They tested their model on a single class during a new term and found that the AUC dropped from 0.75 to 0.63 on the unseen dataset, demonstrating the challenge of maintaining model performance across different contexts. While there is currently no clear standard defining what constitutes ``good'' generalizability in terms of AUC scores, a higher AUC on unseen datasets would indicate a model's stronger applicability in varied real-world educational environments.  

\subsection{Summary}

To summarize, prior research on academic performance prediction typically focuses on predicting end-of-term GPA using data from OLS. A key limitation of these studies is the delayed identification of at-risk students, as most predictions only occur after collecting data four weeks into the term. Recent studies from outside the area of academic performance have highlighted the potential of passive behavioral data that could be used to complement OLS data, providing continuous insights that allow for earlier predictions and interventions. In real-world educational settings, the CSCW and HCI communities emphasize the development of systems using a HCML approach, advocating for models that are not only accurate but also explainable, fair, and generalizable. Despite the importance of these considerations, prior work in academic performance prediction has yet to fully address these human-centered challenges. Our work seeks to fill this gap by exploring three approaches to predict, early in the term, whether a student's end-of-term GPA will fall below 3.2, leveraging passive behavioral data collected no later than the first week. We further assess these approaches' explainability, fairness, and generalizability, recognizing that these factors are essential for creating predictive models that align with the human-centered needs of real-world educational environments.
\section{Data Sources}\label{sec:data}

Data used in this work comes from a longitudinal study aimed at understanding the daily behaviors of college students, alongside physical and mental health concerns~\cite{xu2022globem,sefidgar2019passively}. Conducted at a Carnegie-classified R-1 university in the United States, the study collected data during each Spring term from 2018 to 2023. For the purpose of this paper, we specifically used data collected no later than the first week of the 2018 and 2019 Spring terms to develop our predictive models. We focused on data from these two years to avoid the disruptions to grading and student behavior patterns caused by the COVID-19 pandemic starting in 2020, ensuring consistency and reliability in our modeling approach.  Below, we summarize the participants' details, data sources---including passive behavioral data and self-reports---and the outcome measures used in the modeling approaches.

\subsection{Participants}

Participants in the 2018 study were all first-year college students. Some of these students continued into the 2019 study, with additional first-year students added in 2019. GPA data was available for 195 students in 2018 and 201 students in 2019. Retention within each year was high---96.2\% in 2018 and 98.0\% in 2019---though year-to-year retention was low, at around 21.3\% (detailed in  Table~\ref{tab:study_completion}). The dataset includes various protected groups, such as first-generation college students, students from underrepresented minorities (African-American, Latinx, Native American, and Pacific Islander), gender minorities (non-male students), and those identifying as part of a sexual minority (non-heterosexual students). After excluding participants missing key data, the final analysis included data from 188 students in 2018 and 196 students in 2019. 

\subsection{Passive Behavioral Data}\label{sec:passive_self_collection}
Each year's dataset includes several streams of passive sensing data from smartphones and wearable devices (\ie Fitbits). Key features from the mobile sensing data include physical activity states (\eg stationary, walking, running), application usage (foreground apps and push notifications), battery status (charging/discharging, battery levels),  Bluetooth scans (nearby Bluetooth-enabled devices), call logs (incoming, outgoing, and missed calls), GPS location data, screen status (on/off/lock/unlock), and WiFi interactions (connected and surrounding access points). All these data streams are gathered from both iOS and Android devices to address potential socio-economic bias, as studies suggest that Android users generally have lower socio-economic status compared to Apple users~\cite{jamalova2019comparative}. However, due to privacy restrictions~\cite{AppleATT,AppleUserPrivacy}, application usage data is unavailable for iOS users, impacting 80\% and 70\% of participants in 2018 and 2019, respectively. As a result, we excluded application usage data during pre-processing to ensure consistency in our analysis. Additionally, the dataset includes metrics from wearable activity trackers such as step counts and sleep. 

We followed the feature extraction framework described in prior work~\cite{doryab2018extraction} to extract general, low-level behavioral features, including physical activity, phone usage, travel time, screen time, sleep, and step counts. To capture student behaviors with greater granularity, we grouped the daily behavioral data into five time epochs: morning (6am - 12pm), afternoon (12pm - 6pm), evening (6pm - 12am), night (12am - 6am), and an entire day (24 hours), replicating a similar approach employed in prior work~\cite{wang2015smartgpa}. Features were calculated for each epoch, as well as for the entire day. Sleep features were computed on a daily basis only.  Table~\ref{tab:behavioral-features} provides a detailed summary of these low-level behavioral features, and further implementation details are provided in  Appendix~\ref{feature_implementation}.

Additionally, we replicated high-level academic-related behavioral features identified in prior research~\cite{wang2015smartgpa,lu2010jigsaw,lane2011bewell}, such as student activity duration (\ie non-stationary time), study duration and focus time during study, dorm time, party attendance, indoor and outdoor mobility, class attendance, and changes in behavior patterns. While the calculations of some features were adapted to fit the context and dataset of this study, they remained consistent with the goals of earlier studies. For both the low-level and high-level behavioral features, we computed statistical metrics such as average (avg), standard deviation (std), minimum (min), and maximum (max) values within each epoch.  Further details on how each high-level behavioral feature was calculated can be found in  Appendix~\ref{high-level_feature_implementation}. 

\begin{figure}[t]
    \centering
\includegraphics[scale=0.18]{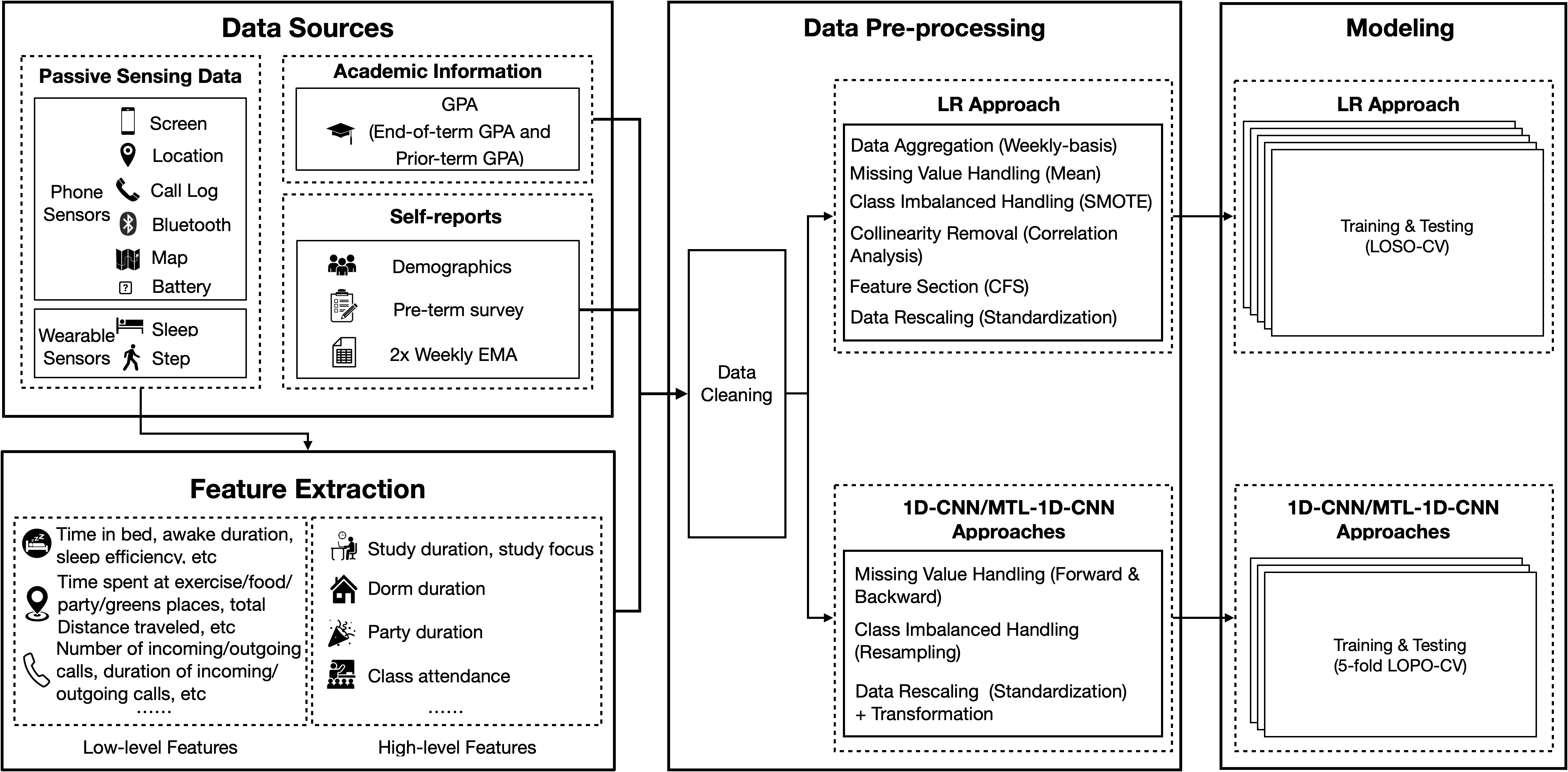}
    \caption{Overview of the whole modeling pipeline for the three approaches. All three approaches utilize the same data sources and extracted features. However, distinct data pre-processing and modeling techniques were applied to the LR approach compared to the 1D-CNN and MTL-1D-CNN approaches.} 
    \label{fig:pipeline}
\end{figure}

\subsection{Self-reports}
Self-report data includes four key sources. First, a demographic survey captured participants' demographics such as race, first-generation college student status, financial background, and personality traits. Second, a \textit{pre}-term survey gathered information on students' mental and physical health concerns, major life adversities, self-assessments of academic performance, social media usage, and mobile service provider. Third, twice-weekly Ecological Momentary Assessment (EMA) surveys, collected on Wednesdays and Sundays, assessing students' experience of unfair treatment, affect, stress, and substance use. Finally, prior Winter term GPA is also captured as a measure of students' previous academic outcome. Note that, the \textit{pre}-term survey consists of a series of well-established scales to evaluate students' mental health states, \eg depression (CES-D~\cite{radloff1977ces}), anxiety (STAI~\cite{kabacoff1997psychometric}), and stress (PSS~\cite{cohen1983global}), as well as physical health (CHIPS~\cite{cohen1983positive}). We included self-reports, as research indicates they provide reliable measures compared to implicit measures, particularly for mental content~\cite{corneille2024self}. In this paper, we used both calculated scale scores and individual items of these scales as  self-report features.  The types of passive data collected remains consistent across 2018 and 2019, though the \textit{pre}-term and EMA surveys are slightly different between the two years due to small changes made during the data collection to improve efficiency. A description of a preliminary data cleaning procedures applied to both the passive behavioral data and self-reports is provided in Appendix~\ref{appx:common_data_cleaning}.

\subsection{Ground Truth}

Students' end-of-term GPA was used as the ground truth for performance classification. The outcome label was binary, distinguishing between students with a GPA above or below a predetermined threshold. Prior research has employed various GPA cutoffs to define at-risk students, such as those predicted to earn a final grade of C+ or lower~\cite{chen2020utilizing}, those failing a course~\cite{yu2018improving,sukhbaatar2019artificial}, or a GPA below 4.0 on a 6-point scale~\cite{sano2015recognizing}. In this study, a GPA cutoff of 3.2 (on a 4-point scale) was selected, corresponding to both the university's reported average GPA, aligning both with typical student performance levels and the minimum required for a B+ grade\footnote{Our data was collected from a smaller subset of students at the university, so the average GPA for each dataset may differ slightly from the overall university average of 3.2.}. The left side of  Figure~\ref{fig:pipeline} visualizes all data sources used in this study.

\begin{table}[htb!]
    \caption{Demographics. The table shows the total number and percentage of individuals in the entire population (EP), along with the number of low performers and corresponding percentages within each group: protected group (PG) and unprotected group (UG). Protected groups include underrepresented minorities, first-generation college students, non-male gender (including female and transgender individuals), and sexual minorities (\eg homosexual and bisexual individuals), based on race, first-generation status, gender, and sexual orientation. ``\# Total'' refers to the total number in each group, and ``\% in EP'' indicates the percentage of that group within the entire population. ``\# Low Performers'' refers to the number of low performers in each group. ``\% in PG'' and ``\% in UG'' represent the percentage of low performers within the protected and unprotected groups, respectively.}
 \renewcommand{\arraystretch}{1.1}
   \resizebox{\textwidth}{!}{
    \begin{tabular}{c|lccccccc}\hline\hline
    \multirow{2}{*}{\textbf{Year}} && \multirow{2}{*}{\textbf{Protected Trait}}&&\multicolumn{2}{c}{\textbf{Protected Group (PG)}}&&\multicolumn{2}{c}{\textbf{Unprotected Group (UG)}} \\
    \cline{5-6}\cline{8-9}
    && && \# Total (\% in EP)  & \# Low Performers (\% in PG) && \# Total (\% in EP)  &\# Low Performers (\% in UG) \\\hline
    \multirow{4}{*}{2018} && \makecell[l]{Race} &&  32 (17.0\%) &  12 (37.5\%)  && 156 (83.0\%)  &  31 (19.9\%)   \\\cline{2-9}
    && \makecell[l]{First-generation} && 57 (30.3\%)  &  19 (33.3\%)  &&  131 (69.7\%) &  24 (18.3\%)    \\\cline{2-9}
    && \makecell[l]{Gender} &&  123 (65.4\%) &  80 (24.4\%)  &&  65 (34.5\%) &   13 (20.0\%) \\\cline{2-9}
    && \makecell[l]{Sexual Orientation} &&  21 (11.2\%) &  3 (14.3\%)  &&  167 (88.8\%) &  40 (24.0\%)    \\\hline
    \multirow{4}{*}{2019} && \makecell[l]{Race} &&  23 (11.7\%) &  12 (52.2\%)  &&  173 (88.3\%) &   51 (29.5\%) \\\cline{2-9}
    && \makecell[l]{First-generation} &&  58 (29.6\%) &   26 (44.8\%) &&  138 (70.4\%) &  37 (26.8\%) \\\cline{2-9}
    && \makecell[l]{Gender} && 100 (51.0\%)  &  35 (35\%)  && 96 (49.0\%)  & 28 (29.2\%) \\\cline{2-9}
    && \makecell[l]{Sexual Orientation} &&  21 (10.7\%) &  5 (23.8\%)  &&  175 (89.3\%) &   58 (33.1\%)  \\\hline\hline                
\end{tabular}}
\label{tab:demographics}
\end{table}
\renewcommand{\arraystretch}{1.2}

Students with a GPA above 3.2 were classified as high performers, while those with a GPA of 3.2 or lower were considered low performers. In Spring 2018, the mean GPA was 3.48, with a standard deviation of 0.47. Of the 188 students, 145 (77\%) were high performers and 43 (23\%) were low performers. In Spring 2019, the average GPA decreased slightly to 3.32, with a standard deviation of 0.60. Of the 196 students, 133 (68\%) were high performers and 66 (32\%) were low performers. Further details on the GPA distribution can be found in Figure~\ref{fig:group_gpa} in Appendix~\ref{appx:gpa_dis}. The right columns of  Table~\ref{tab:demographics} offer a breakdown of the number and percentage of low performers across different demographic groups, while the left columns show the total number and percentage of students in both protected and unprotected groups within the overall population. For instance, in 2018, 32 students (17.0\%) identified as underrepresented minorities, of whom 12 (37.5\%) were categorized as low performers. 

\section{Three Early Academic Prediction Approaches Within Human-Centered Settings}\label{sec:method}

\begin{figure}[!htb]
    \centering
    \subfloat[LR and 1D-CNN approaches.]{\includegraphics[scale=0.52]{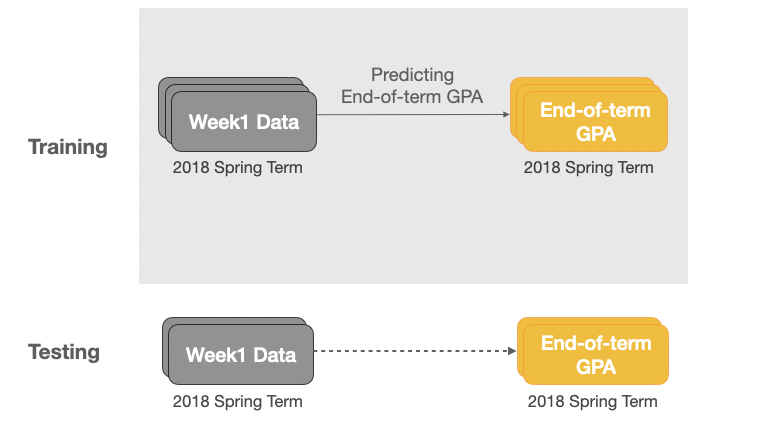}\label{fig:model_setup1}}
     \subfloat[MTL-1D-CNN approach.]{\includegraphics[scale=0.52]{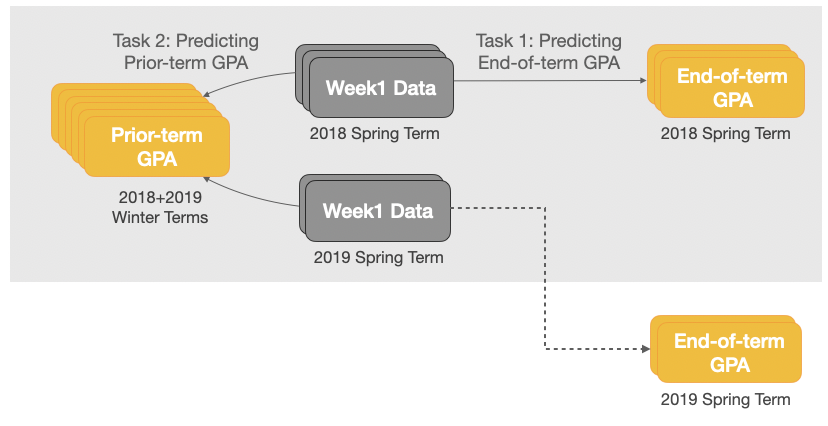}\label{fig:model_setup2}}
    \caption{Overview of the training (highlighted in light gray) and testing process for the three approaches. (a) shows the training and testing process for the LR and 1D-CNN approaches (using 2018 Spring term data as an example), where data collected by the first week is used for training and testing to predict end-of-term GPA for both 2018 and 2019. (b) shows the training and testing process for the MTL-1D-CNN approach, where training includes two tasks: Task 1 uses the first week of data from 2018 to predict end-of-term GPA, while Task 2 combines first-week data from both 2018 and 2019 to predict prior-term Winter GPA. Testing uses data from 2019 to predict end-of-term GPA.
    } 
    \label{fig:modeling_setup}
\end{figure}

In this section, we present three early academic prediction modeling approaches, with a focus on the human-centered principles of \textit{explainability}, \textit{fairness}, and \textit{generalizability}. We begin by detailing the design rationale for each approach, including the reasoning behind model selection and the specific methods employed. This is followed by a description of the data pre-processing steps used to prepare the behavioral data, self-reports, and academic outcomes for input into the models. Finally, we provide an overview of the modeling setup, including the specific configurations and methods. The middle and right sections of  Figure~\ref{fig:pipeline} depict the data pre-processing and modeling steps for each of the three modeling approaches. Additionally,  Figure~\ref{fig:modeling_setup} outlines the training and testing setups specific to these approaches.


\subsection{The LR Approach}\label{sec:lr}

\subsubsection{Design Rationale} Logistic Regression (LR) is a widely used and interpretable machine learning model, often selected when explainability is a key priority~\cite{barredo_arrieta_explainable_2020}. Its linear structure allows for easy interpretation of feature importance, helping users understand how each variable contributes to the model's predictions~\cite{bhatt2020explainable}. Additionally, the simplicity of LR makes it easier to address and correct bias, with methods such as reweighing and post-processing bias correction being more effectively applied to linear models, thus supporting fairness across different student groups~\cite{mehrabi2021survey}.

\subsubsection{Data Pre-processing} We implemented several data pre-processing steps to address missing values, class imbalance, collinearity, and feature selection. For missing values, two imputation methods were considered: assigning a default value (999) or imputing based on the mean of the training set, with the latter chosen due to its better performance in the 2018 data experiments. Using the mean value from the training set for both training and testing ensures that test data does not influence the imputation process, preventing data leakage by avoiding any knowledge of the test distribution~\cite{geron2022hands,bishop2006pattern}. To address class imbalance, we employed SMOTE~\cite{chawla2002smote}, which oversamples the minority class to balance the dataset. Features exhibiting collinearity, indicated by correlations exceeding 0.7~\cite{dormann2013collinearity}, were removed from both the training and test sets to avoid distortion in model estimation. Finally, we applied correlation-based feature selection (CFS)~\cite{hall1999correlation}, conducting a grid search to identify the optimal correlation cutoff value, \textit{r}, ensuring generalization from training to unseen data. Additional details on these processes can be found in Appendix~\ref{appx:data_prep_lr}.

\subsubsection{Modeling Setup} We employed Leave-One-Subject-Out Cross-Validation (LOSO-CV) to minimize overfitting and ensure robust model performance. In each iteration of LOSO-CV, feature scaling was applied to the training set, with standardization performed to enhance convergence. All features were aggregated at a weekly level, represented by means and standard deviations, to create a unified data structure. 

To fine-tune the details of the LR modeling pipeline, we treated the 2018 data as an ``experimental'' dataset, testing various methods for missing data imputation, class imbalance handling, different values for the regularization parameter \textit{r}, and cutoff values for feature selection. Since these experiments could introduce a risk of overfitting to the 2018 data, the final pipeline developed from this experimentation was applied without any modification to the 2019 dataset.  Figure~\ref{fig:model_setup1} shows the training and testing process for the LR approach.

\subsection{The 1D-CNN Approach}\label{sec:1d-cnn}

\subsubsection{Design Rationale} One-Dimensional Convolutional Neural Networks (1D-CNN) are a deep learning model that are highly effective for analyzing time-series data, making them particularly advantageous when working with behavioral data collected from sensors~\cite{xu2023globem,kiranyaz20211d}. Unlike linear models, which may struggle to capture complex patterns, 1D-CNNs excel in detecting subtle temporal dependencies and intricate relationships across features over time~\cite{kiranyaz20191}. This is especially useful for analyzing sequential data, where the ability to recognize patterns evolving over time is essential. Furthermore, 1D-CNNs' capacity to learn localized patterns within sequences enables them to generalize effectively across varying contexts~\cite{wang2017time}, making them suitable for real-world applications where data variability and complexity are common.

\subsubsection{Data Pre-processing} Since feature selection is not required in deep learning models like 1D-CNN, we employed different methods for missing value handling, class imbalance handling, and data standardization and transformation. To handle missing values in our time series data, we used forward filling followed by backward filling within each participant's data, a common imputation technique for time series data~\cite{che2018recurrent}. For class imbalance, we balanced the training set by duplicating instances from the minority class (low performers), ensuring an equal 1:1 ratio between classes. Additionally, we standardized all features and transformed the data into a three-dimensional structure, with dimensions representing participants, days, and features. For further details on these processes, refer to  Appendix~\ref{appx:data_prep_1d-cnn}.

\subsubsection{Modeling Setup} To reduce computational costs, we split the data into an 80\% training set and a 20\% testing set. For the 1D-CNN approach pipeline, we used 5-fold Leave-Participants-Out Cross-Validation (LOPO-CV) on the training set, replacing the LOSO-CV used in the LR approach pipeline due to the computational cost. The entire modeling process was repeated five times, and the average model performance was reported to mitigate stochastic influences. The best model was selected using only the training data, ensuring no information leakage to the test data set. In contrast to the LR pipeline, where data was aggregated at a weekly level, the 1D-CNN approach maintained the input data as daily time series, allowing the model to capture finer temporal patterns. More details about the 1D-CNN model architecture can be found in  Appendix~\ref{appx:1d-cnn_arch}.

\subsection{The MTL-1D-CNN Approach}\label{sec:mtl-1d-cnn}

\subsubsection{Design Rationale}

Multi-task learning (MTL)~\cite{ruder2017overview} enhances models by enabling them to perform multiple related tasks simultaneously. This approach allows a model to learn shared representations across tasks, improving its ability to generalize to new data~\cite{caruana1997multitask,caruana1993multitask}. Additionally, when designed appropriately, MTL can support real-world early prediction settings by leveraging knowledge from the secondary task to refine predictions in the primary task, thereby enhancing practical implementation. In line with this, we introduce a third modeling approach that extends the 1D-CNN approach with a secondary task---predicting prior term GPA---resulting in the MTL-1D-CNN approach. The pre-processing steps for this approach remain the same as those used for 1D-CNN. Below, we detail the modeling setup for this extension.

\subsubsection{Modeling Setup} 

Our MTL approach extends the 1D-CNN network by adding a secondary task: predicting prior term GPA, which is known to correlate with current academic success~\cite{paschall2003does}. The primary task, predicting end-of-term GPA, is trained on data from the first week of 2018 and tested on 2019 data, while the secondary task is trained on the combined data from the first week of both years. Note that, for the secondary task, we only used prior term GPA labels, which is available before the new term begins, ensuring no data leakage and maintaining the ability to make predictions at the end of week one without needing end-of-term GPAs from the 2019 data.  Figure~\ref{fig:model_setup2} visualizes the training and testing process for this approach. 

This approach is an extension of hard parameter sharing, where both tasks typically use the same dataset and input format. However, since our tasks use different datasets---training data for the primary task and a combination of training and test data for the secondary task---this creates a challenge. Soft parameter sharing is often used in such cases, where each task has its own model and data, but in our approach, both tasks share the same model while using different datasets. To address the imbalance in sample size between the primary and secondary tasks, we resampled the smaller dataset (from the primary task) to match the size of the larger dataset and continued using hard parameter sharing.

\section{Evaluations}\label{sec:approach_eval}

In this section, we first evaluate the three approaches' effectiveness in making early predictions about academic performance. We then assess their explainability, focusing on how easily we can extract interpretable insights into the factors influencing students' academic outcomes. Afterward, we examine the fairness of these approachs, evaluating whether particular student groups are disproportionately impacted by the predictions. Lastly, we explore these approaches' generalizability, assessing how well they maintain performance when applied to new, unseen data. 

\subsection{Effectiveness in Early Predicting Academic Performance}\label{subsec:effectiveness_eval}

Since data and code from the reviewed prior research were inaccessible, we defined and compared our approaches against three baselines. The first baseline, 0R (Zero Rule), naively predicted that all students were high performers, reflecting the majority class. The second baseline, 1R-SVM (One Rule), was trained on a single feature---prior term GPA---using a Support Vector Machine (SVM), which was selected as the best-performing model from a comparison of eight classical machine learning models\footnote{Logistic Regression, Support Vector Machine (SVM), K-Nearest Neighbor~\cite{aha1991instance}, AdaBoost~\cite{freund1997decision}, Random Forest~\cite{breiman2001random}, XGBoost~\cite{chen2016xgboost}, Gradient Boosting~\cite{friedman2001greedy}, and Decision Tree.}. Lastly, we re-implemented the Long Short-Term Memory (LSTM) model from prior work~\cite{chen2020utilizing}, which is the only  previous research that evaluated the generalizability of their academic performance prediction model. All approaches were evaluated using seven performance metrics: accuracy, precision, recall, F1 score, and AUC, alongside two metrics tailored for imbalanced data: kappa~\cite{mchugh2012interrater} and balanced accuracy~\cite{brodersen2010balanced}.

To recap, both our LR and 1D-CNN approaches used the 2018 dataset as an ``experimental'' set to fine-tune their modeling pipelines, addressing factors such as missing value imputation and class imbalance handling. Once the pipelines were finalized, they were applied to the 2019 dataset without modification, ensuring consistent testing across both years. The MTL-1D-CNN approach, however, differs in its setup, as it utilized both the 2018 and 2019 datasets to train and test two tasks, focusing primarily on improving generalizability. Given this distinction, we focus our performance evaluation on the LR and 1D-CNN approaches. 

\begin{table}[htb!]
 \caption{Performance of the LR and 1D-CNN approaches on two years of data, with 0R (Zero Rule), 1R-SVM (One Rule), and a re-implemented LSTM model as baselines. Results are sorted by \textbf{Balanced accuracy}. The high performance of the two approaches on both the 2018 and 2019 datasets demonstrates that accurate early prediction is \textit{possible}. The highest-performing approach, based on Balanced Accuracy, is highlighted in \textbf{bold}.}
 \renewcommand{\arraystretch}{1.1}
   \resizebox{\textwidth}{!}{
    \begin{tabular}{c l l r r r r r r r}
        \hline
        \hline
         \textbf{Year} &  \makecell[c]{\textbf{Earliest} \\ \textbf{predictable time}} & \makecell[c]{\textbf{Approach}} &  \textbf{Accuracy} & \textbf{Precision} & \textbf{Recall} & \textbf{F1} & \textbf{AUC} & \textbf{Kappa} & \makecell[c]{\textbf{Balanced}\\ \textbf{ accuracy}}\\\hline
         2018 & any time in Spring term & 0R (Zero Rule) & 0.771 & 0.771& 1.000 & 0.871& 0.500& 0.000 & 0.500  \\
         2018 & wk1 in Spring term & LSTM (\cite{chen2020utilizing}) & 0.737 & 0.833 & 0.769 & 0.800 & 0.417 & 0.417 & 0.718  \\
        2018 & before Spring term & 1R-SVM (One Rule) & 0.766 & 0.955 & 0.731 & 0.828 & 0.834 & 0.481 & 0.807\\\hdashline 
         2018 & wk1 in Spring term & LR (Our Approach) & 0.915 & 0.901 & \textbf{1.000} & 0.948 & 0.962 & 0.722 & 0.814\\
         2018 & wk1 in Spring term & \textbf{1D-CNN (Our Approach)} & \textbf{0.948} & \textbf{0.958} & 0.975 & \textbf{0.966} & \textbf{0.987} & \textbf{0.852} & \textbf{0.918}\\
         \hline
         2019 & wk1 in Spring term & LSTM (\cite{chen2020utilizing}) & 0.611 &  0.769 & 0.714 & 0.741 & 0.482 & -0.033 & 0.482 \\
         2019 & any time in Spring term & 0R (Zero Rule) & 0.679 & 0.679 & 1.000 & 0.809 & 0.500 & 0.000 & 0.500 \\
         2019 & before Spring term & 1R-SVM (One Rule) & 0.668 & 0.815 & 0.662 & 0.730 & 0.682 & 0.312 & 0.672 \\\hdashline
         2019 & wk1 in Spring term & LR (Our Approach) & 0.893  & 0.894 & \textbf{0.955} & 0.924 & 0.796 & 0.745 & 0.858  \\
        2019 & wk1 in Spring term & \textbf{1D-CNN (Our Approach)} & \textbf{0.898}  & \textbf{0.901} & 0.955 & \textbf{0.927} & \textbf{0.866} & \textbf{0.758} & \textbf{0.866}  \\\hline
         \hline
    \end{tabular}}
    \label{tab:wk1_metrics}
\end{table}
\renewcommand{\arraystretch}{1.0}

\subsubsection{Evaluation Results} As shown in  Table~\ref{tab:wk1_metrics}, both the LR and 1D-CNN approaches demonstrated high performance in predicting academic outcomes, with similar results observed across both 2018 and 2019 data, based on information collected no later than the first week of the Spring term. Our results from the 1D-CNN approach are comparable to the previous earliest prediction study~\cite{sano2015recognizing} that used data collected over the first four weeks of the term, achieving a similar average accuracy of 92\% but with predictions made \textit{three weeks earlier}.
Both approaches outperformed the 0R, 1R-SVM, and LSTM baselines. 

The robustness of both the LR and 1D-CNN approaches across the 2018 and 2019 datasets provides strong evidence that early predictions of student performance, using data available no later than the first week of the term, are \textit{possible}. However, as emphasized earlier, predictive models designed for real-world applications in academic performance must also account for their social and ethical implications. In the following sections, we evaluate the explainability, fairness, and generalizability of all three approaches.

\subsection{Explainability Evaluation}\label{subsec:explainability_eval}
We assess the explainability of the three approaches by examining how effectively they provide information that is interpretable and useful in understanding academic performance, especially their potential to offer actionable insights.

\subsubsection{Evaluation Results} By its nature, the LR approach offers high interpretability, as it directly maps features to outcomes through easily interpretable coefficients. To aid in understanding which features most influenced academic performance predictions, we applied feature importance ranking~\cite{bhatt2020explainable}. Given that feature selection in the LR pipeline occurred iteratively during LOSO-CV, each selected feature was assigned multiple importance scores across iterations. We summed these importance scores to generate final rankings for each feature. In total, the model selected 49 features for the 2018 dataset and 35 features for the 2019 dataset, associated with end-of-term GPA in the first week of the Spring term. We report the top 30 features based on their final importance scores for each year, as shown in  Tables~\ref{tab:top30_features_2018} and \ref{tab:top30_features_2019}. 

An example to illustrate reading these tables: the top-ranked feature in  Table~\ref{tab:top30_features_2018} (first row) represents the change in \textit{number of unlock screen events per minute} (in the \textbf{Feature} column) at night (in the \textbf{Epoch} column) during the second half of the week (in the \textbf{Behavioral Change Indicator} column), with the strongest association with end-of-term GPA (in the \textbf{Rank} column). Specifically, higher variation in screen unlocks during this period positively impacts a student's GPA (in the \textbf{Impact on GPA} column).

\begin{table}[htb!]
\caption{Top 30 selected features in the first week of the 2018 Spring term. The \textbf{Feature} column lists the top 30 features selected by the LR approach, ranked by their importance in the \textbf{Rank} column. The \textbf{Impact on GPA} column shows each feature’s weight, indicating its influence on GPA prediction: (+) for a positive association and (-) for a negative one. The \textbf{Agg.} column specifies the statistical metric (\eg average, standard deviation) used to compute each feature within an \textbf{Epoch}. The \textbf{Behavioral Change Indicator} column identifies if the feature represents a behavioral change associated with GPA, detailing when this change occurs (\eg in the first or second half of the week, across the full week, or around a breakpoint). The \textbf{Bkp.} column marks the specific day indicating a change in weekly behavioral trends, with additional slopes noted for changes before and after the breakpoint, where relevant.}
\renewcommand{\arraystretch}{1.1}
\footnotesize
   \resizebox{\textwidth}{!}{
\begin{tabular}{c|lllllcc}
\hline
 \hline
 & \multicolumn{1}{c}{\textbf{Feature}} &
 \multicolumn{1}{l}{\textbf{Agg.}}  & \multicolumn{1}{l}{\textbf{Epoch}} & \multicolumn{1}{c}{\makecell[c]{\textbf{Behavioral Change}\\  \textbf{Indicator}}} & 
 \multicolumn{1}{c}{\textbf{Bkp.}} &
 \multicolumn{1}{c}{\makecell[c]{\textbf{Impact}\\  \textbf{on GPA}}} & \multicolumn{1}{c}{\textbf{Rank}}  \\ \hline
 \parbox[t]{2mm}{\multirow{23}{*}{\rotatebox[origin=c]{90}{\textbf{Passive behavioral data}}}}
 &  num of unlock screen events per minute      &                      & night     & second-half slope    &  & 1.941 (+) & 1                        \\
 & num of steps taken per bout among all active bouts  & min  & night    & slope all &    &  1.414 (-) & 2  \\
& duration of phone remaining unlocked & min               & evening   & breakpoint  &  Wed   & 1.217 (-)                    & 5 \\
&  shortest duration of staying awake & avg  & 24hr    &    &    & 1.205 (+)                     & 6          \\
&  duration of all phone interactions & sum             & night     & slope before breakpoint  &  Th  & 1.181 (-)                    & 7                        \\
&  duration of sedentary bouts & std &   24hr    & breakpoint   &  Th  & 1.103 (+)                     & 8                        \\
&  time spent at living places & avg &   24hr    & slope after breakpoint   &  Th  &       1.028 (-)               & 9                        \\
&  duration of awake &             & 24hr    & breakpoint  &  Th  &                  1.014 (-)   &   10                      \\ 
&  time spent at third-ranked location cluster   &        & 24hr & slope before breakpoint  &  Wed  &            0.992 (-)        & 11                        \\ 
&  percentage of time in class & avg   & 24hr    &        &          & 0.952 (+)                     & 14                         \\ 
&  num of scans of bluetooth devices owned by others  &   & evening   & second-half slope &    &    0.940 (+)                  & 15                        \\
& duration of restless & max              & 24hr   & slope after breakpoint  &   Th  &              0.889 (-)       &          18  \\ 
&  time spent at local location clusters & std & morning    & second-half slope  &   &           0.883 (-)          & 19           \\ 
& regularity in circadian movement    &  std  &    evening      &     second-half slope   &        &    0.878 (+)       & 20                       \\
&  time spent at exercise places in minutes &  &   24hr    & first-half slope   &     &             0.831 (+)         & 22                        \\
& normalized entropy of local location clusters &  &   24hr    & slope after breakpoint   &  Th  &   0.829 (+)    & 23                        \\
&  time spent at first-ranked location cluster  &  &   evening    & slope after breakpoint   &  Th  &   0.819 (-)   & 24                       \\
&  time spent at living places & min &   24hr    & slope after breakpoint   &   Wed  &             0.818 (-)      & 25                       \\
&  num of active bouts & std &   morning    &   &    &             0.807 (-)         & 26                      \\
&  time spent at local location clusters & max & morning    & breakpoint  &   Th  &          0.801 (+)           & 27          \\ 
&  time spent at second-ranked location cluster &  & morning    & slope after breakpoint  & Th  &        0.784 (-)             & 28         \\ 
&  duration of time spent at living places in minutes &  &   24hr    & second-half slope   &    &             0.775 (-)        & 29                        \\
&  time spent at second-ranked location cluster &  & morning    & breakpoint  & Th  &           0.750 (-)        & 30         \\\hline
\parbox[t]{2mm}{\multirow{7}{*}{\rotatebox[origin=c]{90}{\textbf{Self-report}}}} 
& got lower grades than expected      &                             &          &      &  & 1.319 (-)   & 3                         \\
 & type of service provider &                                       &          &        &          & 1.315 (-)                   & 4 \\
& experienced helplessness in a difficult situation  &    &     &      &              & 0.982 (-)                    & 12                        \\
& had problems with partners       &                            &         &       &    &          0.962 (-)           & 13                         \\
& how unhappy if actual GPA is lower than expected   &           &          &     &     & 0.913 (-)                     & 16                        \\ 
& understand less than others  about their school    &    &          &        &        &   0.896 (-)                    & 17                        \\
& felt weak all over   &    &          &        &        &  0.850 (-)    & 21                        \\\hline\hline
    \end{tabular}}
    \label{tab:top30_features_2018}
\end{table}
\begin{table}[htb!]
\caption{Top 30 selected features in the first week of the 2019 Spring term. 
}
\renewcommand{\arraystretch}{1.1}
\footnotesize
   \resizebox{\textwidth}{!}{
\begin{tabular}{c|lllllcc}
 \hline\hline
 & \multicolumn{1}{c}{\textbf{Feature}} &
 \multicolumn{1}{l}{\textbf{Agg.}}  & \multicolumn{1}{l}{\textbf{Epoch}} & \multicolumn{1}{c}{\makecell[c]{\textbf{Behavioral Change}\\  \textbf{Indicator}}} & 
 \multicolumn{1}{c}{\textbf{Bkp.}} &
 \multicolumn{1}{c}{\makecell[c]{\textbf{Impact}\\  \textbf{on GPA}}}& \multicolumn{1}{c}{\textbf{Rank}}  \\ \hline
 \parbox[t]{2mm}{\multirow{28}{*}{\rotatebox[origin=c]{90}{\textbf{Passive behavioral data}}}}
 & percentage of time spent at outlier locations   &                        & night     & slope before breakpoint  & Fri &                  1.502 (-)   & 1                         \\
 & num of location bouts at greens   &                        & evening     & slope before breakpoint  & Th &                    1.271 (+)  & 2                         \\
& num of location bouts with duration >= 10 mins at living places   &                        & evening     & slope before breakpoint  & Th &                  1.203 (+)    & 3                        \\ 
& num of location bouts with duration >= 30 mins at exercise places   &             & evening     & second-half slope  &  &                    1.138 (-)  & 4                       \\
& num of location bouts at exercise places   &   avg          & evening     & second-half slope  &  & 1.030 (-) & 5       \\
& num of location bouts with duration >= 30 mins at living places   &                        & 24hr     & second-half slope  &  &                  1.028 (-)    & 6                        \\ 
& time spent at second-ranked global location cluster   &                  & 24hr     & slope after breakpoint       &    Fri   &                    0.959 (-)  & 7                        \\
& time spent at second-ranked location cluster   &                  & night    & breakpoint      &   Fri    &                    0.959 (-)  & 8                        \\
& party duration in minutes   &                  & night    & breakpoint      &    Fri   &                   0.949 (-)   & 9                        \\
& num of scans of all self-owned bluetooth devices  & std               & afternoon & first-half slope     &      &                    0.937 (-)  & 10 \\
& duration of location bouts at living places  & min               & morning & second-half slope     &       &                   0.897 (-) & 11 \\
& num of location bouts with duration >= 10 mins at food places   &             & evening     & second-half slope  &   &                  0.864 (+)    & 12                   \\
& duration of location bouts at exercise places  & std               & morning & breakpoint    &    Th   &                   0.849 (+) & 13 \\
& normalized entropy of local location clusters  &                         & morning   & second-half slope     &     &                  0.798 (-)   & 14                         \\
& duration of location bouts at Greek houses  & max             & night & slope after breakpoint    &   Fri    &                  0.787 (+)  & 15 \\
& duration of location bouts at living places  & max             & night & first-half slope    &       &    0.765 (+)   & 16 \\
& percentage of time spent at greens   &                        & afternoon    & slope before breakpoint  & Th &                  0.764 (+)    & 17                        \\
& duration of awake in minutes   &       avg                 & 24hr    &  &  &                  0.740 (-)   & 18                        \\
& num of scans of all self-owned bluetooth devices  & avg               & morning & first-half slope     &      &                    0.734 (-)  & 19 \\
& duration of phone remaining unlocked  & min              & afternoon & breakpoint    &   Th    &     0.727 (-)                 & 20 \\
& percentage of time spent near home (within 10m)   &              & morning & first-half slope    &       &                   0.708 (-)   & 21 \\
& party duration in minutes   &       avg           & night    &      &       &                  0.692 (-)    & 22                       \\
& num of scans of least frequently scanned bluetooth device of others &  & 24hr    & breakpoint        &   Fri      &                   0.683 (-)   & 23                        \\ 
& num of location bouts at food places   &             & evening     &  slope after breakpoint & Th  & 0.682 (+) & 24      \\
& num of location bouts with duration >= 10 mins outside   &                        & evening     & slope all  &  &                    0.636 (+)  & 26                        \\  
& percentage of time spent near home (within 100m)   &              & 24hr & breakpoint   &    Th   &                    0.626 (+)  & 27 \\
& num of location bouts with duration >= 30 mins at exercise places   &             & evening     & slope all &  &                   0.620 (-)   & 28                       \\ 
& time spent at second-ranked cluster in minutes  &             & evening     & slope before breakpoint & Th &                 0.605 (-)     & 30                       \\ \hline
 \parbox[t]{6mm}{\multirow{2}{*}{\rotatebox[origin=c]{90}{\makecell[c]{\textbf{Self-}\\  \textbf{report}}}}} 
 
 & had trouble sleeping because of pain    &                                  &          &          &                  &                      0.640 (-) &      25                   \\
  & had traumatic experiences    &                                  &          &          &                  &                    0.619 (-)  & 29   \\
 \hline\hline
    \end{tabular}}%
    \label{tab:top30_features_2019}
\end{table}

In contrast to the LR approach, the 1D-CNN and MTL-1D-CNN approaches, which are based on deep learning techniques, capture more intricate and complex patterns in the data. This capability allows them to model the sequential and temporal dependencies that are essential in behavioral data. However, this comes at the cost of reduced transparency, which makes it challenging to interpret the specific contribution of individual features to the predictions. This ``black box'' nature of deep learning models introduces difficulties in explaining how certain behaviors influence the outcomes, thereby complicating the interpretability and explainability of these approaches.

\subsubsection{Behavioral Patterns Associated with Academic Performance}\label{subsubsec:behavioral_patterns}  We analyzed key in-class and outside-classroom behavioral patterns and self-reported factors associated with academic performance by grouping the top features in  Tables~\ref{tab:top30_features_2018} and \ref{tab:top30_features_2019}. Interestingly, we observed a greater number of relevant outside-classroom behaviors than in-class behaviors. Additionally, behavioral shifts frequently occurred on Thursdays in both years (see \textbf{Bkp.} column in  Tables~\ref{tab:top30_features_2018} and \ref{tab:top30_features_2019}), indicating that students' weekend behaviors may begin on Fridays rather than Saturdays for a considerable portion of the population. For this analysis, we thus distinguish between ``weekday'' and ``weekend'' behaviors starting on Fridays. Throughout this paper, features are referenced by year and rank for consistency. For example, 2018-R14 (\textit{percentage of time in class}) refers to a feature ranked 14th in 2018 ( Table~\ref{tab:top30_features_2018}). While associations noted here are not causal, we summarize the observed academic-related behavioral patterns below, along with implications for early prediction. Future research should continue to investigate these patterns to clarify their role in early academic intervention. 

Among in-class behaviors, class attendance during the first week of the Spring term is associated with end-of-term GPA, showing that higher attendance correlates positively with academic outcomes (2018-R14). This emphasizes the importance of early engagement in academic activities and suggests that supporting students in establishing consistent attendance patterns could be an effective early intervention strategy. Interestingly, although study duration and study focus time were included in the training process, these factors did not appear among the top predictors. This absence suggests that attendance might capture a broader engagement factor, while study-specific metrics may require more context or extended observation to reveal their impact on academic performance.

Among outside-classroom behaviors, several patterns were significantly associated with end-of-term GPA. For instance, phone usage shows contrasting effects depending on timing: weekday phone use is negatively associated with GPA (2018-R7, 2019-R19), possibly reflecting distractions during school times, while phone use on weekends shows a positive association with GPA (2018-R1, 2018-R15), perhaps serving as a way for students to unwind after the week. Similarly, time spent at exercise locations during weekdays positively correlates with academic performance (2018-R22), aligning with findings that physical activity supports academic performance~\cite{al2016physical}, but this association turns negative when exercise occurs in the evenings on weekends (2019-R4, 2019-R5), possibly suggesting that late-weekend exercise may disrupt academic focus for the upcoming week. Sleep patterns are also crucial; poor quality, frequent wakefulness, and all-nighters predict lower GPAs (2018-R18, 2019-R18), consistent with research linking sleep quality to academic success. Notably, short wakefulness periods positively associate with GPA (2018-R6), a finding that warrants further exploration as it contrasts with studies emphasizing sleep consistency and quality~\cite{okano2019sleep}. 

In addition to behaviors, several serious self-reported stressors, including relationship issues (2018-R13), health concerns (2018-R21, 2019-R25), and traumatic experiences (2019-R29), were strongly linked to lower GPAs. Academic-related stressors like underperforming in a prior term (2018-R3) or achieving a lower GPA than expected (2018-R16) were also associated with academic decline, aligning with previous studies on the negative impact of stress on academic outcomes~\cite{de2016relationship,pereira2018types}. These findings suggest that early mental health support and academic counseling at the start of the term could help students better manage stress and improve resilience, benefiting their academic performance. For a deeper exploration of these patterns, refer to  Appendix~\ref{appx:behavioral_patterns}.

\subsection{Fairness Evaluation} 
We assessed fairness, specifically algorithmic bias, across four protected traits: race, first-generation status, gender, and sexual orientation. To evaluate whether the three approaches exhibit biases against these protected groups, we employed three widely used fairness metrics: First, \textit{demographic parity}, which requires that the likelihood of a positive prediction is the same across protected and unprotected groups. Second, \textit{equalized odds}, which ensures that both groups have equal true positive and false positive rates. This metric addresses the limitation of demographic parity, where even a fully accurate classifier may be seen as biased if the actual ratio of positive outcomes differs between groups. Third, \textit{equal opportunity}, a less strict measure than equalized odds, which only requires that the true positive rates be equal across groups. The fairness evaluation was carried out using \textit{Fairlearn}~\cite{bird2020fairlearn}, a Python toolkit designed to assess and mitigate bias in machine learning models.

\begin{figure}
    \centering
    \includegraphics[width=1\linewidth]{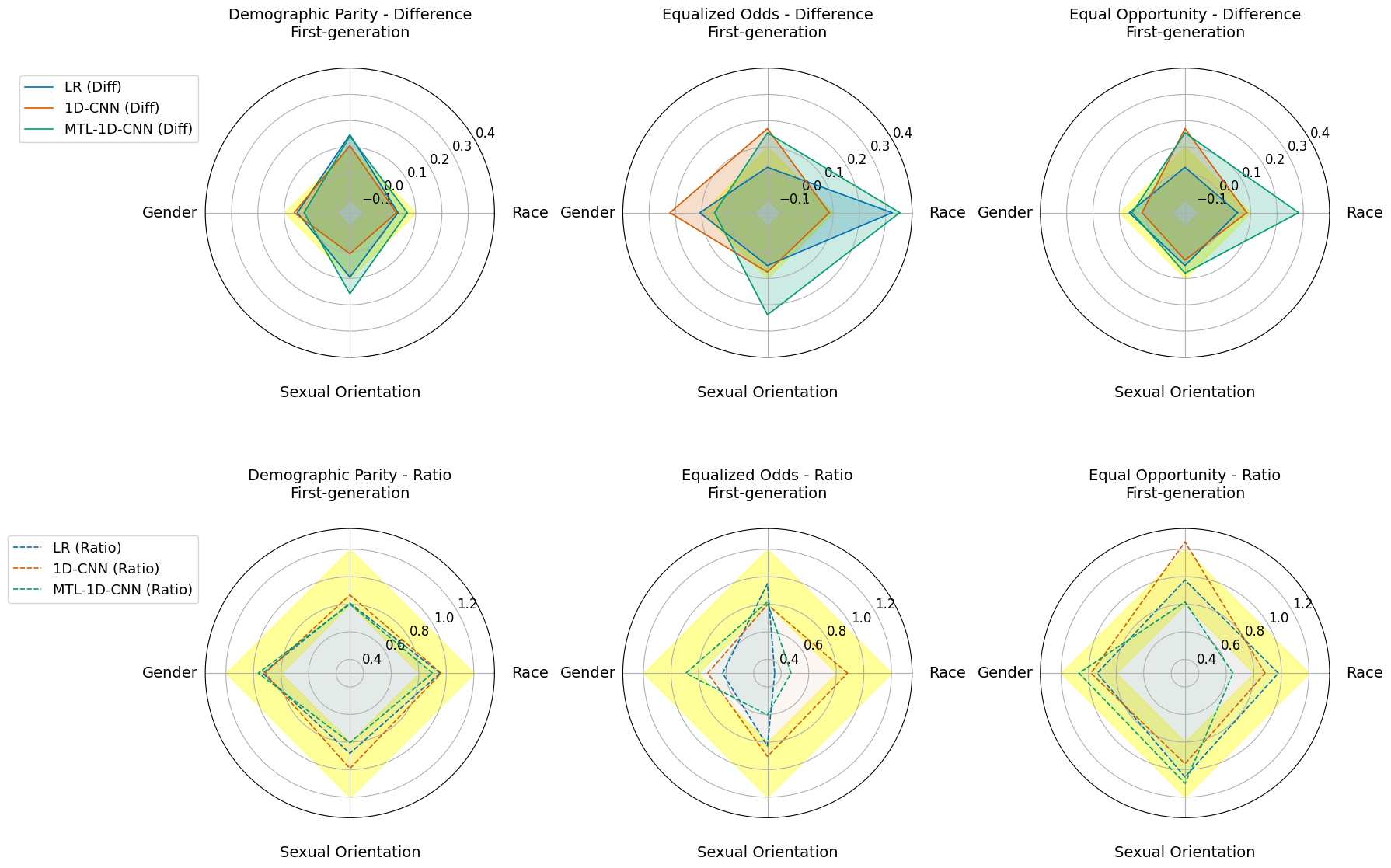}
    \caption{Radar charts comparing the fairness performance of three approaches (LR, 1D-CNN, MTL-1D-CNN) across four protected traits (race, first-generation, gender, sexual orientation) using three fairness metrics: demographic parity, equalized odds, and equal opportunity. The first row shows the difference between the protected traits, where the light yellow shaded regions indicate values between -0.1 and 0.1, representing a \textit{reasonable} fair difference. The second row shows the ratio, where the light yellow shaded regions highlight ratio values between 0.8 and 1.2, indicating a \textit{reasonable} fair performance.}
    \label{fig:fairness_eval}
\end{figure}

While much of the existing literature offers theoretical frameworks for fairness, practical guidelines on acceptable bias thresholds are less established. One exception is demographic parity, where a difference between -0.1 and 0.1, and a ratio between 0.8 and 1.2, is considered \textit{reasonable}~\cite{kobayashi2021one,feldman2015certifying,pessach2022review}. We extended these thresholds to our assessments of equalized odds and equal opportunity. Given that the 2018 dataset served primarily as an ``experimental'' dataset for refining our modeling pipelines, our fairness assessment focused on all three approaches on the 2019 dataset.  Figure~\ref{fig:fairness_eval} visualizes the fairness of each of the three approaches for each group, with values within the \textit{reasonable} range defined above highlighted in light yellow. Detailed fairness evaluation results can be found in  Table~\ref{tab:fairness_evaluation} in  Appendix~\ref{appx:fairness_eval}. 

\subsubsection{Evaluation Results Based on Demographic Parity} Both the LR and 1D-CNN approaches demonstrate generally fair performance for race, gender, and sexual orientation, with small differences and ratios close to 1, which are considered within a reasonable range for fairness. However, for first-generation status, these two approaches show larger differences, suggesting potential biases against this group. In contrast, the MTL-1D-CNN approach consistently shows larger differences for first-generation and sexual orientation, indicating that this approach might have more fairness issues for these traits. 

\subsubsection{Evaluation Results Based on Equalized Odds} The 1D-CNN approach again demonstrates reasonably fair performance for race and sexual orientation, with differences below 0.1 and ratios close to or above 0.9, suggesting that this approach predicts these traits fairly. However, for first-generation status and gender, the differences and ratios are less favorable, indicating potential fairness issues for these groups. The LR approach shows a similar pattern, with relatively better fairness for sexual orientation and first-generation status, but lower fairness for race and gender. In contrast, the MTL-1D-CNN approach, while performing better for gender, displays the largest differences and lowest ratios across most other traits, emphasizing more substantial biases in its predictions.

\subsubsection{Evaluation Results Based on Equal Opportunity} The LR approach demonstrates relatively fair treatment for all protected traits, with differences below 0.1 and ratios close to 1. The 1D-CNN approach also shows good performance for race, gender, and sexual orientation, with ratios approaching 1. However, MTL-1D-CNN continues to exhibit larger differences and lower ratios for race and first-generation, suggesting fairness issues for these traits.


\subsection{Generalizability Evaluation}

In  Section~\ref{subsec:effectiveness_eval}, we demonstrated that when training and testing an approach, including data pre-processing and modeling, using data from different students within the same year (a concept referred to as \textit{pipeline-level} generalizability~\cite{xu2019leveraging}), the LR and 1D-CNN approaches showed robust performance. However, for real-world applications, it is essential to assess whether a pre-trained model can generalize across different contexts (a concept reffered to as \textit{model-level} generalizability~\cite{xu2019leveraging}), such as applying the model to data from a different year or institution in academic performance prediction. In addition, for real-world academic performance prediction, the most useful models are those that can accurately predict not only students who consistently perform well or poorly but also those who may experience significant changes in performance, such as transitioning from high to low performers. 

To assess these aspects of generalizability, we tested our three approaches in two ways. First, we evaluated \textit{model-level} generalizability by training the LR and 1D-CNN approaches on 2018 data and testing them on unseen 2019 data. The MTL-1D-CNN approach, by its design, was assessed on both the 2018 and 2019 datasets, leveraging the multi-task learning framework to test its generalization ability across different years. Second, we analyzed the approaches' accuracy in making predictions for students whose performance remained stable from the Winter term to the following Spring term (\ie consistently high or low performers) and those whose performance shifted (\ie transitioning from high to low performers, or vice versa). Specifically, we categorized students into four categories: those who remained high performers (111 participants), those who remained low performers (29 participants), those who improved to high performers (22 participants), and those who declined to low performers (34 participants). For each approach, we then calculated the percentage of accurately predicted outcomes within these four categories.

\renewcommand{\arraystretch}{1.2}
\begin{table}[h!]
\caption{Performance of approaches trained on 2018 data and tested on 2019 data compared to three baselines (seperated by the dashline) to predict end-of-term GPA. Results are sorted by \textbf{Balanced accuracy}. The results can indicate \textit{model-level} generalizability of each approach. The MTL-1D-CNN approach significantly outperformed the LSTM baseline. The highest-performing approach, based on Balanced Accuracy, is highlighted in \textbf{bold}.}
\scalebox{0.85}{
    \centering
    \begin{tabular}{l r r r r r r r}
        \hline
        \hline
         \makecell[c]{\textbf{Approach}} &  \textbf{Accuracy} & \textbf{Precision} & \textbf{Recall} & \textbf{F1} & \textbf{AUC} & \textbf{Kappa} & \makecell[c]{\textbf{Balanced}\\ \textbf{ accuracy}}\\\hline
         0R (Zero Rule) & 0.679 & 0.679 & 1.000 & 0.809 & 0.500 & 0.000 & 0.500 \\
          LSTM (\cite{chen2020utilizing}) & 0.633 & 0.719 & 0.752 & 0.735 & 0.566 & 0.136 & 0.566 \\
         1R-SVM (One Rule) &  0.668 & 0.815 & 0.662 & 0.730 & 0.677 & 0.312 & 0.672 \\\hdashline
         LR (Our Approach) & 0.679 & 0.679 & 1.000 &  0.809 & 0.652 & 0.000 & 0.500 \\
         1D-CNN (Our Approach) &  0.673 &  0.732 & 0.820 &  0.773 &  0.592 & 0.592 & 0.559 \\
         \textbf{MTL-1D-CNN (Our Approach)}  & \textbf{0.745} & \textbf{0.817} & \textbf{0.805} & \textbf{0.811} & \textbf{0.712} & \textbf{0.420} & \textbf{0.712} \\\hline
         \hline
    \end{tabular}}
    \label{tab:model_generalizability}
\end{table}
\renewcommand{\arraystretch}{1.2}

\subsubsection{Model-level Generalizability Evaluation Results}
 Table~\ref{tab:model_generalizability} presents the \textit{model-level} generalizability performance of our three approaches across different years, with the MTL-1D-CNN approach achieving the highest scores across all evaluated metrics. However, neither the LR nor 1D-CNN approaches outperformed the three baseline models, highlighting the challenges in achieving strong generalizability across multiple contexts.

\subsubsection{Consistency and Transition Evaluation Results}

\begin{figure}[htb!]
\scalebox{0.70}{
    \centering
\includegraphics[scale=0.55]{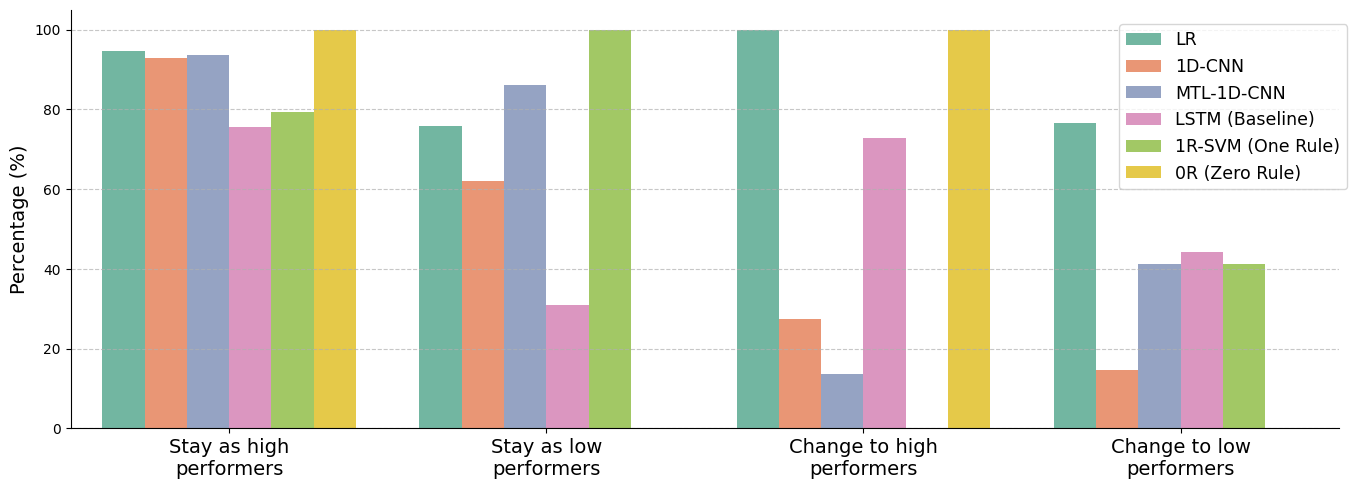}}
\caption{Accuracy of three approaches as well as the baselines in predicting academic performance consistency and transitions. It presents the percentage of each approach accurately predicting four categories: remained a high performer, remained a low performer, improved from low to high performer, and declined from high to low performer.} 
\label{fig:model_comparison}
\end{figure}

 Figure~\ref{fig:model_comparison} presents a comparison of the approaches' performance in predicting both consistent and transitioning student outcomes. For the ``Stay as high performers'' category, the 0R baseline, which naively predicts all students as high performers, achieved 100\% accuracy as expected. Assessing our approaches, all three performed quite well in identifying students who remained high performers, with accuracy rates exceeding 90\%. In the ``Stay as low performers'' category, all three approaches outperformed the LSTM and 0R baselines, with the MTL-1D-CNN approach achieving an accuracy above 80\%. Interestingly, the prior term GPA (1R-SVM) baseline was especially effective for this group, achieving 100\% accuracy, indicating that prior academic performance serves as a strong predictor for students who consistently perform poorly. 

For the ``Change to high performers'' category, the 0R baseline once again achieved perfect accuracy, by its design. Surprisingly, the LR approach also achieved perfect accuracy in this category, demonstrating its effectiveness in identifying students who improved their performance. However, both deep learning approaches, 1D-CNN and MTL-1D-CNN, showed weaker performance, with accuracy below 35\%. When predicting students who transitioned from high to low performance, the LR approach performed the best, with accuracy above 75\%, indicating its strength in detecting drops in performance. However, the deep learning approaches and the baselines performed less effectively, with most accuracies around 40\%, reinforcing our earlier observations that deep learning models struggle with predicting performance transitions compared to simpler models.

\section{Discussion}\label{sec:discussion}

Our study examines how predictive models can be integrated into collaborative student support systems. Effective academic performance prediction is not just a technical challenge but a socio-technical problem that involves multiple stakeholders. Below, we discuss the insights gained from developing and evaluating early academic performance prediction models, focusing on the importance of HCML principles, the trade-offs involved in balancing these aspects, and the scenarios where each approach may be most beneficial for different stakeholders. We also highlight the need for data that supports short-term interventions, summarizing insights learned from analyzing behavioral patterns along with their implications for early interventions. Following this, we discuss the technical insights gained from our study. Finally, we examine the broader implications of our findings and propose future directions for designing predictive models in collaborative student support systems.  

\subsection{HCML Principles and Trade-offs in Academic Performance Prediction Models}\label{subsec:tradeoffs}

The development of academic performance prediction models requires careful attention to HCML principles, particularly their social and ethical implications, to ensure successful implementation in real-world educational settings. Addressing key aspects such as explainability, fairness, and generalizability is essential for fostering trust and usability among key stakeholders, including educators, students, and policymakers~\cite{doshi2017towards,ehsan2020human,mehrabi2021survey,holstein2019improving,caruana1997multitask,xu2023globem}. While our initial goal was to develop a single model that could balance these three aspects, we found that achieving such a balance is more complex than anticipated. To better understand the inherent trade-offs, we explored three approaches (LR, 1D-CNN, and MTL-1D-CNN) using existing ML/DL techniques. 

The LR approach offers the highest level of explainability and demonstrates reasonable fairness, particularly for protected groups such as sexual minorities and first-generation college students. However, its generalizability shows mixed results across different contexts. It performs reliably when predicting students whose academic performance remains stable or changes significantly from one term to the next, especially in identifying students who experience a significant drop in performance (high performers transitioning to low performers). However, its performance declines when applied to unseen data---a common requirement in real-world applications. This approach may be most beneficial in scenarios where interpretability and fairness are prioritized over generalizability, such as single-institution studies identifying actionable early intervention factors. Its transparency enables students and educators to recognize relevant academic behaviors, supporting proactive interventions, while its fairness helps ensure interventions are designed equitably, benefiting diverse student groups without unintended bias.

The 1D-CNN approach also demonstrates reasonable fairness, but its fairness benefits different protected groups compared to the LR approach. When predicting academic performance within the same dataset, it achieves the highest prediction performance, with an average balanced accuracy of 89.2\% across 2018 and 2019. However, this approach struggles with generalizability, showing the worst performance when predicting both consistent and transitioning students and performing comparably worse than the LR approach when applied to new, unseen data. Additionally, its reliance on deep learning techniques reduces explainability, making it more difficult to interpret the influence of individual features on its predictions. Given its strengths and limitations, the 1D-CNN approach may be most suitable in scenarios that prioritize high prediction accuracy within a single, well-defined dataset, particularly when explainability and adaptability are not the primary concern. 

The MTL-1D-CNN approach demonstrates the highest generalizability among the three approaches, effectively adapting to new, unseen datasets and consistently predicting students experiencing sustained academic challenges across terms. However, similar to the 1D-CNN approach, its reliance on deep learning reduces explainability, limiting insights into specific behavioral factors influencing predictions. Furthermore, MTL-1D-CNN shows the lowest fairness performance, raising equity concerns for diverse student demographics. Given its strengths and limitations, the MTL-1D-CNN approach may be especially suited to broad, scalable applications where adaptability is crucial, such as multi-institution or multi-term deployments.

\subsection{Behavioral Data and Early Intervention Implications}\label{subsec:behavioral_patterns}

As briefly discussed in Section~\ref{subsubsec:explainability}, while some prior work has provided insights into factors such as personality traits~\cite{wang2015smartgpa,yao2019predicting} and students' interactions with OLS that contribute to academic performance~\cite{bravo2021early,waheed2023early}, these factors are often difficult to address through short-term interventions. Traits like personality tend to be stable over time, making them less actionable for immediate intervention efforts aimed at improving academic outcomes. Similarly, online behaviors such as visits to forums or number of attempts to access questionnaires, which are collected over extended periods, are challenging to intervene in without a deeper understanding of the motivations driving these behaviors. Therefore, focusing on more dynamic and modifiable factors, such as daily behaviors, could offer greater potential for timely and effective interventions.

The broader HCI community has leveraged passively sensed behavioral data to predict student academic performance and provide insights into daily academic-related behaviors~\cite{wang2015smartgpa}. However, previous research has generally relied on data collected across the entire term, which can delay the timeliness of interventions. In contrast, our work focuses on early-term data to identify key in-class and outside-classroom behavioral patterns, as well as self-reported factors, associated with student academic performance. Our findings (see  Section~\ref{subsubsec:behavioral_patterns}) reveal that a greater proportion of academic-related factors are linked to outside-classroom behaviors, highlighting the importance of capturing broader aspects of student life. Below, we further discuss the insights gained from analyzing these behavioral patterns and propose additional considerations for designing early interventions.

As described in  Section~\ref{subsubsec:behavioral_patterns}, we observed that students' academic-related weekend behaviors tend to begin on Fridays rather than Saturdays, suggesting that routines are generally stable from Monday to Thursday, with shifts starting as early as Friday. Interestingly, this pattern aligns with findings in mood-related studies, where a significant mood shift often occurs on Friday evenings~\cite{stone2012day}, signaling a transition into weekend-related behavior and attitudes. This similarity suggests that academic and mood-related routines might both be influenced by the anticipation of the weekend, highlighting Fridays as a potentially critical point for early interventions. Recognizing this shift allows educators and support programs to consider targeted strategies at the end of the week, when students may benefit from reminders to maintain academic habits or engage in well-being activities before the weekend begins.

Additionally, our analysis shows the importance of time granularity in understanding how behaviors impact academic performance. The effects of some  behaviors, such as phone usage or time spent exercising, vary depending on the day and time: for instance, phone usage negatively correlates with academic outcomes on weekdays but shows a positive association over the weekend. Similarly, exercise during weekday daytime aligns positively with academic performance, whereas evening exercise on weekends shows a negative relationship. These findings on the timing of behaviors provide critical insights that can enhance short-term interventions by addressing when particular behaviors are more or less beneficial for students. This also highlights the need for collecting continuous data to accurately capture and interpret these time-sensitive patterns.

\subsection{Technical Insights Learned from Our Experiments}

For real-world academic performance prediction applications to be meaningful and ethical, they must address explainability, fairness, and generalizability simultaneously. Our study highlights the complexity of achieving this balance in a single model, underscoring the need for further research in each. Explainability remains a challenge, particularly with DL models that work with high-dimensional behavioral data. Existing interpretability techniques, such as permutation feature importance~\cite{nepal2022covid}, have shown potential but are computationally costly for datasets with thousands of features. As DL models see broader adoption in educational prediction contexts, advancing methods, such as SHAP~\cite{lundberg2017unified} and LIME~\cite{ribeiro2016should}, their interpretability is crucial to enabling actionable insights.

Fairness remains an essential yet under-explored consideration in educational predictive models. Our results show that none of the approaches fully achieves fairness across all protected groups, highlighting the need for effective mitigation strategies. Future work should consider three main types of fairness techniques: pre-processing, which modifies data before model training to reduce bias~\cite{kamiran2012data,feldman2015certifying}; in-processing, which adjusts the learning algorithm itself to enhance fairness~\cite{zhang2018mitigating,zafar2017fairness}; and post-processing, which corrects biases in model predictions after training~\cite{hardt2016equality,kamiran2012decision}. Moreover, a critical gap persists in the absence of clear, practical guidelines for validating fairness. Although some efforts have evaluated fairness on real datasets~\cite{awasthi2021evaluating}, concrete standards for what constitutes reasonable fairness are still lacking. Prior work mentions ``slack approximations,'' or ranges of acceptable demographic parity~\cite{delobelle2021ethical}, but more robust, actionable criteria are essential for practical research and deployment.

Generalizability also poses a challenge. Although the ideal goal is to develop a robustly generalizable model, our findings reveal a considerable gap in achieving reliable generalizability. A possible limitation is the exclusion of features closely tied to GPA that were inconsistent across different datasets, either because they were unique to one dataset or completely missing in the other. For instance, \textit{type of service provider} (2018-R4) was excluded from our most generalizable approach (MTL-1D-CNN) as it was not collected in 2019. Lacking a top selected feature like this could significantly affect the model's performance and generalizability, and could be addressed at data collection time without much burden to participants. In addition, introducing other tasks to our MTL approach to either replace the prediction of prior term GPA or to learn more than two tasks in parallel may be of interest for future work, to improve generalizability. Furthermore, we acknowledge the presence of returning students from 2018 to 2019, which may influence result reliability. Future work should consider testing their models/approaches in different contexts to validate trustworthiness and applicability.


\subsection{Implications for Collaborative Student Support Systems}

Predictive models for academic performance are not just technical tools; they function within broader systems where students, educators, advisors, and institutions collaborate to support student success. For these models to be effective in real-world educational settings, they must go beyond accuracy and consider how different stakeholders interpret and act on their outputs. The integration of predictive models into student support systems presents not only technical considerations but also social, material, and theoretical challenges. This study contributes to addressing these challenges by evaluating different modeling approaches, assessing their trade-offs, and exploring how behavioral data can improve early academic interventions.

A central technical challenge in predictive modeling is balancing predictive accuracy with HCML principles to ensure that models are both effective and usable. While deep learning models, such as 1D-CNN, achieve higher predictive performance, their complexity reduces transparency, making it difficult for stakeholders to interpret and apply the predictions. In contrast, the LR approach offers greater interpretability, allowing students and educators to understand how different factors contribute to predictions, but it struggles with generalizability when applied to new data. This trade-off between explainability and performance highlights a key technical challenge: how can predictive models be designed to optimize both accuracy and usability in multi-stakeholder decision-making? Future work should explore co-design approaches, where educators and institutional decision-makers are actively involved in defining explainability requirements for predictive models.

The integration of predictive models into educational support systems raises social challenges related to fairness and trust. Our results reveal variations in fairness performance across different student demographics, raising concerns about potential bias in AI-driven decision-making. While fairness-aware interventions exist~\cite{selbst2019fairness,zhang2023framework,aird2024dynamic}, they often involve trade-offs between model accuracy and equitable outcomes. Additionally, clear guidance on fairness metrics is still missing~\cite{zhang2024illuminating}. From a CSCW perspective, trust in predictive systems cannot be achieved solely through technical bias mitigation—it requires participatory approaches that involve students, educators, and policymakers in defining fairness criteria. Future research should explore human-in-the-loop approaches, where stakeholders collaborate to evaluate, interpret, and refine predictive insights to ensure they align with institutional values and student needs.

Existing academic prediction models rely heavily on sporadic, classroom-derived data, limiting their ability to capture holistic student behaviors~\cite{lu2018applying,qu2019predicting,sukhbaatar2019artificial}. Our study demonstrates that behavioral data, including sleep patterns, physical activity, and weekend study routines, can provide valuable early signals of academic performance. The findings suggest that integrating behavioral data allows for earlier and more context-aware interventions than traditional models relying solely on classroom engagement. However, the use of behavioral data introduces challenges regarding privacy, data ethics, and responsible data collection. We argue that future research should explore privacy-preserving machine learning techniques and participatory data governance models to ensure responsible data use in educational settings.

A key theoretical challenge in CSCW is how AI-driven insights can be effectively integrated into human-centered decision-making workflows. Additionally, given that different stakeholders often have varying needs, a broader question arises: how can AI systems align with diverse stakeholder goals? In Section~\ref{subsec:tradeoffs}, we discuss the potential use cases for different predictive models---for example, educators may prioritize interpretable models, while institutional administrators may favor generalizable solutions. Future research should foster close collaboration between model builders and stakeholders to ensure predictive models are designed with real-world needs in mind. Additionally, future work should explore collaborative AI frameworks that embed predictive models within student support ecosystems, enabling stakeholders to co-define intervention strategies informed by model predictions.
\section{Conclusion}

In this paper, we highlight three critical gaps in current academic performance prediction models that limit their applicability in real-world educational contexts. First, many models lack a human-centered approach that integrates social values alongside technical robustness. Second, these models frequently rely on data collected mid- or late-term, restricting timely opportunities for early intervention that could assist at-risk students before they encounter significant academic challenges. Finally, most models overlook continuous, actionable behavioral data that reflects students' broader daily activities, both within and beyond the classroom environment. To address these gaps, we explored three modeling approaches designed to address HCML principles of \textit{explainability}, \textit{fairness}, and \textit{generalizability} while providing early predictions and actionable insights for intervention. Our findings show that it is possible to accurately identify academic risks as early as the first week of the term and provide insights into academic-related behaviors and factors. However, achieving a balanced model across all three HCML principles remains challenging, underscoring the need for further research. We encourage continued exploration of these aspects---and additional dimensions beyond this study---to better serve diverse educational needs and support ethical, practical deployment.


\bibliographystyle{ACM-Reference-Format}
\bibliography{main} 


\begin{thebibliography}{171}


\ifx \showCODEN    \undefined \def \showCODEN     #1{\unskip}     \fi
\ifx \showDOI      \undefined \def \showDOI       #1{#1}\fi
\ifx \showISBNx    \undefined \def \showISBNx     #1{\unskip}     \fi
\ifx \showISBNxiii \undefined \def \showISBNxiii  #1{\unskip}     \fi
\ifx \showISSN     \undefined \def \showISSN      #1{\unskip}     \fi
\ifx \showLCCN     \undefined \def \showLCCN      #1{\unskip}     \fi
\ifx \shownote     \undefined \def \shownote      #1{#1}          \fi
\ifx \showarticletitle \undefined \def \showarticletitle #1{#1}   \fi
\ifx \showURL      \undefined \def \showURL       {\relax}        \fi
\providecommand\bibfield[2]{#2}
\providecommand\bibinfo[2]{#2}
\providecommand\natexlab[1]{#1}
\providecommand\showeprint[2][]{arXiv:#2}

\bibitem[Adler et~al\mbox{.}(2022a)]%
        {adler2022burnout}
\bibfield{author}{\bibinfo{person}{Daniel~A Adler}, \bibinfo{person}{Emily Tseng}, \bibinfo{person}{Khatiya~C Moon}, \bibinfo{person}{John~Q Young}, \bibinfo{person}{John~M Kane}, \bibinfo{person}{Emanuel Moss}, \bibinfo{person}{David~C Mohr}, {and} \bibinfo{person}{Tanzeem Choudhury}.} \bibinfo{year}{2022}\natexlab{a}.
\newblock \showarticletitle{Burnout and the quantified workplace: Tensions around personal sensing interventions for stress in resident physicians}.
\newblock \bibinfo{journal}{\emph{Proceedings of the ACM on Human-computer Interaction}} \bibinfo{volume}{6}, \bibinfo{number}{CSCW2} (\bibinfo{year}{2022}), \bibinfo{pages}{1--48}.
\newblock


\bibitem[Adler et~al\mbox{.}(2022b)]%
        {adler2022machine}
\bibfield{author}{\bibinfo{person}{Daniel~A Adler}, \bibinfo{person}{Fei Wang}, \bibinfo{person}{David~C Mohr}, {and} \bibinfo{person}{Tanzeem Choudhury}.} \bibinfo{year}{2022}\natexlab{b}.
\newblock \showarticletitle{Machine learning for passive mental health symptom prediction: Generalization across different longitudinal mobile sensing studies}.
\newblock \bibinfo{journal}{\emph{Plos one}} \bibinfo{volume}{17}, \bibinfo{number}{4} (\bibinfo{year}{2022}), \bibinfo{pages}{e0266516}.
\newblock


\bibitem[Aha et~al\mbox{.}(1991)]%
        {aha1991instance}
\bibfield{author}{\bibinfo{person}{David~W Aha}, \bibinfo{person}{Dennis Kibler}, {and} \bibinfo{person}{Marc~K Albert}.} \bibinfo{year}{1991}\natexlab{}.
\newblock \showarticletitle{Instance-based learning algorithms}.
\newblock \bibinfo{journal}{\emph{Machine learning}} \bibinfo{volume}{6}, \bibinfo{number}{1} (\bibinfo{year}{1991}), \bibinfo{pages}{37--66}.
\newblock


\bibitem[Ahern(2024)]%
        {ahern2024potential}
\bibfield{author}{\bibinfo{person}{Samantha~J Ahern}.} \bibinfo{year}{2024}\natexlab{}.
\newblock \showarticletitle{The potential and pitfalls of learning analytics as a tool for supporting student wellbeing}.
\newblock  (\bibinfo{year}{2024}).
\newblock


\bibitem[Ahmad et~al\mbox{.}(2020)]%
        {ahmad2020fairness}
\bibfield{author}{\bibinfo{person}{Muhammad~Aurangzeb Ahmad}, \bibinfo{person}{Arpit Patel}, \bibinfo{person}{Carly Eckert}, \bibinfo{person}{Vikas Kumar}, {and} \bibinfo{person}{Ankur Teredesai}.} \bibinfo{year}{2020}\natexlab{}.
\newblock \showarticletitle{Fairness in machine learning for healthcare}. In \bibinfo{booktitle}{\emph{Proceedings of the 26th acm sigkdd international conference on knowledge discovery \& data mining}}. \bibinfo{pages}{3529--3530}.
\newblock


\bibitem[Aird et~al\mbox{.}(2024)]%
        {aird2024dynamic}
\bibfield{author}{\bibinfo{person}{Amanda Aird}, \bibinfo{person}{Paresha Farastu}, \bibinfo{person}{Joshua Sun}, \bibinfo{person}{Elena Stefancov{\'a}}, \bibinfo{person}{Cassidy All}, \bibinfo{person}{Amy Voida}, \bibinfo{person}{Nicholas Mattei}, {and} \bibinfo{person}{Robin Burke}.} \bibinfo{year}{2024}\natexlab{}.
\newblock \showarticletitle{Dynamic fairness-aware recommendation through multi-agent social choice}.
\newblock \bibinfo{journal}{\emph{ACM Transactions on Recommender Systems}} \bibinfo{volume}{3}, \bibinfo{number}{2} (\bibinfo{year}{2024}), \bibinfo{pages}{1--35}.
\newblock


\bibitem[Al-Drees et~al\mbox{.}(2016)]%
        {al2016physical}
\bibfield{author}{\bibinfo{person}{Abdulmajeed Al-Drees}, \bibinfo{person}{Hamza Abdulghani}, \bibinfo{person}{Mohammad Irshad}, \bibinfo{person}{Abdulsalam~Ali Baqays}, \bibinfo{person}{Abdulaziz~Ali Al-Zhrani}, \bibinfo{person}{Sulaiman~Abdullah Alshammari}, {and} \bibinfo{person}{Norah~Ibrahim Alturki}.} \bibinfo{year}{2016}\natexlab{}.
\newblock \showarticletitle{Physical activity and academic achievement among the medical students: A cross-sectional study}.
\newblock \bibinfo{journal}{\emph{Medical teacher}} \bibinfo{volume}{38}, \bibinfo{number}{sup1} (\bibinfo{year}{2016}), \bibinfo{pages}{S66--S72}.
\newblock


\bibitem[Amershi et~al\mbox{.}(2019)]%
        {amershi2019guidelines}
\bibfield{author}{\bibinfo{person}{Saleema Amershi}, \bibinfo{person}{Dan Weld}, \bibinfo{person}{Mihaela Vorvoreanu}, \bibinfo{person}{Adam Fourney}, \bibinfo{person}{Besmira Nushi}, \bibinfo{person}{Penny Collisson}, \bibinfo{person}{Jina Suh}, \bibinfo{person}{Shamsi Iqbal}, \bibinfo{person}{Paul~N Bennett}, \bibinfo{person}{Kori Inkpen}, {et~al\mbox{.}}} \bibinfo{year}{2019}\natexlab{}.
\newblock \showarticletitle{Guidelines for human-AI interaction}. In \bibinfo{booktitle}{\emph{Proceedings of the 2019 chi conference on human factors in computing systems}}. \bibinfo{pages}{1--13}.
\newblock


\bibitem[Andersen et~al\mbox{.}(2023)]%
        {andersen2023introduction}
\bibfield{author}{\bibinfo{person}{Tariq~Osman Andersen}, \bibinfo{person}{Francisco Nunes}, \bibinfo{person}{Lauren Wilcox}, \bibinfo{person}{Enrico Coiera}, {and} \bibinfo{person}{Yvonne Rogers}.} \bibinfo{year}{2023}\natexlab{}.
\newblock \bibinfo{title}{Introduction to the special issue on human-centred AI in healthcare: Challenges appearing in the wild}.
\newblock , \bibinfo{numpages}{12}~pages.
\newblock


\bibitem[Arrieta et~al\mbox{.}(2020)]%
        {arrieta2020explainable}
\bibfield{author}{\bibinfo{person}{Alejandro~Barredo Arrieta}, \bibinfo{person}{Natalia D{\'\i}az-Rodr{\'\i}guez}, \bibinfo{person}{Javier Del~Ser}, \bibinfo{person}{Adrien Bennetot}, \bibinfo{person}{Siham Tabik}, \bibinfo{person}{Alberto Barbado}, \bibinfo{person}{Salvador Garc{\'\i}a}, \bibinfo{person}{Sergio Gil-L{\'o}pez}, \bibinfo{person}{Daniel Molina}, \bibinfo{person}{Richard Benjamins}, {et~al\mbox{.}}} \bibinfo{year}{2020}\natexlab{}.
\newblock \showarticletitle{Explainable Artificial Intelligence (XAI): Concepts, taxonomies, opportunities and challenges toward responsible AI}.
\newblock \bibinfo{journal}{\emph{Information fusion}}  \bibinfo{volume}{58} (\bibinfo{year}{2020}), \bibinfo{pages}{82--115}.
\newblock


\bibitem[Assari et~al\mbox{.}(2017)]%
        {assari2017racial}
\bibfield{author}{\bibinfo{person}{Shervin Assari}, \bibinfo{person}{Ehsan Moazen-Zadeh}, \bibinfo{person}{Cleopatra~Howard Caldwell}, {and} \bibinfo{person}{Marc~A Zimmerman}.} \bibinfo{year}{2017}\natexlab{}.
\newblock \showarticletitle{Racial discrimination during adolescence predicts mental health deterioration in adulthood: Gender differences among Blacks}.
\newblock \bibinfo{journal}{\emph{Frontiers in public health}}  \bibinfo{volume}{5} (\bibinfo{year}{2017}), \bibinfo{pages}{104}.
\newblock


\bibitem[Augner and Hacker(2012)]%
        {augner2012associations}
\bibfield{author}{\bibinfo{person}{Christoph Augner} {and} \bibinfo{person}{Gerhard~W Hacker}.} \bibinfo{year}{2012}\natexlab{}.
\newblock \showarticletitle{Associations between problematic mobile phone use and psychological parameters in young adults}.
\newblock \bibinfo{journal}{\emph{International journal of public health}} \bibinfo{volume}{57}, \bibinfo{number}{2} (\bibinfo{year}{2012}), \bibinfo{pages}{437--441}.
\newblock


\bibitem[Awasthi et~al\mbox{.}(2021)]%
        {awasthi2021evaluating}
\bibfield{author}{\bibinfo{person}{Pranjal Awasthi}, \bibinfo{person}{Alex Beutel}, \bibinfo{person}{Matth{\"a}us Kleindessner}, \bibinfo{person}{Jamie Morgenstern}, {and} \bibinfo{person}{Xuezhi Wang}.} \bibinfo{year}{2021}\natexlab{}.
\newblock \showarticletitle{Evaluating fairness of machine learning models under uncertain and incomplete information}. In \bibinfo{booktitle}{\emph{Proceedings of the 2021 ACM Conference on Fairness, Accountability, and Transparency}}. \bibinfo{pages}{206--214}.
\newblock


\bibitem[Banovic et~al\mbox{.}(2023)]%
        {banovic2023being}
\bibfield{author}{\bibinfo{person}{Nikola Banovic}, \bibinfo{person}{Zhuoran Yang}, \bibinfo{person}{Aditya Ramesh}, {and} \bibinfo{person}{Alice Liu}.} \bibinfo{year}{2023}\natexlab{}.
\newblock \showarticletitle{Being trustworthy is not enough: How untrustworthy artificial intelligence (AI) can deceive the end-users and gain their trust}.
\newblock \bibinfo{journal}{\emph{Proceedings of the ACM on Human-Computer Interaction}} \bibinfo{volume}{7}, \bibinfo{number}{CSCW1} (\bibinfo{year}{2023}), \bibinfo{pages}{1--17}.
\newblock


\bibitem[Barocas and Selbst(2016)]%
        {barocas2016big}
\bibfield{author}{\bibinfo{person}{Solon Barocas} {and} \bibinfo{person}{Andrew~D Selbst}.} \bibinfo{year}{2016}\natexlab{}.
\newblock \showarticletitle{Big data's disparate impact}.
\newblock \bibinfo{journal}{\emph{Calif. L. Rev.}}  \bibinfo{volume}{104} (\bibinfo{year}{2016}), \bibinfo{pages}{671}.
\newblock


\bibitem[Barredo~Arrieta et~al\mbox{.}(2020)]%
        {barredo_arrieta_explainable_2020}
\bibfield{author}{\bibinfo{person}{Alejandro Barredo~Arrieta}, \bibinfo{person}{Natalia Díaz-Rodríguez}, \bibinfo{person}{Javier Del~Ser}, \bibinfo{person}{Adrien Bennetot}, \bibinfo{person}{Siham Tabik}, \bibinfo{person}{Alberto Barbado}, \bibinfo{person}{Salvador Garcia}, \bibinfo{person}{Sergio Gil-Lopez}, \bibinfo{person}{Daniel Molina}, \bibinfo{person}{Richard Benjamins}, \bibinfo{person}{Raja Chatila}, {and} \bibinfo{person}{Francisco Herrera}.} \bibinfo{year}{2020}\natexlab{}.
\newblock \showarticletitle{Explainable {Artificial} {Intelligence} ({XAI}): {Concepts}, taxonomies, opportunities and challenges toward responsible {AI}}.
\newblock \bibinfo{journal}{\emph{Information Fusion}}  \bibinfo{volume}{58} (\bibinfo{date}{June} \bibinfo{year}{2020}), \bibinfo{pages}{82--115}.
\newblock
\showISSN{15662535}
\urldef\tempurl%
\url{https://doi.org/10.1016/j.inffus.2019.12.012}
\showDOI{\tempurl}


\bibitem[Bhatt et~al\mbox{.}(2020)]%
        {bhatt2020explainable}
\bibfield{author}{\bibinfo{person}{Umang Bhatt}, \bibinfo{person}{Alice Xiang}, \bibinfo{person}{Shubham Sharma}, \bibinfo{person}{Adrian Weller}, \bibinfo{person}{Ankur Taly}, \bibinfo{person}{Yunhan Jia}, \bibinfo{person}{Joydeep Ghosh}, \bibinfo{person}{Ruchir Puri}, \bibinfo{person}{Jos{\'e}~MF Moura}, {and} \bibinfo{person}{Peter Eckersley}.} \bibinfo{year}{2020}\natexlab{}.
\newblock \showarticletitle{Explainable machine learning in deployment}. In \bibinfo{booktitle}{\emph{Proceedings of the 2020 conference on fairness, accountability, and transparency}}. \bibinfo{pages}{648--657}.
\newblock


\bibitem[Bird et~al\mbox{.}(2020)]%
        {bird2020fairlearn}
\bibfield{author}{\bibinfo{person}{Sarah Bird}, \bibinfo{person}{Miro Dud{\'\i}k}, \bibinfo{person}{Richard Edgar}, \bibinfo{person}{Brandon Horn}, \bibinfo{person}{Roman Lutz}, \bibinfo{person}{Vanessa Milan}, \bibinfo{person}{Mehrnoosh Sameki}, \bibinfo{person}{Hanna Wallach}, {and} \bibinfo{person}{Kathleen Walker}.} \bibinfo{year}{2020}\natexlab{}.
\newblock \showarticletitle{Fairlearn: A toolkit for assessing and improving fairness in AI}.
\newblock \bibinfo{journal}{\emph{Microsoft, Tech. Rep. MSR-TR-2020-32}} (\bibinfo{year}{2020}).
\newblock


\bibitem[Bishop and Nasrabadi(2006)]%
        {bishop2006pattern}
\bibfield{author}{\bibinfo{person}{Christopher~M Bishop} {and} \bibinfo{person}{Nasser~M Nasrabadi}.} \bibinfo{year}{2006}\natexlab{}.
\newblock \bibinfo{booktitle}{\emph{Pattern recognition and machine learning}}. Vol.~\bibinfo{volume}{4}.
\newblock \bibinfo{publisher}{Springer}.
\newblock


\bibitem[Boumi and Vela(2021)]%
        {boumi2021quantifying}
\bibfield{author}{\bibinfo{person}{Shahab Boumi} {and} \bibinfo{person}{Adan~Ernesto Vela}.} \bibinfo{year}{2021}\natexlab{}.
\newblock \showarticletitle{Quantifying the Impact of Students' Semester Course Load on Their Academic Performance}. In \bibinfo{booktitle}{\emph{2021 ASEE Virtual Annual Conference Content Access}}.
\newblock


\bibitem[Bravo-Agapito et~al\mbox{.}(2021)]%
        {bravo2021early}
\bibfield{author}{\bibinfo{person}{Javier Bravo-Agapito}, \bibinfo{person}{Sonia~J Romero}, {and} \bibinfo{person}{Sonia Pamplona}.} \bibinfo{year}{2021}\natexlab{}.
\newblock \showarticletitle{Early prediction of undergraduate Student's academic performance in completely online learning: A five-year study}.
\newblock \bibinfo{journal}{\emph{Computers in Human Behavior}}  \bibinfo{volume}{115} (\bibinfo{year}{2021}), \bibinfo{pages}{106595}.
\newblock


\bibitem[Breiman(2001)]%
        {breiman2001random}
\bibfield{author}{\bibinfo{person}{Leo Breiman}.} \bibinfo{year}{2001}\natexlab{}.
\newblock \showarticletitle{Random forests}.
\newblock \bibinfo{journal}{\emph{Machine learning}} \bibinfo{volume}{45}, \bibinfo{number}{1} (\bibinfo{year}{2001}), \bibinfo{pages}{5--32}.
\newblock


\bibitem[Brodersen et~al\mbox{.}(2010)]%
        {brodersen2010balanced}
\bibfield{author}{\bibinfo{person}{Kay~Henning Brodersen}, \bibinfo{person}{Cheng~Soon Ong}, \bibinfo{person}{Klaas~Enno Stephan}, {and} \bibinfo{person}{Joachim~M Buhmann}.} \bibinfo{year}{2010}\natexlab{}.
\newblock \showarticletitle{The balanced accuracy and its posterior distribution}. In \bibinfo{booktitle}{\emph{2010 20th international conference on pattern recognition}}. IEEE, \bibinfo{pages}{3121--3124}.
\newblock


\bibitem[Buyl and De~Bie(2024)]%
        {buyl2024inherent}
\bibfield{author}{\bibinfo{person}{Maarten Buyl} {and} \bibinfo{person}{Tijl De~Bie}.} \bibinfo{year}{2024}\natexlab{}.
\newblock \showarticletitle{Inherent limitations of AI fairness}.
\newblock \bibinfo{journal}{\emph{Commun. ACM}} \bibinfo{volume}{67}, \bibinfo{number}{2} (\bibinfo{year}{2024}), \bibinfo{pages}{48--55}.
\newblock


\bibitem[Caruana(1993)]%
        {caruana1993multitask}
\bibfield{author}{\bibinfo{person}{R Caruana}.} \bibinfo{year}{1993}\natexlab{}.
\newblock \showarticletitle{Multitask learning: A knowledge-based source of inductive bias1}. In \bibinfo{booktitle}{\emph{Proceedings of the Tenth International Conference on Machine Learning}}. Citeseer, \bibinfo{pages}{41--48}.
\newblock


\bibitem[Caruana(1997)]%
        {caruana1997multitask}
\bibfield{author}{\bibinfo{person}{Rich Caruana}.} \bibinfo{year}{1997}\natexlab{}.
\newblock \showarticletitle{Multitask learning}.
\newblock \bibinfo{journal}{\emph{Machine learning}}  \bibinfo{volume}{28} (\bibinfo{year}{1997}), \bibinfo{pages}{41--75}.
\newblock


\bibitem[Carvalho et~al\mbox{.}(2019)]%
        {carvalho2019machine}
\bibfield{author}{\bibinfo{person}{Diogo~V Carvalho}, \bibinfo{person}{Eduardo~M Pereira}, {and} \bibinfo{person}{Jaime~S Cardoso}.} \bibinfo{year}{2019}\natexlab{}.
\newblock \showarticletitle{Machine learning interpretability: A survey on methods and metrics}.
\newblock \bibinfo{journal}{\emph{Electronics}} \bibinfo{volume}{8}, \bibinfo{number}{8} (\bibinfo{year}{2019}), \bibinfo{pages}{832}.
\newblock


\bibitem[Cassells(2018)]%
        {cassells2018effectiveness}
\bibfield{author}{\bibinfo{person}{Laetitia Cassells}.} \bibinfo{year}{2018}\natexlab{}.
\newblock \showarticletitle{The effectiveness of early identification of ‘at risk’students in higher education institutions}.
\newblock \bibinfo{journal}{\emph{Assessment \& Evaluation in Higher Education}} \bibinfo{volume}{43}, \bibinfo{number}{4} (\bibinfo{year}{2018}), \bibinfo{pages}{515--526}.
\newblock


\bibitem[Chancellor(2023)]%
        {chancellor2023toward}
\bibfield{author}{\bibinfo{person}{Stevie Chancellor}.} \bibinfo{year}{2023}\natexlab{}.
\newblock \showarticletitle{Toward practices for human-centered machine learning}.
\newblock \bibinfo{journal}{\emph{Commun. ACM}} \bibinfo{volume}{66}, \bibinfo{number}{3} (\bibinfo{year}{2023}), \bibinfo{pages}{78--85}.
\newblock


\bibitem[Chawla et~al\mbox{.}(2002)]%
        {chawla2002smote}
\bibfield{author}{\bibinfo{person}{Nitesh~V Chawla}, \bibinfo{person}{Kevin~W Bowyer}, \bibinfo{person}{Lawrence~O Hall}, {and} \bibinfo{person}{W~Philip Kegelmeyer}.} \bibinfo{year}{2002}\natexlab{}.
\newblock \showarticletitle{SMOTE: synthetic minority over-sampling technique}.
\newblock \bibinfo{journal}{\emph{Journal of artificial intelligence research}}  \bibinfo{volume}{16} (\bibinfo{year}{2002}), \bibinfo{pages}{321--357}.
\newblock


\bibitem[Che et~al\mbox{.}(2018)]%
        {che2018recurrent}
\bibfield{author}{\bibinfo{person}{Zhengping Che}, \bibinfo{person}{Sanjay Purushotham}, \bibinfo{person}{Kyunghyun Cho}, \bibinfo{person}{David Sontag}, {and} \bibinfo{person}{Yan Liu}.} \bibinfo{year}{2018}\natexlab{}.
\newblock \showarticletitle{Recurrent neural networks for multivariate time series with missing values}.
\newblock \bibinfo{journal}{\emph{Scientific reports}} \bibinfo{volume}{8}, \bibinfo{number}{1} (\bibinfo{year}{2018}), \bibinfo{pages}{1--12}.
\newblock


\bibitem[Chen and Cui(2020)]%
        {chen2020utilizing}
\bibfield{author}{\bibinfo{person}{Fu Chen} {and} \bibinfo{person}{Ying Cui}.} \bibinfo{year}{2020}\natexlab{}.
\newblock \showarticletitle{Utilizing Student Time Series Behaviour in Learning Management Systems for Early Prediction of Course Performance.}
\newblock \bibinfo{journal}{\emph{Journal of Learning Analytics}} \bibinfo{volume}{7}, \bibinfo{number}{2} (\bibinfo{year}{2020}), \bibinfo{pages}{1--17}.
\newblock


\bibitem[Chen and Guestrin(2016)]%
        {chen2016xgboost}
\bibfield{author}{\bibinfo{person}{Tianqi Chen} {and} \bibinfo{person}{Carlos Guestrin}.} \bibinfo{year}{2016}\natexlab{}.
\newblock \showarticletitle{Xgboost: A scalable tree boosting system}. In \bibinfo{booktitle}{\emph{Proceedings of the 22nd acm sigkdd international conference on knowledge discovery and data mining}}. \bibinfo{pages}{785--794}.
\newblock


\bibitem[Cohen and Hoberman(1983)]%
        {cohen1983positive}
\bibfield{author}{\bibinfo{person}{Sheldon Cohen} {and} \bibinfo{person}{Harry~M Hoberman}.} \bibinfo{year}{1983}\natexlab{}.
\newblock \showarticletitle{Positive events and social supports as buffers of life change stress 1}.
\newblock \bibinfo{journal}{\emph{Journal of applied social psychology}} \bibinfo{volume}{13}, \bibinfo{number}{2} (\bibinfo{year}{1983}), \bibinfo{pages}{99--125}.
\newblock


\bibitem[Cohen et~al\mbox{.}(1983)]%
        {cohen1983global}
\bibfield{author}{\bibinfo{person}{Sheldon Cohen}, \bibinfo{person}{Tom Kamarck}, {and} \bibinfo{person}{Robin Mermelstein}.} \bibinfo{year}{1983}\natexlab{}.
\newblock \showarticletitle{A global measure of perceived stress}.
\newblock \bibinfo{journal}{\emph{Journal of health and social behavior}} (\bibinfo{year}{1983}), \bibinfo{pages}{385--396}.
\newblock


\bibitem[Coleman et~al\mbox{.}(2023)]%
        {coleman2023reconsidering}
\bibfield{author}{\bibinfo{person}{Toshka Coleman}, \bibinfo{person}{Sarina Till}, \bibinfo{person}{Jaydon Farao}, \bibinfo{person}{Londiwe Shandu}, \bibinfo{person}{Nonkululeko Khuzwayo}, \bibinfo{person}{Livhuwani Muthelo}, \bibinfo{person}{Masenyani Mbombi}, \bibinfo{person}{Mamare Bopape}, \bibinfo{person}{Alastair van Heerden}, \bibinfo{person}{Tebogo Mothiba}, {et~al\mbox{.}}} \bibinfo{year}{2023}\natexlab{}.
\newblock \showarticletitle{Reconsidering priorities for digital maternal and child health: community-centered perspectives from South Africa}.
\newblock \bibinfo{journal}{\emph{Proceedings of the ACM on Human-Computer Interaction}} \bibinfo{volume}{7}, \bibinfo{number}{CSCW2} (\bibinfo{year}{2023}), \bibinfo{pages}{1--31}.
\newblock


\bibitem[Corneille and Gawronski(2024)]%
        {corneille2024self}
\bibfield{author}{\bibinfo{person}{Olivier Corneille} {and} \bibinfo{person}{Bertram Gawronski}.} \bibinfo{year}{2024}\natexlab{}.
\newblock \showarticletitle{Self-reports are better measurement instruments than implicit measures}.
\newblock \bibinfo{journal}{\emph{Nature Reviews Psychology}} (\bibinfo{year}{2024}), \bibinfo{pages}{1--12}.
\newblock


\bibitem[Costa et~al\mbox{.}(2017)]%
        {costa2017evaluating}
\bibfield{author}{\bibinfo{person}{Evandro~B Costa}, \bibinfo{person}{Baldoino Fonseca}, \bibinfo{person}{Marcelo~Almeida Santana}, \bibinfo{person}{Fabr{\'\i}sia~Ferreira de Ara{\'u}jo}, {and} \bibinfo{person}{Joilson Rego}.} \bibinfo{year}{2017}\natexlab{}.
\newblock \showarticletitle{Evaluating the effectiveness of educational data mining techniques for early prediction of students' academic failure in introductory programming courses}.
\newblock \bibinfo{journal}{\emph{Computers in Human Behavior}}  \bibinfo{volume}{73} (\bibinfo{year}{2017}), \bibinfo{pages}{247--256}.
\newblock


\bibitem[Cox et~al\mbox{.}(2007)]%
        {cox2007academic}
\bibfield{author}{\bibinfo{person}{Reagan~G Cox}, \bibinfo{person}{Lei Zhang}, \bibinfo{person}{William~D Johnson}, {and} \bibinfo{person}{Daniel~R Bender}.} \bibinfo{year}{2007}\natexlab{}.
\newblock \showarticletitle{Academic performance and substance use: findings from a state survey of public high school students}.
\newblock \bibinfo{journal}{\emph{Journal of school health}} \bibinfo{volume}{77}, \bibinfo{number}{3} (\bibinfo{year}{2007}), \bibinfo{pages}{109--115}.
\newblock


\bibitem[Cred{\'e} et~al\mbox{.}(2010)]%
        {crede2010class}
\bibfield{author}{\bibinfo{person}{Marcus Cred{\'e}}, \bibinfo{person}{Sylvia~G Roch}, {and} \bibinfo{person}{Urszula~M Kieszczynka}.} \bibinfo{year}{2010}\natexlab{}.
\newblock \showarticletitle{Class attendance in college: A meta-analytic review of the relationship of class attendance with grades and student characteristics}.
\newblock \bibinfo{journal}{\emph{Review of Educational Research}} \bibinfo{volume}{80}, \bibinfo{number}{2} (\bibinfo{year}{2010}), \bibinfo{pages}{272--295}.
\newblock


\bibitem[Das~Swain et~al\mbox{.}(2022)]%
        {das2022semantic}
\bibfield{author}{\bibinfo{person}{Vedant Das~Swain}, \bibinfo{person}{Victor Chen}, \bibinfo{person}{Shrija Mishra}, \bibinfo{person}{Stephen~M Mattingly}, \bibinfo{person}{Gregory~D Abowd}, {and} \bibinfo{person}{Munmun De~Choudhury}.} \bibinfo{year}{2022}\natexlab{}.
\newblock \showarticletitle{Semantic gap in predicting mental wellbeing through passive sensing}. In \bibinfo{booktitle}{\emph{Proceedings of the 2022 CHI conference on human factors in computing systems}}. \bibinfo{pages}{1--16}.
\newblock


\bibitem[Das~Swain et~al\mbox{.}(2024)]%
        {das2024sensible}
\bibfield{author}{\bibinfo{person}{Vedant Das~Swain}, \bibinfo{person}{Lan Gao}, \bibinfo{person}{Abhirup Mondal}, \bibinfo{person}{Gregory~D Abowd}, {and} \bibinfo{person}{Munmun De~Choudhury}.} \bibinfo{year}{2024}\natexlab{}.
\newblock \showarticletitle{Sensible and Sensitive AI for Worker Wellbeing: Factors that Inform Adoption and Resistance for Information Workers}. In \bibinfo{booktitle}{\emph{Proceedings of the CHI Conference on Human Factors in Computing Systems}}. \bibinfo{pages}{1--30}.
\newblock


\bibitem[Das~Swain et~al\mbox{.}(2023)]%
        {das2023algorithmic}
\bibfield{author}{\bibinfo{person}{Vedant Das~Swain}, \bibinfo{person}{Lan Gao}, \bibinfo{person}{William~A Wood}, \bibinfo{person}{Srikruthi~C Matli}, \bibinfo{person}{Gregory~D Abowd}, {and} \bibinfo{person}{Munmun De~Choudhury}.} \bibinfo{year}{2023}\natexlab{}.
\newblock \showarticletitle{Algorithmic power or punishment: Information worker perspectives on passive sensing enabled ai phenotyping of performance and wellbeing}. In \bibinfo{booktitle}{\emph{Proceedings of the 2023 CHI Conference on Human Factors in Computing Systems}}. \bibinfo{pages}{1--17}.
\newblock


\bibitem[Daud et~al\mbox{.}(2017)]%
        {daud2017predicting}
\bibfield{author}{\bibinfo{person}{Ali Daud}, \bibinfo{person}{Naif~Radi Aljohani}, \bibinfo{person}{Rabeeh~Ayaz Abbasi}, \bibinfo{person}{Miltiadis~D Lytras}, \bibinfo{person}{Farhat Abbas}, {and} \bibinfo{person}{Jalal~S Alowibdi}.} \bibinfo{year}{2017}\natexlab{}.
\newblock \showarticletitle{Predicting student performance using advanced learning analytics}. In \bibinfo{booktitle}{\emph{Proceedings of the 26th international conference on world wide web companion}}. \bibinfo{pages}{415--421}.
\newblock


\bibitem[De~Luca et~al\mbox{.}(2016)]%
        {de2016relationship}
\bibfield{author}{\bibinfo{person}{Susan~M De~Luca}, \bibinfo{person}{Cynthia Franklin}, \bibinfo{person}{Yan Yueqi}, \bibinfo{person}{Shannon Johnson}, {and} \bibinfo{person}{Chris Brownson}.} \bibinfo{year}{2016}\natexlab{}.
\newblock \showarticletitle{The relationship between suicide ideation, behavioral health, and college academic performance}.
\newblock \bibinfo{journal}{\emph{Community mental health journal}} \bibinfo{volume}{52}, \bibinfo{number}{5} (\bibinfo{year}{2016}), \bibinfo{pages}{534--540}.
\newblock


\bibitem[Delmonaco et~al\mbox{.}(2024)]%
        {delmonaco2024you}
\bibfield{author}{\bibinfo{person}{Daniel Delmonaco}, \bibinfo{person}{Samuel Mayworm}, \bibinfo{person}{Hibby Thach}, \bibinfo{person}{Josh Guberman}, \bibinfo{person}{Aurelia Augusta}, {and} \bibinfo{person}{Oliver~L Haimson}.} \bibinfo{year}{2024}\natexlab{}.
\newblock \showarticletitle{" What are you doing, TikTok?": How Marginalized Social Media Users Perceive, Theorize, and" Prove" Shadowbanning}.
\newblock \bibinfo{journal}{\emph{Proceedings of the ACM on Human-Computer Interaction}} \bibinfo{volume}{8}, \bibinfo{number}{CSCW1} (\bibinfo{year}{2024}), \bibinfo{pages}{1--39}.
\newblock


\bibitem[Delobelle et~al\mbox{.}(2021)]%
        {delobelle2021ethical}
\bibfield{author}{\bibinfo{person}{Pieter Delobelle}, \bibinfo{person}{Paul Temple}, \bibinfo{person}{Gilles Perrouin}, \bibinfo{person}{Beno{\^\i}t Fr{\'e}nay}, \bibinfo{person}{Patrick Heymans}, {and} \bibinfo{person}{Bettina Berendt}.} \bibinfo{year}{2021}\natexlab{}.
\newblock \showarticletitle{Ethical adversaries: Towards mitigating unfairness with adversarial machine learning}.
\newblock \bibinfo{journal}{\emph{ACM SIGKDD Explorations Newsletter}} \bibinfo{volume}{23}, \bibinfo{number}{1} (\bibinfo{year}{2021}), \bibinfo{pages}{32--41}.
\newblock


\bibitem[Dormann et~al\mbox{.}(2013)]%
        {dormann2013collinearity}
\bibfield{author}{\bibinfo{person}{Carsten~F Dormann}, \bibinfo{person}{Jane Elith}, \bibinfo{person}{Sven Bacher}, \bibinfo{person}{Carsten Buchmann}, \bibinfo{person}{Gudrun Carl}, \bibinfo{person}{Gabriel Carr{\'e}}, \bibinfo{person}{Jaime R~Garc{\'\i}a Marqu{\'e}z}, \bibinfo{person}{Bernd Gruber}, \bibinfo{person}{Bruno Lafourcade}, \bibinfo{person}{Pedro~J Leit{\~a}o}, {et~al\mbox{.}}} \bibinfo{year}{2013}\natexlab{}.
\newblock \showarticletitle{Collinearity: a review of methods to deal with it and a simulation study evaluating their performance}.
\newblock \bibinfo{journal}{\emph{Ecography}} \bibinfo{volume}{36}, \bibinfo{number}{1} (\bibinfo{year}{2013}), \bibinfo{pages}{27--46}.
\newblock


\bibitem[Doryab et~al\mbox{.}(2018)]%
        {doryab2018extraction}
\bibfield{author}{\bibinfo{person}{Afsaneh Doryab}, \bibinfo{person}{Prerna Chikarsel}, \bibinfo{person}{Xinwen Liu}, {and} \bibinfo{person}{Anind~K Dey}.} \bibinfo{year}{2018}\natexlab{}.
\newblock \showarticletitle{Extraction of behavioral features from smartphone and wearable data}.
\newblock \bibinfo{journal}{\emph{arXiv preprint arXiv:1812.10394}} (\bibinfo{year}{2018}).
\newblock


\bibitem[Doshi-Velez and Kim(2017)]%
        {doshi2017towards}
\bibfield{author}{\bibinfo{person}{Finale Doshi-Velez} {and} \bibinfo{person}{Been Kim}.} \bibinfo{year}{2017}\natexlab{}.
\newblock \showarticletitle{Towards a rigorous science of interpretable machine learning}.
\newblock \bibinfo{journal}{\emph{arXiv preprint arXiv:1702.08608}} (\bibinfo{year}{2017}).
\newblock


\bibitem[Du et~al\mbox{.}(2019)]%
        {du2019techniques}
\bibfield{author}{\bibinfo{person}{Mengnan Du}, \bibinfo{person}{Ninghao Liu}, {and} \bibinfo{person}{Xia Hu}.} \bibinfo{year}{2019}\natexlab{}.
\newblock \showarticletitle{Techniques for interpretable machine learning}.
\newblock \bibinfo{journal}{\emph{Commun. ACM}} \bibinfo{volume}{63}, \bibinfo{number}{1} (\bibinfo{year}{2019}), \bibinfo{pages}{68--77}.
\newblock


\bibitem[Ehsan et~al\mbox{.}(2024)]%
        {ehsan2024xai}
\bibfield{author}{\bibinfo{person}{Upol Ehsan}, \bibinfo{person}{Samir Passi}, \bibinfo{person}{Q~Vera Liao}, \bibinfo{person}{Larry Chan}, \bibinfo{person}{I-Hsiang Lee}, \bibinfo{person}{Michael Muller}, {and} \bibinfo{person}{Mark~O Riedl}.} \bibinfo{year}{2024}\natexlab{}.
\newblock \showarticletitle{The Who in XAI: How AI Background Shapes Perceptions of AI Explanations}. In \bibinfo{booktitle}{\emph{Proceedings of the CHI Conference on Human Factors in Computing Systems}}. \bibinfo{pages}{1--32}.
\newblock


\bibitem[Ehsan and Riedl(2020)]%
        {ehsan2020human}
\bibfield{author}{\bibinfo{person}{Upol Ehsan} {and} \bibinfo{person}{Mark~O Riedl}.} \bibinfo{year}{2020}\natexlab{}.
\newblock \showarticletitle{Human-centered explainable ai: Towards a reflective sociotechnical approach}. In \bibinfo{booktitle}{\emph{HCI International 2020-Late Breaking Papers: Multimodality and Intelligence: 22nd HCI International Conference, HCII 2020, Copenhagen, Denmark, July 19--24, 2020, Proceedings 22}}. Springer, \bibinfo{pages}{449--466}.
\newblock


\bibitem[Ehsan et~al\mbox{.}(2023)]%
        {ehsan2023charting}
\bibfield{author}{\bibinfo{person}{Upol Ehsan}, \bibinfo{person}{Koustuv Saha}, \bibinfo{person}{Munmun De~Choudhury}, {and} \bibinfo{person}{Mark~O Riedl}.} \bibinfo{year}{2023}\natexlab{}.
\newblock \showarticletitle{Charting the sociotechnical gap in explainable ai: A framework to address the gap in xai}.
\newblock \bibinfo{journal}{\emph{Proceedings of the ACM on human-computer interaction}} \bibinfo{volume}{7}, \bibinfo{number}{CSCW1} (\bibinfo{year}{2023}), \bibinfo{pages}{1--32}.
\newblock


\bibitem[Ester et~al\mbox{.}(1996)]%
        {ester1996density}
\bibfield{author}{\bibinfo{person}{Martin Ester}, \bibinfo{person}{Hans-Peter Kriegel}, \bibinfo{person}{J{\"o}rg Sander}, \bibinfo{person}{Xiaowei Xu}, {et~al\mbox{.}}} \bibinfo{year}{1996}\natexlab{}.
\newblock \showarticletitle{A density-based algorithm for discovering clusters in large spatial databases with noise.}. In \bibinfo{booktitle}{\emph{kdd}}, Vol.~\bibinfo{volume}{96}. \bibinfo{pages}{226--231}.
\newblock


\bibitem[{FairLearn Contributors}(2022)]%
        {fairLearnPackage}
\bibfield{author}{\bibinfo{person}{{FairLearn Contributors}}.} \bibinfo{year}{2022}\natexlab{}.
\newblock \bibinfo{title}{Fairlearn Metrics Package}.
\newblock \bibinfo{howpublished}{\url{https://fairlearn.org/v0.7.0/api_reference/fairlearn.metrics.html}}.
\newblock


\bibitem[Feldman et~al\mbox{.}(2015)]%
        {feldman2015certifying}
\bibfield{author}{\bibinfo{person}{Michael Feldman}, \bibinfo{person}{Sorelle~A Friedler}, \bibinfo{person}{John Moeller}, \bibinfo{person}{Carlos Scheidegger}, {and} \bibinfo{person}{Suresh Venkatasubramanian}.} \bibinfo{year}{2015}\natexlab{}.
\newblock \showarticletitle{Certifying and removing disparate impact}. In \bibinfo{booktitle}{\emph{proceedings of the 21th ACM SIGKDD international conference on knowledge discovery and data mining}}. \bibinfo{pages}{259--268}.
\newblock


\bibitem[Felez-Nobrega et~al\mbox{.}(2018)]%
        {felez2018activpal}
\bibfield{author}{\bibinfo{person}{Mireia Felez-Nobrega}, \bibinfo{person}{Charles~H Hillman}, \bibinfo{person}{Kieran~P Dowd}, \bibinfo{person}{Eva Cirera}, {and} \bibinfo{person}{Anna Puig-Ribera}.} \bibinfo{year}{2018}\natexlab{}.
\newblock \showarticletitle{ActivPAL™ determined sedentary behaviour, physical activity and academic achievement in college students}.
\newblock \bibinfo{journal}{\emph{Journal of sports sciences}} \bibinfo{volume}{36}, \bibinfo{number}{20} (\bibinfo{year}{2018}), \bibinfo{pages}{2311--2316}.
\newblock


\bibitem[Felisoni and Godoi(2018)]%
        {felisoni2018cell}
\bibfield{author}{\bibinfo{person}{Daniel~Darghan Felisoni} {and} \bibinfo{person}{Alexandra~Strommer Godoi}.} \bibinfo{year}{2018}\natexlab{}.
\newblock \showarticletitle{Cell phone usage and academic performance: An experiment}.
\newblock \bibinfo{journal}{\emph{Computers \& Education}}  \bibinfo{volume}{117} (\bibinfo{year}{2018}), \bibinfo{pages}{175--187}.
\newblock


\bibitem[{Fitbit Team}(2023)]%
        {fitbitsleep}
\bibfield{author}{\bibinfo{person}{{Fitbit Team}}.} \bibinfo{year}{2023}\natexlab{}.
\newblock \bibinfo{title}{Fitbit development: Sleep logs}.
\newblock \bibinfo{howpublished}{\url{https://dev.fitbit.com/build/reference/web-api/sleep/}}.
\newblock


\bibitem[Fredrickson(2000)]%
        {fredrickson2000extracting}
\bibfield{author}{\bibinfo{person}{Barbara~L Fredrickson}.} \bibinfo{year}{2000}\natexlab{}.
\newblock \showarticletitle{Extracting meaning from past affective experiences: The importance of peaks, ends, and specific emotions}.
\newblock \bibinfo{journal}{\emph{Cognition \& Emotion}} \bibinfo{volume}{14}, \bibinfo{number}{4} (\bibinfo{year}{2000}), \bibinfo{pages}{577--606}.
\newblock


\bibitem[Freund and Schapire(1997)]%
        {freund1997decision}
\bibfield{author}{\bibinfo{person}{Yoav Freund} {and} \bibinfo{person}{Robert~E Schapire}.} \bibinfo{year}{1997}\natexlab{}.
\newblock \showarticletitle{A decision-theoretic generalization of on-line learning and an application to boosting}.
\newblock \bibinfo{journal}{\emph{Journal of computer and system sciences}} \bibinfo{volume}{55}, \bibinfo{number}{1} (\bibinfo{year}{1997}), \bibinfo{pages}{119--139}.
\newblock


\bibitem[Friedler et~al\mbox{.}(2016)]%
        {friedler2016possibility}
\bibfield{author}{\bibinfo{person}{Sorelle~A Friedler}, \bibinfo{person}{Carlos Scheidegger}, {and} \bibinfo{person}{Suresh Venkatasubramanian}.} \bibinfo{year}{2016}\natexlab{}.
\newblock \showarticletitle{On the (im) possibility of fairness}.
\newblock \bibinfo{journal}{\emph{arXiv preprint arXiv:1609.07236}} (\bibinfo{year}{2016}).
\newblock


\bibitem[Friedman(2001)]%
        {friedman2001greedy}
\bibfield{author}{\bibinfo{person}{Jerome~H Friedman}.} \bibinfo{year}{2001}\natexlab{}.
\newblock \showarticletitle{Greedy function approximation: a gradient boosting machine}.
\newblock \bibinfo{journal}{\emph{Annals of statistics}} (\bibinfo{year}{2001}), \bibinfo{pages}{1189--1232}.
\newblock


\bibitem[Gal and Ghahramani(2016)]%
        {gal2016theoretically}
\bibfield{author}{\bibinfo{person}{Yarin Gal} {and} \bibinfo{person}{Zoubin Ghahramani}.} \bibinfo{year}{2016}\natexlab{}.
\newblock \showarticletitle{A theoretically grounded application of dropout in recurrent neural networks}.
\newblock \bibinfo{journal}{\emph{Advances in neural information processing systems}}  \bibinfo{volume}{29} (\bibinfo{year}{2016}).
\newblock


\bibitem[Gallagher(2006)]%
        {gallagher2006national}
\bibfield{author}{\bibinfo{person}{Robert~P Gallagher}.} \bibinfo{year}{2006}\natexlab{}.
\newblock \showarticletitle{National survey of counseling center directors 2005}.
\newblock  (\bibinfo{year}{2006}).
\newblock


\bibitem[Geiser and Santelices(2007)]%
        {geiser2007validity}
\bibfield{author}{\bibinfo{person}{Saul Geiser} {and} \bibinfo{person}{Maria~Veronica Santelices}.} \bibinfo{year}{2007}\natexlab{}.
\newblock \showarticletitle{Validity of high-school grades in predicting student success beyond the freshman year: High-school record vs. standardized tests as indicators of four-year college outcomes}.
\newblock  (\bibinfo{year}{2007}).
\newblock


\bibitem[G{\'e}ron(2022)]%
        {geron2022hands}
\bibfield{author}{\bibinfo{person}{Aur{\'e}lien G{\'e}ron}.} \bibinfo{year}{2022}\natexlab{}.
\newblock \bibinfo{booktitle}{\emph{Hands-on machine learning with Scikit-Learn, Keras, and TensorFlow}}.
\newblock \bibinfo{publisher}{" O'Reilly Media, Inc."}.
\newblock


\bibitem[Giunchiglia et~al\mbox{.}(2018)]%
        {giunchiglia2018mobile}
\bibfield{author}{\bibinfo{person}{Fausto Giunchiglia}, \bibinfo{person}{Mattia Zeni}, \bibinfo{person}{Elisa Gobbi}, \bibinfo{person}{Enrico Bignotti}, {and} \bibinfo{person}{Ivano Bison}.} \bibinfo{year}{2018}\natexlab{}.
\newblock \showarticletitle{Mobile social media usage and academic performance}.
\newblock \bibinfo{journal}{\emph{Computers in Human Behavior}}  \bibinfo{volume}{82} (\bibinfo{year}{2018}), \bibinfo{pages}{177--185}.
\newblock


\bibitem[Gomes et~al\mbox{.}(2011)]%
        {gomes2011sleep}
\bibfield{author}{\bibinfo{person}{Ana~Allen Gomes}, \bibinfo{person}{Jos{\'e} Tavares}, {and} \bibinfo{person}{Maria Helena~P de Azevedo}.} \bibinfo{year}{2011}\natexlab{}.
\newblock \showarticletitle{Sleep and academic performance in undergraduates: a multi-measure, multi-predictor approach}.
\newblock \bibinfo{journal}{\emph{Chronobiology international}} \bibinfo{volume}{28}, \bibinfo{number}{9} (\bibinfo{year}{2011}), \bibinfo{pages}{786--801}.
\newblock


\bibitem[Grubb(2006)]%
        {grubb2006does}
\bibfield{author}{\bibinfo{person}{Farley Grubb}.} \bibinfo{year}{2006}\natexlab{}.
\newblock \showarticletitle{Does going Greek impair undergraduate academic performance? A case study}.
\newblock \bibinfo{journal}{\emph{American Journal of Economics and Sociology}} \bibinfo{volume}{65}, \bibinfo{number}{5} (\bibinfo{year}{2006}), \bibinfo{pages}{1085--1110}.
\newblock


\bibitem[Hall(1999)]%
        {hall1999correlation}
\bibfield{author}{\bibinfo{person}{Mark~Andrew Hall}.} \bibinfo{year}{1999}\natexlab{}.
\newblock \showarticletitle{Correlation-based feature selection for machine learning}.
\newblock  (\bibinfo{year}{1999}).
\newblock


\bibitem[Hamaideh(2011)]%
        {hamaideh2011stressors}
\bibfield{author}{\bibinfo{person}{Shaher~H Hamaideh}.} \bibinfo{year}{2011}\natexlab{}.
\newblock \showarticletitle{Stressors and reactions to stressors among university students}.
\newblock \bibinfo{journal}{\emph{International journal of social psychiatry}} \bibinfo{volume}{57}, \bibinfo{number}{1} (\bibinfo{year}{2011}), \bibinfo{pages}{69--80}.
\newblock


\bibitem[Hardt et~al\mbox{.}(2016)]%
        {hardt2016equality}
\bibfield{author}{\bibinfo{person}{Moritz Hardt}, \bibinfo{person}{Eric Price}, {and} \bibinfo{person}{Nati Srebro}.} \bibinfo{year}{2016}\natexlab{}.
\newblock \showarticletitle{Equality of opportunity in supervised learning}.
\newblock \bibinfo{journal}{\emph{Advances in neural information processing systems}}  \bibinfo{volume}{29} (\bibinfo{year}{2016}).
\newblock


\bibitem[Hlosta et~al\mbox{.}(2017)]%
        {hlosta2017ouroboros}
\bibfield{author}{\bibinfo{person}{Martin Hlosta}, \bibinfo{person}{Zdenek Zdrahal}, {and} \bibinfo{person}{Jaroslav Zendulka}.} \bibinfo{year}{2017}\natexlab{}.
\newblock \showarticletitle{Ouroboros: early identification of at-risk students without models based on legacy data}. In \bibinfo{booktitle}{\emph{Proceedings of the seventh international learning analytics \& knowledge conference}}. \bibinfo{pages}{6--15}.
\newblock


\bibitem[Holstein et~al\mbox{.}(2019)]%
        {holstein2019improving}
\bibfield{author}{\bibinfo{person}{Kenneth Holstein}, \bibinfo{person}{Jennifer Wortman~Vaughan}, \bibinfo{person}{Hal Daum{\'e}~III}, \bibinfo{person}{Miro Dudik}, {and} \bibinfo{person}{Hanna Wallach}.} \bibinfo{year}{2019}\natexlab{}.
\newblock \showarticletitle{Improving fairness in machine learning systems: What do industry practitioners need?}. In \bibinfo{booktitle}{\emph{Proceedings of the 2019 CHI conference on human factors in computing systems}}. \bibinfo{pages}{1--16}.
\newblock


\bibitem[Holt et~al\mbox{.}(2007)]%
        {holt2007multiple}
\bibfield{author}{\bibinfo{person}{Melissa~K Holt}, \bibinfo{person}{David Finkelhor}, {and} \bibinfo{person}{Glenda~Kaufman Kantor}.} \bibinfo{year}{2007}\natexlab{}.
\newblock \showarticletitle{Multiple victimization experiences of urban elementary school students: Associations with psychosocial functioning and academic performance}.
\newblock \bibinfo{journal}{\emph{Child abuse \& neglect}} \bibinfo{volume}{31}, \bibinfo{number}{5} (\bibinfo{year}{2007}), \bibinfo{pages}{503--515}.
\newblock


\bibitem[Hong et~al\mbox{.}(2020)]%
        {hong2020human}
\bibfield{author}{\bibinfo{person}{Sungsoo~Ray Hong}, \bibinfo{person}{Jessica Hullman}, {and} \bibinfo{person}{Enrico Bertini}.} \bibinfo{year}{2020}\natexlab{}.
\newblock \showarticletitle{Human factors in model interpretability: Industry practices, challenges, and needs}.
\newblock \bibinfo{journal}{\emph{Proceedings of the ACM on Human-Computer Interaction}} \bibinfo{volume}{4}, \bibinfo{number}{CSCW1} (\bibinfo{year}{2020}), \bibinfo{pages}{1--26}.
\newblock


\bibitem[Hooker(2021)]%
        {hooker2021moving}
\bibfield{author}{\bibinfo{person}{Sara Hooker}.} \bibinfo{year}{2021}\natexlab{}.
\newblock \showarticletitle{Moving beyond “algorithmic bias is a data problem”}.
\newblock \bibinfo{journal}{\emph{Patterns}} \bibinfo{volume}{2}, \bibinfo{number}{4} (\bibinfo{year}{2021}), \bibinfo{pages}{100241}.
\newblock


\bibitem[Hunt and Eisenberg(2010)]%
        {hunt2010mental}
\bibfield{author}{\bibinfo{person}{Justin Hunt} {and} \bibinfo{person}{Daniel Eisenberg}.} \bibinfo{year}{2010}\natexlab{}.
\newblock \showarticletitle{Mental health problems and help-seeking behavior among college students}.
\newblock \bibinfo{journal}{\emph{Journal of adolescent health}} \bibinfo{volume}{46}, \bibinfo{number}{1} (\bibinfo{year}{2010}), \bibinfo{pages}{3--10}.
\newblock


\bibitem[Huynh et~al\mbox{.}(2017)]%
        {huynh2017not}
\bibfield{author}{\bibinfo{person}{Virginia~W Huynh}, \bibinfo{person}{Que-Lam Huynh}, {and} \bibinfo{person}{Mary-Patricia Stein}.} \bibinfo{year}{2017}\natexlab{}.
\newblock \showarticletitle{Not just sticks and stones: Indirect ethnic discrimination leads to greater physiological reactivity.}
\newblock \bibinfo{journal}{\emph{Cultural Diversity and Ethnic Minority Psychology}} \bibinfo{volume}{23}, \bibinfo{number}{3} (\bibinfo{year}{2017}), \bibinfo{pages}{425}.
\newblock


\bibitem[Inc.(2024)]%
        {AppleATT}
\bibfield{author}{\bibinfo{person}{Apple Inc.}} \bibinfo{year}{2024}\natexlab{}.
\newblock \bibinfo{title}{If an app asks to track your activity}.
\newblock \bibinfo{howpublished}{\url{https://support.apple.com/en-us/102420}}.
\newblock


\bibitem[Inc.(2025)]%
        {AppleUserPrivacy}
\bibfield{author}{\bibinfo{person}{Apple Inc.}} \bibinfo{year}{2025}\natexlab{}.
\newblock \bibinfo{title}{User privacy and data use}.
\newblock \bibinfo{howpublished}{\url{https://developer.apple.com/app-store/user-privacy-and-data-use/}}.
\newblock


\bibitem[Jamalova and Constantinovits(2019)]%
        {jamalova2019comparative}
\bibfield{author}{\bibinfo{person}{Maral Jamalova} {and} \bibinfo{person}{M Constantinovits}.} \bibinfo{year}{2019}\natexlab{}.
\newblock \showarticletitle{The comparative study of the relationship between smartphone choice and socio-economic indicators}.
\newblock \bibinfo{journal}{\emph{Int. J. Mark. Stud}} \bibinfo{volume}{11}, \bibinfo{number}{11} (\bibinfo{year}{2019}), \bibinfo{pages}{10--5539}.
\newblock


\bibitem[Jayaprakash et~al\mbox{.}(2014)]%
        {jayaprakash2014early}
\bibfield{author}{\bibinfo{person}{Sandeep~M Jayaprakash}, \bibinfo{person}{Erik~W Moody}, \bibinfo{person}{Eitel~JM Laur{\'\i}a}, \bibinfo{person}{James~R Regan}, {and} \bibinfo{person}{Joshua~D Baron}.} \bibinfo{year}{2014}\natexlab{}.
\newblock \showarticletitle{Early alert of academically at-risk students: An open source analytics initiative}.
\newblock \bibinfo{journal}{\emph{Journal of Learning Analytics}} \bibinfo{volume}{1}, \bibinfo{number}{1} (\bibinfo{year}{2014}), \bibinfo{pages}{6--47}.
\newblock


\bibitem[Kabacoff et~al\mbox{.}(1997)]%
        {kabacoff1997psychometric}
\bibfield{author}{\bibinfo{person}{Robert~I Kabacoff}, \bibinfo{person}{Daniel~L Segal}, \bibinfo{person}{Michel Hersen}, {and} \bibinfo{person}{Vincent~B Van~Hasselt}.} \bibinfo{year}{1997}\natexlab{}.
\newblock \showarticletitle{Psychometric properties and diagnostic utility of the Beck Anxiety Inventory and the State-Trait Anxiety Inventory with older adult psychiatric outpatients}.
\newblock \bibinfo{journal}{\emph{Journal of anxiety disorders}} \bibinfo{volume}{11}, \bibinfo{number}{1} (\bibinfo{year}{1997}), \bibinfo{pages}{33--47}.
\newblock


\bibitem[Kadapatti and Vijayalaxmi(2012)]%
        {kadapatti2012stressors}
\bibfield{author}{\bibinfo{person}{Manjula~G Kadapatti} {and} \bibinfo{person}{AHM Vijayalaxmi}.} \bibinfo{year}{2012}\natexlab{}.
\newblock \showarticletitle{Stressors of academic stress-a study on pre-university students}.
\newblock \bibinfo{journal}{\emph{Indian Journal of Scientific Research}} \bibinfo{volume}{3}, \bibinfo{number}{1} (\bibinfo{year}{2012}), \bibinfo{pages}{171--175}.
\newblock


\bibitem[Kamiran and Calders(2012)]%
        {kamiran2012data}
\bibfield{author}{\bibinfo{person}{Faisal Kamiran} {and} \bibinfo{person}{Toon Calders}.} \bibinfo{year}{2012}\natexlab{}.
\newblock \showarticletitle{Data preprocessing techniques for classification without discrimination}.
\newblock \bibinfo{journal}{\emph{Knowledge and information systems}} \bibinfo{volume}{33}, \bibinfo{number}{1} (\bibinfo{year}{2012}), \bibinfo{pages}{1--33}.
\newblock


\bibitem[Kamiran et~al\mbox{.}(2012)]%
        {kamiran2012decision}
\bibfield{author}{\bibinfo{person}{Faisal Kamiran}, \bibinfo{person}{Asim Karim}, {and} \bibinfo{person}{Xiangliang Zhang}.} \bibinfo{year}{2012}\natexlab{}.
\newblock \showarticletitle{Decision theory for discrimination-aware classification}. In \bibinfo{booktitle}{\emph{2012 IEEE 12th international conference on data mining}}. IEEE, \bibinfo{pages}{924--929}.
\newblock


\bibitem[Katamei and Omwono(2015)]%
        {katamei2015intervention}
\bibfield{author}{\bibinfo{person}{Jacob~Merew Katamei} {and} \bibinfo{person}{Gedion~A Omwono}.} \bibinfo{year}{2015}\natexlab{}.
\newblock \showarticletitle{Intervention strategies to improve students' academic performance in public secondary schools in arid and semi-arid lands in Kenya}.
\newblock \bibinfo{journal}{\emph{Int'l J. Soc. Sci. Stud.}}  \bibinfo{volume}{3} (\bibinfo{year}{2015}), \bibinfo{pages}{107}.
\newblock


\bibitem[Kawakami et~al\mbox{.}(2023)]%
        {kawakami2023sensing}
\bibfield{author}{\bibinfo{person}{Anna Kawakami}, \bibinfo{person}{Shreya Chowdhary}, \bibinfo{person}{Shamsi~T Iqbal}, \bibinfo{person}{Q~Vera Liao}, \bibinfo{person}{Alexandra Olteanu}, \bibinfo{person}{Jina Suh}, {and} \bibinfo{person}{Koustuv Saha}.} \bibinfo{year}{2023}\natexlab{}.
\newblock \showarticletitle{Sensing wellbeing in the workplace, why and for whom? envisioning impacts with organizational stakeholders}.
\newblock \bibinfo{journal}{\emph{Proceedings of the ACM on Human-Computer Interaction}} \bibinfo{volume}{7}, \bibinfo{number}{CSCW2} (\bibinfo{year}{2023}), \bibinfo{pages}{1--33}.
\newblock


\bibitem[Khan and Ghosh(2021)]%
        {khan2021student}
\bibfield{author}{\bibinfo{person}{Anupam Khan} {and} \bibinfo{person}{Soumya~K Ghosh}.} \bibinfo{year}{2021}\natexlab{}.
\newblock \showarticletitle{Student performance analysis and prediction in classroom learning: A review of educational data mining studies}.
\newblock \bibinfo{journal}{\emph{Education and information technologies}} \bibinfo{volume}{26}, \bibinfo{number}{1} (\bibinfo{year}{2021}), \bibinfo{pages}{205--240}.
\newblock


\bibitem[Khan et~al\mbox{.}(2017)]%
        {khan2017multifactorial}
\bibfield{author}{\bibinfo{person}{Mariam Khan}, \bibinfo{person}{Misja Ilcisin}, {and} \bibinfo{person}{Katherine Saxton}.} \bibinfo{year}{2017}\natexlab{}.
\newblock \showarticletitle{Multifactorial discrimination as a fundamental cause of mental health inequities}.
\newblock \bibinfo{journal}{\emph{International Journal for Equity in Health}}  \bibinfo{volume}{16} (\bibinfo{year}{2017}), \bibinfo{pages}{1--12}.
\newblock


\bibitem[Kim et~al\mbox{.}(2021)]%
        {kim2021human}
\bibfield{author}{\bibinfo{person}{Seunghyun Kim}, \bibinfo{person}{Afsaneh Razi}, \bibinfo{person}{Gianluca Stringhini}, \bibinfo{person}{Pamela~J Wisniewski}, {and} \bibinfo{person}{Munmun De~Choudhury}.} \bibinfo{year}{2021}\natexlab{}.
\newblock \showarticletitle{A human-centered systematic literature review of cyberbullying detection algorithms}.
\newblock \bibinfo{journal}{\emph{Proceedings of the ACM on Human-Computer Interaction}} \bibinfo{volume}{5}, \bibinfo{number}{CSCW2} (\bibinfo{year}{2021}), \bibinfo{pages}{1--34}.
\newblock


\bibitem[Kinkel and Henke(2006)]%
        {kinkel2006impact}
\bibfield{author}{\bibinfo{person}{Doreen~H Kinkel} {and} \bibinfo{person}{Scott~E Henke}.} \bibinfo{year}{2006}\natexlab{}.
\newblock \showarticletitle{Impact of undergraduate research on academic performance, educational planning, and career development}.
\newblock \bibinfo{journal}{\emph{Journal of Natural Resources and Life Sciences Education}} \bibinfo{volume}{35}, \bibinfo{number}{1} (\bibinfo{year}{2006}), \bibinfo{pages}{194--201}.
\newblock


\bibitem[Kiranyaz et~al\mbox{.}(2021)]%
        {kiranyaz20211d}
\bibfield{author}{\bibinfo{person}{Serkan Kiranyaz}, \bibinfo{person}{Onur Avci}, \bibinfo{person}{Osama Abdeljaber}, \bibinfo{person}{Turker Ince}, \bibinfo{person}{Moncef Gabbouj}, {and} \bibinfo{person}{Daniel~J Inman}.} \bibinfo{year}{2021}\natexlab{}.
\newblock \showarticletitle{1D convolutional neural networks and applications: A survey}.
\newblock \bibinfo{journal}{\emph{Mechanical systems and signal processing}}  \bibinfo{volume}{151} (\bibinfo{year}{2021}), \bibinfo{pages}{107398}.
\newblock


\bibitem[Kiranyaz et~al\mbox{.}(2019)]%
        {kiranyaz20191}
\bibfield{author}{\bibinfo{person}{Serkan Kiranyaz}, \bibinfo{person}{Turker Ince}, \bibinfo{person}{Osama Abdeljaber}, \bibinfo{person}{Onur Avci}, {and} \bibinfo{person}{Moncef Gabbouj}.} \bibinfo{year}{2019}\natexlab{}.
\newblock \showarticletitle{1-D convolutional neural networks for signal processing applications}. In \bibinfo{booktitle}{\emph{ICASSP 2019-2019 IEEE International Conference on Acoustics, Speech and Signal Processing (ICASSP)}}. IEEE, \bibinfo{pages}{8360--8364}.
\newblock


\bibitem[Kobayashi and Nakao(2021)]%
        {kobayashi2021one}
\bibfield{author}{\bibinfo{person}{Kenji Kobayashi} {and} \bibinfo{person}{Yuri Nakao}.} \bibinfo{year}{2021}\natexlab{}.
\newblock \showarticletitle{One-vs.-One Mitigation of Intersectional Bias: A General Method for Extending Fairness-Aware Binary Classification}. In \bibinfo{booktitle}{\emph{International Conference on Disruptive Technologies, Tech Ethics and Artificial Intelligence}}. Springer, \bibinfo{pages}{43--54}.
\newblock


\bibitem[Lane et~al\mbox{.}(2011)]%
        {lane2011bewell}
\bibfield{author}{\bibinfo{person}{Nicholas~D Lane}, \bibinfo{person}{Mashfiqui Mohammod}, \bibinfo{person}{Mu Lin}, \bibinfo{person}{Xiaochao Yang}, \bibinfo{person}{Hong Lu}, \bibinfo{person}{Shahid Ali}, \bibinfo{person}{Afsaneh Doryab}, \bibinfo{person}{Ethan Berke}, \bibinfo{person}{Tanzeem Choudhury}, {and} \bibinfo{person}{Andrew Campbell}.} \bibinfo{year}{2011}\natexlab{}.
\newblock \showarticletitle{Bewell: A smartphone application to monitor, model and promote wellbeing}. In \bibinfo{booktitle}{\emph{5th international ICST conference on pervasive computing technologies for healthcare}}, Vol.~\bibinfo{volume}{10}.
\newblock


\bibitem[Lara et~al\mbox{.}(2014)]%
        {lara2014system}
\bibfield{author}{\bibinfo{person}{Juan~A Lara}, \bibinfo{person}{David Lizcano}, \bibinfo{person}{Mar{\'\i}a~A Mart{\'\i}nez}, \bibinfo{person}{Juan Pazos}, {and} \bibinfo{person}{Teresa Riera}.} \bibinfo{year}{2014}\natexlab{}.
\newblock \showarticletitle{A system for knowledge discovery in e-learning environments within the European Higher Education Area--Application to student data from Open University of Madrid, UDIMA}.
\newblock \bibinfo{journal}{\emph{Computers \& Education}}  \bibinfo{volume}{72} (\bibinfo{year}{2014}), \bibinfo{pages}{23--36}.
\newblock


\bibitem[Larrabee~S{\o}nderlund et~al\mbox{.}(2019)]%
        {larrabee2019efficacy}
\bibfield{author}{\bibinfo{person}{Anders Larrabee~S{\o}nderlund}, \bibinfo{person}{Emily Hughes}, {and} \bibinfo{person}{Joanne Smith}.} \bibinfo{year}{2019}\natexlab{}.
\newblock \showarticletitle{The efficacy of learning analytics interventions in higher education: A systematic review}.
\newblock \bibinfo{journal}{\emph{British Journal of Educational Technology}} \bibinfo{volume}{50}, \bibinfo{number}{5} (\bibinfo{year}{2019}), \bibinfo{pages}{2594--2618}.
\newblock


\bibitem[Lepp et~al\mbox{.}(2015)]%
        {lepp2015relationship}
\bibfield{author}{\bibinfo{person}{Andrew Lepp}, \bibinfo{person}{Jacob~E Barkley}, {and} \bibinfo{person}{Aryn~C Karpinski}.} \bibinfo{year}{2015}\natexlab{}.
\newblock \showarticletitle{The relationship between cell phone use and academic performance in a sample of US college students}.
\newblock \bibinfo{journal}{\emph{Sage Open}} \bibinfo{volume}{5}, \bibinfo{number}{1} (\bibinfo{year}{2015}), \bibinfo{pages}{2158244015573169}.
\newblock


\bibitem[L{\'o}pez~Zambrano et~al\mbox{.}(2021)]%
        {lopez2021early}
\bibfield{author}{\bibinfo{person}{Javier L{\'o}pez~Zambrano}, \bibinfo{person}{Juan~Alfonso Lara~Torralbo}, \bibinfo{person}{Crist{\'o}bal Romero~Morales}, {et~al\mbox{.}}} \bibinfo{year}{2021}\natexlab{}.
\newblock \showarticletitle{Early prediction of student learning performance through data mining: A systematic review}.
\newblock \bibinfo{journal}{\emph{Psicothema}} (\bibinfo{year}{2021}).
\newblock


\bibitem[Lu et~al\mbox{.}(2010)]%
        {lu2010jigsaw}
\bibfield{author}{\bibinfo{person}{Hong Lu}, \bibinfo{person}{Jun Yang}, \bibinfo{person}{Zhigang Liu}, \bibinfo{person}{Nicholas~D Lane}, \bibinfo{person}{Tanzeem Choudhury}, {and} \bibinfo{person}{Andrew~T Campbell}.} \bibinfo{year}{2010}\natexlab{}.
\newblock \showarticletitle{The jigsaw continuous sensing engine for mobile phone applications}. In \bibinfo{booktitle}{\emph{Proceedings of the 8th ACM conference on embedded networked sensor systems}}. \bibinfo{pages}{71--84}.
\newblock


\bibitem[Lu et~al\mbox{.}(2018)]%
        {lu2018applying}
\bibfield{author}{\bibinfo{person}{Owen~HT Lu}, \bibinfo{person}{Anna~YQ Huang}, \bibinfo{person}{Jeff~CH Huang}, \bibinfo{person}{Albert~JQ Lin}, \bibinfo{person}{Hiroaki Ogata}, {and} \bibinfo{person}{Stephen~JH Yang}.} \bibinfo{year}{2018}\natexlab{}.
\newblock \showarticletitle{Applying learning analytics for the early prediction of Students' academic performance in blended learning}.
\newblock \bibinfo{journal}{\emph{Journal of Educational Technology \& Society}} \bibinfo{volume}{21}, \bibinfo{number}{2} (\bibinfo{year}{2018}), \bibinfo{pages}{220--232}.
\newblock


\bibitem[Lundberg(2017)]%
        {lundberg2017unified}
\bibfield{author}{\bibinfo{person}{Scott Lundberg}.} \bibinfo{year}{2017}\natexlab{}.
\newblock \showarticletitle{A unified approach to interpreting model predictions}.
\newblock \bibinfo{journal}{\emph{arXiv preprint arXiv:1705.07874}} (\bibinfo{year}{2017}).
\newblock


\bibitem[Marques et~al\mbox{.}(2018)]%
        {marques2018does}
\bibfield{author}{\bibinfo{person}{Adilson Marques}, \bibinfo{person}{Diana~A Santos}, \bibinfo{person}{Charles~H Hillman}, {and} \bibinfo{person}{Lu{\'\i}s~B Sardinha}.} \bibinfo{year}{2018}\natexlab{}.
\newblock \showarticletitle{How does academic achievement relate to cardiorespiratory fitness, self-reported physical activity and objectively reported physical activity: a systematic review in children and adolescents aged 6--18 years}.
\newblock \bibinfo{journal}{\emph{British Journal of Sports Medicine}} \bibinfo{volume}{52}, \bibinfo{number}{16} (\bibinfo{year}{2018}), \bibinfo{pages}{1039--1039}.
\newblock


\bibitem[McHugh(2012)]%
        {mchugh2012interrater}
\bibfield{author}{\bibinfo{person}{Mary~L McHugh}.} \bibinfo{year}{2012}\natexlab{}.
\newblock \showarticletitle{Interrater reliability: the kappa statistic}.
\newblock \bibinfo{journal}{\emph{Biochemia medica}} \bibinfo{volume}{22}, \bibinfo{number}{3} (\bibinfo{year}{2012}), \bibinfo{pages}{276--282}.
\newblock


\bibitem[Meegahapola et~al\mbox{.}(2024)]%
        {meegahapola2024faircomp}
\bibfield{author}{\bibinfo{person}{Lakmal Meegahapola}, \bibinfo{person}{Dimitris Spathis}, \bibinfo{person}{Marios Constantinides}, \bibinfo{person}{Han Zhang}, \bibinfo{person}{Sofia Yfantidou}, \bibinfo{person}{Niels van Berkel}, {and} \bibinfo{person}{Anind~K Dey}.} \bibinfo{year}{2024}\natexlab{}.
\newblock \showarticletitle{FairComp: 2nd International Workshop on Fairness and Robustness in Machine Learning for Ubiquitous Computing}. In \bibinfo{booktitle}{\emph{Companion of the 2024 on ACM International Joint Conference on Pervasive and Ubiquitous Computing}}. \bibinfo{pages}{996--999}.
\newblock


\bibitem[Mehrabi et~al\mbox{.}(2021)]%
        {mehrabi2021survey}
\bibfield{author}{\bibinfo{person}{Ninareh Mehrabi}, \bibinfo{person}{Fred Morstatter}, \bibinfo{person}{Nripsuta Saxena}, \bibinfo{person}{Kristina Lerman}, {and} \bibinfo{person}{Aram Galstyan}.} \bibinfo{year}{2021}\natexlab{}.
\newblock \showarticletitle{A survey on bias and fairness in machine learning}.
\newblock \bibinfo{journal}{\emph{ACM Computing Surveys (CSUR)}} \bibinfo{volume}{54}, \bibinfo{number}{6} (\bibinfo{year}{2021}), \bibinfo{pages}{1--35}.
\newblock


\bibitem[Mendez et~al\mbox{.}(2021)]%
        {mendez2021showing}
\bibfield{author}{\bibinfo{person}{Gonzalo Mendez}, \bibinfo{person}{Luis Gal{\'a}rraga}, {and} \bibinfo{person}{Katherine Chiluiza}.} \bibinfo{year}{2021}\natexlab{}.
\newblock \showarticletitle{Showing academic performance predictions during term planning: effects on students’ decisions, behaviors, and preferences}. In \bibinfo{booktitle}{\emph{Proceedings of the 2021 CHI Conference on Human Factors in Computing Systems}}. \bibinfo{pages}{1--17}.
\newblock


\bibitem[Miller(2019)]%
        {miller2019explanation}
\bibfield{author}{\bibinfo{person}{Tim Miller}.} \bibinfo{year}{2019}\natexlab{}.
\newblock \showarticletitle{Explanation in artificial intelligence: Insights from the social sciences}.
\newblock \bibinfo{journal}{\emph{Artificial intelligence}}  \bibinfo{volume}{267} (\bibinfo{year}{2019}), \bibinfo{pages}{1--38}.
\newblock


\bibitem[Molnar(2020)]%
        {molnar2020interpretable}
\bibfield{author}{\bibinfo{person}{Christoph Molnar}.} \bibinfo{year}{2020}\natexlab{}.
\newblock \bibinfo{booktitle}{\emph{Interpretable machine learning}}.
\newblock \bibinfo{publisher}{Lulu. com}.
\newblock


\bibitem[Morshed et~al\mbox{.}(2019)]%
        {morshed2019prediction}
\bibfield{author}{\bibinfo{person}{Mehrab~Bin Morshed}, \bibinfo{person}{Koustuv Saha}, \bibinfo{person}{Richard Li}, \bibinfo{person}{Sidney~K D'Mello}, \bibinfo{person}{Munmun De~Choudhury}, \bibinfo{person}{Gregory~D Abowd}, {and} \bibinfo{person}{Thomas Pl{\"o}tz}.} \bibinfo{year}{2019}\natexlab{}.
\newblock \showarticletitle{Prediction of mood instability with passive sensing}.
\newblock \bibinfo{journal}{\emph{Proceedings of the ACM on Interactive, Mobile, Wearable and Ubiquitous Technologies}} \bibinfo{volume}{3}, \bibinfo{number}{3} (\bibinfo{year}{2019}), \bibinfo{pages}{1--21}.
\newblock


\bibitem[Mwalumbwe and Mtebe(2017)]%
        {mwalumbwe2017using}
\bibfield{author}{\bibinfo{person}{Imani Mwalumbwe} {and} \bibinfo{person}{Joel~S Mtebe}.} \bibinfo{year}{2017}\natexlab{}.
\newblock \showarticletitle{Using learning analytics to predict students’ performance in Moodle learning management system: A case of Mbeya University of Science and Technology}.
\newblock \bibinfo{journal}{\emph{The Electronic Journal of Information Systems in Developing Countries}} \bibinfo{volume}{79}, \bibinfo{number}{1} (\bibinfo{year}{2017}), \bibinfo{pages}{1--13}.
\newblock


\bibitem[Nadal et~al\mbox{.}(2020)]%
        {nadal2020injurious}
\bibfield{author}{\bibinfo{person}{Kevin~L Nadal}, \bibinfo{person}{Katie~E Griffin}, \bibinfo{person}{Yinglee Wong}, \bibinfo{person}{Kristin~C Davidoff}, {and} \bibinfo{person}{Lindsey~S Davis}.} \bibinfo{year}{2020}\natexlab{}.
\newblock \showarticletitle{The injurious relationship between racial microaggressions and physical health: Implications for social work}.
\newblock In \bibinfo{booktitle}{\emph{Microaggressions and Social Work Research, Practice and Education}}. \bibinfo{publisher}{Routledge}, \bibinfo{pages}{7--18}.
\newblock


\bibitem[Namoun and Alshanqiti(2020)]%
        {namoun2020predicting}
\bibfield{author}{\bibinfo{person}{Abdallah Namoun} {and} \bibinfo{person}{Abdullah Alshanqiti}.} \bibinfo{year}{2020}\natexlab{}.
\newblock \showarticletitle{Predicting student performance using data mining and learning analytics techniques: A systematic literature review}.
\newblock \bibinfo{journal}{\emph{Applied Sciences}} \bibinfo{volume}{11}, \bibinfo{number}{1} (\bibinfo{year}{2020}), \bibinfo{pages}{237}.
\newblock


\bibitem[Nepal et~al\mbox{.}(2022)]%
        {nepal2022covid}
\bibfield{author}{\bibinfo{person}{Subigya Nepal}, \bibinfo{person}{Weichen Wang}, \bibinfo{person}{Vlado Vojdanovski}, \bibinfo{person}{Jeremy~F Huckins}, \bibinfo{person}{Alex Dasilva}, \bibinfo{person}{Meghan Meyer}, {and} \bibinfo{person}{Andrew Campbell}.} \bibinfo{year}{2022}\natexlab{}.
\newblock \showarticletitle{COVID student study: A year in the life of college students during the COVID-19 pandemic through the lens of mobile phone sensing}. In \bibinfo{booktitle}{\emph{Proceedings of the 2022 CHI conference on human factors in computing systems}}. \bibinfo{pages}{1--19}.
\newblock


\bibitem[Nghe et~al\mbox{.}(2007)]%
        {Nghe07}
\bibfield{author}{\bibinfo{person}{Nguyen~Thai Nghe}, \bibinfo{person}{Paul Janecek}, {and} \bibinfo{person}{Peter Haddawy}.} \bibinfo{year}{2007}\natexlab{}.
\newblock \showarticletitle{A comparative analysis of techniques for predicting academic performance}. In \bibinfo{booktitle}{\emph{2007 37th Annual Frontiers In Education Conference - Global Engineering: Knowledge Without Borders, Opportunities Without Passports}}. \bibinfo{pages}{T2G--7--T2G--12}.
\newblock
\urldef\tempurl%
\url{https://doi.org/10.1109/FIE.2007.4417993}
\showDOI{\tempurl}


\bibitem[Ojajuni et~al\mbox{.}(2021)]%
        {ojajuni2021predicting}
\bibfield{author}{\bibinfo{person}{Opeyemi Ojajuni}, \bibinfo{person}{Foluso Ayeni}, \bibinfo{person}{Olagunju Akodu}, \bibinfo{person}{Femi Ekanoye}, \bibinfo{person}{Samson Adewole}, \bibinfo{person}{Timothy Ayo}, \bibinfo{person}{Sanjay Misra}, {and} \bibinfo{person}{Victor Mbarika}.} \bibinfo{year}{2021}\natexlab{}.
\newblock \showarticletitle{Predicting student academic performance using machine learning}. In \bibinfo{booktitle}{\emph{Computational Science and Its Applications--ICCSA 2021: 21st International Conference, Cagliari, Italy, September 13--16, 2021, Proceedings, Part IX 21}}. Springer, \bibinfo{pages}{481--491}.
\newblock


\bibitem[Okano et~al\mbox{.}(2019)]%
        {okano2019sleep}
\bibfield{author}{\bibinfo{person}{Kana Okano}, \bibinfo{person}{Jakub~R Kaczmarzyk}, \bibinfo{person}{Neha Dave}, \bibinfo{person}{John~DE Gabrieli}, {and} \bibinfo{person}{Jeffrey~C Grossman}.} \bibinfo{year}{2019}\natexlab{}.
\newblock \showarticletitle{Sleep quality, duration, and consistency are associated with better academic performance in college students}.
\newblock \bibinfo{journal}{\emph{NPJ science of learning}} \bibinfo{volume}{4}, \bibinfo{number}{1} (\bibinfo{year}{2019}), \bibinfo{pages}{1--5}.
\newblock


\bibitem[Olteanu et~al\mbox{.}(2019)]%
        {olteanu2019social}
\bibfield{author}{\bibinfo{person}{Alexandra Olteanu}, \bibinfo{person}{Carlos Castillo}, \bibinfo{person}{Fernando Diaz}, {and} \bibinfo{person}{Emre K{\i}c{\i}man}.} \bibinfo{year}{2019}\natexlab{}.
\newblock \showarticletitle{Social data: Biases, methodological pitfalls, and ethical boundaries}.
\newblock \bibinfo{journal}{\emph{Frontiers in Big Data}}  \bibinfo{volume}{2} (\bibinfo{year}{2019}), \bibinfo{pages}{13}.
\newblock


\bibitem[Paschall and Freisthler(2003)]%
        {paschall2003does}
\bibfield{author}{\bibinfo{person}{Mallie~J Paschall} {and} \bibinfo{person}{Bridget Freisthler}.} \bibinfo{year}{2003}\natexlab{}.
\newblock \showarticletitle{Does heavy drinking affect academic performance in college? Findings from a prospective study of high achievers.}
\newblock \bibinfo{journal}{\emph{Journal of Studies on Alcohol}} \bibinfo{volume}{64}, \bibinfo{number}{4} (\bibinfo{year}{2003}), \bibinfo{pages}{515--519}.
\newblock


\bibitem[Pereira et~al\mbox{.}(2018)]%
        {pereira2018types}
\bibfield{author}{\bibinfo{person}{Juliana~L Pereira}, \bibinfo{person}{Gisela~Maria Guedes-Carneiro}, \bibinfo{person}{Liana~R Netto}, \bibinfo{person}{Patr{\'\i}cia Cavalcanti-Ribeiro}, \bibinfo{person}{Sidnei Lira}, \bibinfo{person}{Jos{\'e}~F Nogueira}, \bibinfo{person}{Carlos~A Teles}, \bibinfo{person}{Karestan~C Koenen}, \bibinfo{person}{Aline~S Sampaio}, \bibinfo{person}{Lucas~C Quarantini}, {et~al\mbox{.}}} \bibinfo{year}{2018}\natexlab{}.
\newblock \showarticletitle{Types of trauma, posttraumatic stress disorder, and academic performance in a population of university students}.
\newblock \bibinfo{journal}{\emph{The Journal of Nervous and Mental Disease}} \bibinfo{volume}{206}, \bibinfo{number}{7} (\bibinfo{year}{2018}), \bibinfo{pages}{507--512}.
\newblock


\bibitem[Pessach and Shmueli(2022)]%
        {pessach2022review}
\bibfield{author}{\bibinfo{person}{Dana Pessach} {and} \bibinfo{person}{Erez Shmueli}.} \bibinfo{year}{2022}\natexlab{}.
\newblock \showarticletitle{A Review on Fairness in Machine Learning}.
\newblock \bibinfo{journal}{\emph{ACM Computing Surveys (CSUR)}} \bibinfo{volume}{55}, \bibinfo{number}{3} (\bibinfo{year}{2022}), \bibinfo{pages}{1--44}.
\newblock


\bibitem[Pluut et~al\mbox{.}(2015)]%
        {pluut2015social}
\bibfield{author}{\bibinfo{person}{Helen Pluut}, \bibinfo{person}{Petru~Lucian Cur{\c{s}}eu}, {and} \bibinfo{person}{Remus Ilies}.} \bibinfo{year}{2015}\natexlab{}.
\newblock \showarticletitle{Social and study related stressors and resources among university entrants: Effects on well-being and academic performance}.
\newblock \bibinfo{journal}{\emph{Learning and Individual Differences}}  \bibinfo{volume}{37} (\bibinfo{year}{2015}), \bibinfo{pages}{262--268}.
\newblock


\bibitem[Pyle et~al\mbox{.}(2023)]%
        {pyle2023social}
\bibfield{author}{\bibinfo{person}{Cassidy Pyle}, \bibinfo{person}{Nicole~B Ellison}, {and} \bibinfo{person}{Nazanin Andalibi}.} \bibinfo{year}{2023}\natexlab{}.
\newblock \showarticletitle{Social Media and College-Related Social Support Exchange for First-Generation, Low-Income Students: The Role of Identity Disclosures}.
\newblock \bibinfo{journal}{\emph{Proceedings of the ACM on Human-Computer Interaction}} \bibinfo{volume}{7}, \bibinfo{number}{CSCW2} (\bibinfo{year}{2023}), \bibinfo{pages}{1--36}.
\newblock


\bibitem[Qu et~al\mbox{.}(2019)]%
        {qu2019predicting}
\bibfield{author}{\bibinfo{person}{Shaojie Qu}, \bibinfo{person}{Kan Li}, \bibinfo{person}{Bo Wu}, \bibinfo{person}{Xuri Zhang}, {and} \bibinfo{person}{Kaihao Zhu}.} \bibinfo{year}{2019}\natexlab{}.
\newblock \showarticletitle{Predicting student performance and deficiency in mastering knowledge points in MOOCs using multi-task learning}.
\newblock \bibinfo{journal}{\emph{Entropy}} \bibinfo{volume}{21}, \bibinfo{number}{12} (\bibinfo{year}{2019}), \bibinfo{pages}{1216}.
\newblock


\bibitem[Radloff(1977)]%
        {radloff1977ces}
\bibfield{author}{\bibinfo{person}{Lenore~Sawyer Radloff}.} \bibinfo{year}{1977}\natexlab{}.
\newblock \showarticletitle{The CES-D scale: A self-report depression scale for research in the general population}.
\newblock \bibinfo{journal}{\emph{Applied psychological measurement}} \bibinfo{volume}{1}, \bibinfo{number}{3} (\bibinfo{year}{1977}), \bibinfo{pages}{385--401}.
\newblock


\bibitem[Ribeiro et~al\mbox{.}(2016)]%
        {ribeiro2016should}
\bibfield{author}{\bibinfo{person}{Marco~Tulio Ribeiro}, \bibinfo{person}{Sameer Singh}, {and} \bibinfo{person}{Carlos Guestrin}.} \bibinfo{year}{2016}\natexlab{}.
\newblock \showarticletitle{" Why should i trust you?" Explaining the predictions of any classifier}. In \bibinfo{booktitle}{\emph{Proceedings of the 22nd ACM SIGKDD international conference on knowledge discovery and data mining}}. \bibinfo{pages}{1135--1144}.
\newblock


\bibitem[Rosenfeld and Richardson(2019)]%
        {rosenfeld2019explainability}
\bibfield{author}{\bibinfo{person}{Avi Rosenfeld} {and} \bibinfo{person}{Ariella Richardson}.} \bibinfo{year}{2019}\natexlab{}.
\newblock \showarticletitle{Explainability in human--agent systems}.
\newblock \bibinfo{journal}{\emph{Autonomous agents and multi-agent systems}}  \bibinfo{volume}{33} (\bibinfo{year}{2019}), \bibinfo{pages}{673--705}.
\newblock


\bibitem[Ruder(2017)]%
        {ruder2017overview}
\bibfield{author}{\bibinfo{person}{Sebastian Ruder}.} \bibinfo{year}{2017}\natexlab{}.
\newblock \showarticletitle{An overview of multi-task learning in deep neural networks}.
\newblock \bibinfo{journal}{\emph{arXiv preprint arXiv:1706.05098}} (\bibinfo{year}{2017}).
\newblock


\bibitem[Sano et~al\mbox{.}(2015)]%
        {sano2015recognizing}
\bibfield{author}{\bibinfo{person}{Akane Sano}, \bibinfo{person}{Andrew~J Phillips}, \bibinfo{person}{Z~Yu Amy}, \bibinfo{person}{Andrew~W McHill}, \bibinfo{person}{Sara Taylor}, \bibinfo{person}{Natasha Jaques}, \bibinfo{person}{Charles~A Czeisler}, \bibinfo{person}{Elizabeth~B Klerman}, {and} \bibinfo{person}{Rosalind~W Picard}.} \bibinfo{year}{2015}\natexlab{}.
\newblock \showarticletitle{Recognizing academic performance, sleep quality, stress level, and mental health using personality traits, wearable sensors and mobile phones}. In \bibinfo{booktitle}{\emph{2015 IEEE 12th International Conference on Wearable and Implantable Body Sensor Networks (BSN)}}. IEEE, \bibinfo{pages}{1--6}.
\newblock


\bibitem[Sefidgar et~al\mbox{.}(2019)]%
        {sefidgar2019passively}
\bibfield{author}{\bibinfo{person}{Yasaman~S Sefidgar}, \bibinfo{person}{Woosuk Seo}, \bibinfo{person}{Kevin~S Kuehn}, \bibinfo{person}{Tim Althoff}, \bibinfo{person}{Anne Browning}, \bibinfo{person}{Eve Riskin}, \bibinfo{person}{Paula~S Nurius}, \bibinfo{person}{Anind~K Dey}, {and} \bibinfo{person}{Jennifer Mankoff}.} \bibinfo{year}{2019}\natexlab{}.
\newblock \showarticletitle{Passively-sensed Behavioral Correlates of Discrimination Events in College Students}.
\newblock \bibinfo{journal}{\emph{Proceedings of the ACM on Human-Computer Interaction}} \bibinfo{volume}{3}, \bibinfo{number}{CSCW} (\bibinfo{year}{2019}), \bibinfo{pages}{1--29}.
\newblock


\bibitem[Selbst et~al\mbox{.}(2019)]%
        {selbst2019fairness}
\bibfield{author}{\bibinfo{person}{Andrew~D Selbst}, \bibinfo{person}{Danah Boyd}, \bibinfo{person}{Sorelle~A Friedler}, \bibinfo{person}{Suresh Venkatasubramanian}, {and} \bibinfo{person}{Janet Vertesi}.} \bibinfo{year}{2019}\natexlab{}.
\newblock \showarticletitle{Fairness and abstraction in sociotechnical systems}. In \bibinfo{booktitle}{\emph{Proceedings of the conference on fairness, accountability, and transparency}}. \bibinfo{pages}{59--68}.
\newblock


\bibitem[Silvola et~al\mbox{.}(2021)]%
        {silvola2021expectations}
\bibfield{author}{\bibinfo{person}{Anni Silvola}, \bibinfo{person}{Piia N{\"a}ykki}, \bibinfo{person}{Anceli Kaveri}, {and} \bibinfo{person}{Hanni Muukkonen}.} \bibinfo{year}{2021}\natexlab{}.
\newblock \showarticletitle{Expectations for supporting student engagement with learning analytics: An academic path perspective}.
\newblock \bibinfo{journal}{\emph{Computers \& Education}}  \bibinfo{volume}{168} (\bibinfo{year}{2021}), \bibinfo{pages}{104192}.
\newblock


\bibitem[Sternthal et~al\mbox{.}(2011)]%
        {sternthal2011racial}
\bibfield{author}{\bibinfo{person}{Michelle~J Sternthal}, \bibinfo{person}{Natalie Slopen}, {and} \bibinfo{person}{David~R Williams}.} \bibinfo{year}{2011}\natexlab{}.
\newblock \showarticletitle{Racial disparities in health: how much does stress really matter? 1}.
\newblock \bibinfo{journal}{\emph{Du Bois review: social science research on race}} \bibinfo{volume}{8}, \bibinfo{number}{1} (\bibinfo{year}{2011}), \bibinfo{pages}{95--113}.
\newblock


\bibitem[Stone et~al\mbox{.}(2012)]%
        {stone2012day}
\bibfield{author}{\bibinfo{person}{Arthur~A Stone}, \bibinfo{person}{Stefan Schneider}, {and} \bibinfo{person}{James~K Harter}.} \bibinfo{year}{2012}\natexlab{}.
\newblock \showarticletitle{Day-of-week mood patterns in the United States: On the existence of ‘Blue Monday’,‘Thank God it's Friday’and weekend effects}.
\newblock \bibinfo{journal}{\emph{The Journal of Positive Psychology}} \bibinfo{volume}{7}, \bibinfo{number}{4} (\bibinfo{year}{2012}), \bibinfo{pages}{306--314}.
\newblock


\bibitem[Strahan(2003)]%
        {strahan2003effects}
\bibfield{author}{\bibinfo{person}{Esther~Y Strahan}.} \bibinfo{year}{2003}\natexlab{}.
\newblock \showarticletitle{The effects of social anxiety and social skills on academic performance}.
\newblock \bibinfo{journal}{\emph{Personality and individual differences}} \bibinfo{volume}{34}, \bibinfo{number}{2} (\bibinfo{year}{2003}), \bibinfo{pages}{347--366}.
\newblock


\bibitem[Sukhbaatar et~al\mbox{.}(2019)]%
        {sukhbaatar2019artificial}
\bibfield{author}{\bibinfo{person}{Otgontsetseg Sukhbaatar}, \bibinfo{person}{Tsuyoshi Usagawa}, {and} \bibinfo{person}{Lodoiravsal Choimaa}.} \bibinfo{year}{2019}\natexlab{}.
\newblock \showarticletitle{An artificial neural network based early prediction of failure-prone students in blended learning course}.
\newblock \bibinfo{journal}{\emph{International Journal of Emerging Technologies in Learning (iJET)}} \bibinfo{volume}{14}, \bibinfo{number}{19} (\bibinfo{year}{2019}), \bibinfo{pages}{77--92}.
\newblock


\bibitem[Sumuer(2021)]%
        {sumuer2021effect}
\bibfield{author}{\bibinfo{person}{Evren Sumuer}.} \bibinfo{year}{2021}\natexlab{}.
\newblock \showarticletitle{The effect of mobile phone usage policy on college students’ learning}.
\newblock \bibinfo{journal}{\emph{Journal of Computing in Higher Education}} \bibinfo{volume}{33}, \bibinfo{number}{2} (\bibinfo{year}{2021}), \bibinfo{pages}{281--295}.
\newblock


\bibitem[Suresh and Guttag(2019)]%
        {suresh2019framework}
\bibfield{author}{\bibinfo{person}{Harini Suresh} {and} \bibinfo{person}{John~V Guttag}.} \bibinfo{year}{2019}\natexlab{}.
\newblock \showarticletitle{A framework for understanding unintended consequences of machine learning}.
\newblock \bibinfo{journal}{\emph{arXiv preprint arXiv:1901.10002}}  \bibinfo{volume}{2} (\bibinfo{year}{2019}).
\newblock


\bibitem[Thayer et~al\mbox{.}(2018)]%
        {thayer2018wise}
\bibfield{author}{\bibinfo{person}{Andrew~J Thayer}, \bibinfo{person}{Clayton~R Cook}, \bibinfo{person}{Aria~E Fiat}, \bibinfo{person}{Meghanne~N Bartlett-Chase}, {and} \bibinfo{person}{Jessie~M Kember}.} \bibinfo{year}{2018}\natexlab{}.
\newblock \showarticletitle{Wise feedback as a timely intervention for at-risk students transitioning into high school}.
\newblock \bibinfo{journal}{\emph{School Psychology Review}} \bibinfo{volume}{47}, \bibinfo{number}{3} (\bibinfo{year}{2018}), \bibinfo{pages}{275--290}.
\newblock


\bibitem[Thurber and Walton(2012)]%
        {thurber2012homesickness}
\bibfield{author}{\bibinfo{person}{Christopher~A Thurber} {and} \bibinfo{person}{Edward~A Walton}.} \bibinfo{year}{2012}\natexlab{}.
\newblock \showarticletitle{Homesickness and adjustment in university students}.
\newblock \bibinfo{journal}{\emph{Journal of American college health}} \bibinfo{volume}{60}, \bibinfo{number}{5} (\bibinfo{year}{2012}), \bibinfo{pages}{415--419}.
\newblock


\bibitem[Trockel et~al\mbox{.}(2000)]%
        {trockel2000health}
\bibfield{author}{\bibinfo{person}{Mickey~T Trockel}, \bibinfo{person}{Michael~D Barnes}, {and} \bibinfo{person}{Dennis~L Egget}.} \bibinfo{year}{2000}\natexlab{}.
\newblock \showarticletitle{Health-related variables and academic performance among first-year college students: Implications for sleep and other behaviors}.
\newblock \bibinfo{journal}{\emph{Journal of American college health}} \bibinfo{volume}{49}, \bibinfo{number}{3} (\bibinfo{year}{2000}), \bibinfo{pages}{125--131}.
\newblock


\bibitem[Tudor-Locke et~al\mbox{.}(2018)]%
        {tudor2018fast}
\bibfield{author}{\bibinfo{person}{Catrine Tudor-Locke}, \bibinfo{person}{Ho Han}, \bibinfo{person}{Elroy~J Aguiar}, \bibinfo{person}{Tiago~V Barreira}, \bibinfo{person}{John~M Schuna~Jr}, \bibinfo{person}{Minsoo Kang}, {and} \bibinfo{person}{David~A Rowe}.} \bibinfo{year}{2018}\natexlab{}.
\newblock \showarticletitle{How fast is fast enough? Walking cadence (steps/min) as a practical estimate of intensity in adults: a narrative review}.
\newblock \bibinfo{journal}{\emph{British Journal of Sports Medicine}} \bibinfo{volume}{52}, \bibinfo{number}{12} (\bibinfo{year}{2018}), \bibinfo{pages}{776--788}.
\newblock


\bibitem[US~Equal Employment Opportunity~Commission et~al\mbox{.}(1978)]%
        {us1978uniform}
\bibfield{author}{\bibinfo{person}{Civil Service Commission Department o f~Labor US~Equal Employment Opportunity~Commission}, \bibinfo{person}{Department o~f Justice}, {et~al\mbox{.}}} \bibinfo{year}{1978}\natexlab{}.
\newblock \showarticletitle{Uniform guidelines on employee selection procedures}.
\newblock \bibinfo{journal}{\emph{Federal Register}} \bibinfo{volume}{43}, \bibinfo{number}{166} (\bibinfo{year}{1978}), \bibinfo{pages}{38295--38309}.
\newblock


\bibitem[Van~der Zanden et~al\mbox{.}(2018)]%
        {van2018domains}
\bibfield{author}{\bibinfo{person}{Petrie~JAC Van~der Zanden}, \bibinfo{person}{Eddie Denessen}, \bibinfo{person}{Antonius~HN Cillessen}, {and} \bibinfo{person}{Paulien~C Meijer}.} \bibinfo{year}{2018}\natexlab{}.
\newblock \showarticletitle{Domains and predictors of first-year student success: A systematic review}.
\newblock \bibinfo{journal}{\emph{Educational Research Review}}  \bibinfo{volume}{23} (\bibinfo{year}{2018}), \bibinfo{pages}{57--77}.
\newblock


\bibitem[Verma and Rubin(2018)]%
        {verma2018fairness}
\bibfield{author}{\bibinfo{person}{Sahil Verma} {and} \bibinfo{person}{Julia Rubin}.} \bibinfo{year}{2018}\natexlab{}.
\newblock \showarticletitle{Fairness definitions explained}. In \bibinfo{booktitle}{\emph{2018 ieee/acm international workshop on software fairness (fairware)}}. IEEE, \bibinfo{pages}{1--7}.
\newblock


\bibitem[Vi{\v{s}}nji{\'c} et~al\mbox{.}(2018)]%
        {vivsnjic2018relationship}
\bibfield{author}{\bibinfo{person}{Aleksandar Vi{\v{s}}nji{\'c}}, \bibinfo{person}{Vladica Veli{\v{c}}kovi{\'c}}, \bibinfo{person}{Du{\v{s}}an Sokolovi{\'c}}, \bibinfo{person}{Miodrag Stankovi{\'c}}, \bibinfo{person}{Kristijan Mijatovi{\'c}}, \bibinfo{person}{Miodrag Stojanovi{\'c}}, \bibinfo{person}{Zoran Milo{\v{s}}evi{\'c}}, {and} \bibinfo{person}{Olivera Radulovi{\'c}}.} \bibinfo{year}{2018}\natexlab{}.
\newblock \showarticletitle{Relationship between the manner of mobile phone use and depression, anxiety, and stress in university students}.
\newblock \bibinfo{journal}{\emph{International journal of environmental research and public health}} \bibinfo{volume}{15}, \bibinfo{number}{4} (\bibinfo{year}{2018}), \bibinfo{pages}{697}.
\newblock


\bibitem[Waheed et~al\mbox{.}(2023)]%
        {waheed2023early}
\bibfield{author}{\bibinfo{person}{Hajra Waheed}, \bibinfo{person}{Saeed-Ul Hassan}, \bibinfo{person}{Raheel Nawaz}, \bibinfo{person}{Naif~R Aljohani}, \bibinfo{person}{Guanliang Chen}, {and} \bibinfo{person}{Dragan Gasevic}.} \bibinfo{year}{2023}\natexlab{}.
\newblock \showarticletitle{Early prediction of learners at risk in self-paced education: A neural network approach}.
\newblock \bibinfo{journal}{\emph{Expert Systems with Applications}}  \bibinfo{volume}{213} (\bibinfo{year}{2023}), \bibinfo{pages}{118868}.
\newblock


\bibitem[Wang et~al\mbox{.}(2014)]%
        {wang2014studentlife}
\bibfield{author}{\bibinfo{person}{R. Wang}, \bibinfo{person}{F.~Chenand~Z. Chen}, \bibinfo{person}{T. Li}, \bibinfo{person}{G. Harari}, \bibinfo{person}{S. Tignor}, \bibinfo{person}{X. Zhou}, \bibinfo{person}{D. Ben-Zeev}, {and} \bibinfo{person}{A.~T. Campbell}.} \bibinfo{year}{2014}\natexlab{}.
\newblock \showarticletitle{Studentlife: Assessing mental health, academic performance and behavioral trends of college students using smartphones.}. In \bibinfo{booktitle}{\emph{Proceedings of the 2014 ACM International Joint Conference on Pervasive and Ubiquitous Computing}}. \bibinfo{pages}{3–14}.
\newblock


\bibitem[Wang et~al\mbox{.}(2015)]%
        {wang2015smartgpa}
\bibfield{author}{\bibinfo{person}{Rui Wang}, \bibinfo{person}{Gabriella Harari}, \bibinfo{person}{Peilin Hao}, \bibinfo{person}{Xia Zhou}, {and} \bibinfo{person}{Andrew~T Campbell}.} \bibinfo{year}{2015}\natexlab{}.
\newblock \showarticletitle{SmartGPA: how smartphones can assess and predict academic performance of college students}. In \bibinfo{booktitle}{\emph{Proceedings of the 2015 ACM international joint conference on pervasive and ubiquitous computing}}. \bibinfo{pages}{295--306}.
\newblock


\bibitem[Wang et~al\mbox{.}(2017)]%
        {wang2017time}
\bibfield{author}{\bibinfo{person}{Zhiguang Wang}, \bibinfo{person}{Weizhong Yan}, {and} \bibinfo{person}{Tim Oates}.} \bibinfo{year}{2017}\natexlab{}.
\newblock \showarticletitle{Time series classification from scratch with deep neural networks: A strong baseline}. In \bibinfo{booktitle}{\emph{2017 International joint conference on neural networks (IJCNN)}}. IEEE, \bibinfo{pages}{1578--1585}.
\newblock


\bibitem[Wyatt and Oswalt(2013)]%
        {wyatt2013comparing}
\bibfield{author}{\bibinfo{person}{Tammy Wyatt} {and} \bibinfo{person}{Sara~B Oswalt}.} \bibinfo{year}{2013}\natexlab{}.
\newblock \showarticletitle{Comparing mental health issues among undergraduate and graduate students}.
\newblock \bibinfo{journal}{\emph{American journal of health education}} \bibinfo{volume}{44}, \bibinfo{number}{2} (\bibinfo{year}{2013}), \bibinfo{pages}{96--107}.
\newblock


\bibitem[Xu et~al\mbox{.}(2022a)]%
        {xu2022algorithmic}
\bibfield{author}{\bibinfo{person}{Jie Xu}, \bibinfo{person}{Yunyu Xiao}, \bibinfo{person}{Wendy~Hui Wang}, \bibinfo{person}{Yue Ning}, \bibinfo{person}{Elizabeth~A Shenkman}, \bibinfo{person}{Jiang Bian}, {and} \bibinfo{person}{Fei Wang}.} \bibinfo{year}{2022}\natexlab{a}.
\newblock \showarticletitle{Algorithmic fairness in computational medicine}.
\newblock \bibinfo{journal}{\emph{EBioMedicine}}  \bibinfo{volume}{84} (\bibinfo{year}{2022}).
\newblock


\bibitem[Xu et~al\mbox{.}(2019a)]%
        {xu2019leveraging}
\bibfield{author}{\bibinfo{person}{Xuhai Xu}, \bibinfo{person}{Prerna Chikersal}, \bibinfo{person}{Afsaneh Doryab}, \bibinfo{person}{Daniella~K Villalba}, \bibinfo{person}{Janine~M Dutcher}, \bibinfo{person}{Michael~J Tumminia}, \bibinfo{person}{Tim Althoff}, \bibinfo{person}{Sheldon Cohen}, \bibinfo{person}{Kasey~G Creswell}, \bibinfo{person}{J~David Creswell}, {et~al\mbox{.}}} \bibinfo{year}{2019}\natexlab{a}.
\newblock \showarticletitle{Leveraging routine behavior and contextually-filtered features for depression detection among college students}.
\newblock \bibinfo{journal}{\emph{Proceedings of the ACM on Interactive, Mobile, Wearable and Ubiquitous Technologies}} \bibinfo{volume}{3}, \bibinfo{number}{3} (\bibinfo{year}{2019}), \bibinfo{pages}{1--33}.
\newblock


\bibitem[Xu et~al\mbox{.}(2023)]%
        {xu2023globem}
\bibfield{author}{\bibinfo{person}{Xuhai Xu}, \bibinfo{person}{Xin Liu}, \bibinfo{person}{Han Zhang}, \bibinfo{person}{Weichen Wang}, \bibinfo{person}{Subigya Nepal}, \bibinfo{person}{Yasaman Sefidgar}, \bibinfo{person}{Woosuk Seo}, \bibinfo{person}{Kevin~S Kuehn}, \bibinfo{person}{Jeremy~F Huckins}, \bibinfo{person}{Margaret~E Morris}, {et~al\mbox{.}}} \bibinfo{year}{2023}\natexlab{}.
\newblock \showarticletitle{GLOBEM: cross-dataset generalization of longitudinal human behavior modeling}.
\newblock \bibinfo{journal}{\emph{Proceedings of the ACM on Interactive, Mobile, Wearable and Ubiquitous Technologies}} \bibinfo{volume}{6}, \bibinfo{number}{4} (\bibinfo{year}{2023}), \bibinfo{pages}{1--34}.
\newblock


\bibitem[Xu et~al\mbox{.}(2019b)]%
        {xu2019prediction}
\bibfield{author}{\bibinfo{person}{Xing Xu}, \bibinfo{person}{Jianzhong Wang}, \bibinfo{person}{Hao Peng}, {and} \bibinfo{person}{Ruilin Wu}.} \bibinfo{year}{2019}\natexlab{b}.
\newblock \showarticletitle{Prediction of academic performance associated with internet usage behaviors using machine learning algorithms}.
\newblock \bibinfo{journal}{\emph{Computers in Human Behavior}}  \bibinfo{volume}{98} (\bibinfo{year}{2019}), \bibinfo{pages}{166--173}.
\newblock


\bibitem[Xu et~al\mbox{.}(2022b)]%
        {xu2022globem}
\bibfield{author}{\bibinfo{person}{Xuhai Xu}, \bibinfo{person}{Han Zhang}, \bibinfo{person}{Yasaman Sefidgar}, \bibinfo{person}{Yiyi Ren}, \bibinfo{person}{Xin Liu}, \bibinfo{person}{Woosuk Seo}, \bibinfo{person}{Jennifer Brown}, \bibinfo{person}{Kevin Kuehn}, \bibinfo{person}{Mike Merrill}, \bibinfo{person}{Paula Nurius}, {et~al\mbox{.}}} \bibinfo{year}{2022}\natexlab{b}.
\newblock \showarticletitle{GLOBEM dataset: multi-year datasets for longitudinal human behavior modeling generalization}.
\newblock \bibinfo{journal}{\emph{Advances in Neural Information Processing Systems}}  \bibinfo{volume}{35} (\bibinfo{year}{2022}), \bibinfo{pages}{24655--24692}.
\newblock


\bibitem[Xu et~al\mbox{.}(2022c)]%
        {xuglobem}
\bibfield{author}{\bibinfo{person}{Xuhai Xu}, \bibinfo{person}{Han Zhang}, \bibinfo{person}{Yasaman~S Sefidgar}, \bibinfo{person}{Yiyi Ren}, \bibinfo{person}{Xin Liu}, \bibinfo{person}{Woosuk Seo}, \bibinfo{person}{Jennifer Brown}, \bibinfo{person}{Kevin~Scott Kuehn}, \bibinfo{person}{Mike~A Merrill}, \bibinfo{person}{Paula~S Nurius}, {et~al\mbox{.}}} \bibinfo{year}{2022}\natexlab{c}.
\newblock \showarticletitle{GLOBEM: Multi-Year Datasets for Longitudinal Human Behavior Modeling Generalization}.
\newblock \bibinfo{journal}{\emph{Thirty-sixth Conference on Neural Information Processing Systems Datasets and Benchmarks Track (Accepted)}} (\bibinfo{year}{2022}).
\newblock


\bibitem[Ya{\u{g}}c{\i}(2022)]%
        {yaugci2022educational}
\bibfield{author}{\bibinfo{person}{Mustafa Ya{\u{g}}c{\i}}.} \bibinfo{year}{2022}\natexlab{}.
\newblock \showarticletitle{Educational data mining: prediction of students' academic performance using machine learning algorithms}.
\newblock \bibinfo{journal}{\emph{Smart Learning Environments}} \bibinfo{volume}{9}, \bibinfo{number}{1} (\bibinfo{year}{2022}), \bibinfo{pages}{11}.
\newblock


\bibitem[Yao et~al\mbox{.}(2019)]%
        {yao2019predicting}
\bibfield{author}{\bibinfo{person}{Huaxiu Yao}, \bibinfo{person}{Defu Lian}, \bibinfo{person}{Yi Cao}, \bibinfo{person}{Yifan Wu}, {and} \bibinfo{person}{Tao Zhou}.} \bibinfo{year}{2019}\natexlab{}.
\newblock \showarticletitle{Predicting academic performance for college students: a campus behavior perspective}.
\newblock \bibinfo{journal}{\emph{ACM Transactions on Intelligent Systems and Technology (TIST)}} \bibinfo{volume}{10}, \bibinfo{number}{3} (\bibinfo{year}{2019}), \bibinfo{pages}{1--21}.
\newblock


\bibitem[Yeboah and Ewur(2014)]%
        {yeboah2014impact}
\bibfield{author}{\bibinfo{person}{Johnson Yeboah} {and} \bibinfo{person}{George~Dominic Ewur}.} \bibinfo{year}{2014}\natexlab{}.
\newblock \showarticletitle{The impact of WhatsApp messenger usage on students performance in Tertiary Institutions in Ghana}.
\newblock \bibinfo{journal}{\emph{Journal of Education and practice}} \bibinfo{volume}{5}, \bibinfo{number}{6} (\bibinfo{year}{2014}), \bibinfo{pages}{157--164}.
\newblock


\bibitem[Yoo et~al\mbox{.}(2024)]%
        {yoo2024missed}
\bibfield{author}{\bibinfo{person}{Dong~Whi Yoo}, \bibinfo{person}{Hayoung Woo}, \bibinfo{person}{Sachin~R Pendse}, \bibinfo{person}{Nathaniel~Young Lu}, \bibinfo{person}{Michael~L Birnbaum}, \bibinfo{person}{Gregory~D Abowd}, {and} \bibinfo{person}{Munmun De~Choudhury}.} \bibinfo{year}{2024}\natexlab{}.
\newblock \showarticletitle{Missed Opportunities for Human-Centered AI Research: Understanding Stakeholder Collaboration in Mental Health AI Research}.
\newblock \bibinfo{journal}{\emph{Proceedings of the ACM on Human-Computer Interaction}} \bibinfo{volume}{8}, \bibinfo{number}{CSCW1} (\bibinfo{year}{2024}), \bibinfo{pages}{1--24}.
\newblock


\bibitem[Yu et~al\mbox{.}(2018)]%
        {yu2018improving}
\bibfield{author}{\bibinfo{person}{Liang-Chih Yu}, \bibinfo{person}{Cheng-Wei Lee}, \bibinfo{person}{HI Pan}, \bibinfo{person}{Chih-Yueh Chou}, \bibinfo{person}{Po-Yao Chao}, \bibinfo{person}{ZH Chen}, \bibinfo{person}{SF Tseng}, \bibinfo{person}{CL Chan}, {and} \bibinfo{person}{K~Robert Lai}.} \bibinfo{year}{2018}\natexlab{}.
\newblock \showarticletitle{Improving early prediction of academic failure using sentiment analysis on self-evaluated comments}.
\newblock \bibinfo{journal}{\emph{Journal of Computer Assisted Learning}} \bibinfo{volume}{34}, \bibinfo{number}{4} (\bibinfo{year}{2018}), \bibinfo{pages}{358--365}.
\newblock


\bibitem[Zafar et~al\mbox{.}(2017)]%
        {zafar2017fairness}
\bibfield{author}{\bibinfo{person}{Muhammad~Bilal Zafar}, \bibinfo{person}{Isabel Valera}, \bibinfo{person}{Manuel Gomez~Rodriguez}, {and} \bibinfo{person}{Krishna~P Gummadi}.} \bibinfo{year}{2017}\natexlab{}.
\newblock \showarticletitle{Fairness beyond disparate treatment \& disparate impact: Learning classification without disparate mistreatment}. In \bibinfo{booktitle}{\emph{Proceedings of the 26th international conference on world wide web}}. \bibinfo{pages}{1171--1180}.
\newblock


\bibitem[Zhang et~al\mbox{.}(2018)]%
        {zhang2018mitigating}
\bibfield{author}{\bibinfo{person}{Brian~Hu Zhang}, \bibinfo{person}{Blake Lemoine}, {and} \bibinfo{person}{Margaret Mitchell}.} \bibinfo{year}{2018}\natexlab{}.
\newblock \showarticletitle{Mitigating unwanted biases with adversarial learning}. In \bibinfo{booktitle}{\emph{Proceedings of the 2018 AAAI/ACM Conference on AI, Ethics, and Society}}. \bibinfo{pages}{335--340}.
\newblock


\bibitem[Zhang et~al\mbox{.}(2022)]%
        {zhang2022impact}
\bibfield{author}{\bibinfo{person}{Han Zhang}, \bibinfo{person}{Margaret~E Morris}, \bibinfo{person}{Paula~S Nurius}, \bibinfo{person}{Kelly Mack}, \bibinfo{person}{Jennifer Brown}, \bibinfo{person}{Kevin~S Kuehn}, \bibinfo{person}{Yasaman~S Sefidgar}, \bibinfo{person}{Xuhai Xu}, \bibinfo{person}{Eve~A Riskin}, \bibinfo{person}{Anind~K Dey}, {et~al\mbox{.}}} \bibinfo{year}{2022}\natexlab{}.
\newblock \showarticletitle{Impact of Online Learning in the Context of COVID-19 on Undergraduates with Disabilities and Mental Health Concerns}.
\newblock \bibinfo{journal}{\emph{ACM Transactions on Accessible Computing (TACCESS)}} (\bibinfo{year}{2022}).
\newblock


\bibitem[Zhang et~al\mbox{.}(2024)]%
        {zhang2024illuminating}
\bibfield{author}{\bibinfo{person}{Han Zhang}, \bibinfo{person}{Vedant~Das Swain}, \bibinfo{person}{Leijie Wang}, \bibinfo{person}{Nan Gao}, \bibinfo{person}{Yilun Sheng}, \bibinfo{person}{Xuhai Xu}, \bibinfo{person}{Flora~D Salim}, \bibinfo{person}{Koustuv Saha}, \bibinfo{person}{Anind~K Dey}, {and} \bibinfo{person}{Jennifer Mankoff}.} \bibinfo{year}{2024}\natexlab{}.
\newblock \showarticletitle{Illuminating the Unseen: A Framework for Designing and Mitigating Context-induced Harms in Behavioral Sensing}.
\newblock \bibinfo{journal}{\emph{arXiv preprint arXiv:2404.14665}} (\bibinfo{year}{2024}).
\newblock


\bibitem[Zhang et~al\mbox{.}(2023)]%
        {zhang2023framework}
\bibfield{author}{\bibinfo{person}{Han Zhang}, \bibinfo{person}{Leijie Wang}, \bibinfo{person}{Yilun Sheng}, \bibinfo{person}{Xuhai Xu}, \bibinfo{person}{Jennifer Mankoff}, {and} \bibinfo{person}{Anind~K Dey}.} \bibinfo{year}{2023}\natexlab{}.
\newblock \showarticletitle{A framework for designing fair ubiquitous computing systems}. In \bibinfo{booktitle}{\emph{Adjunct Proceedings of the 2023 ACM International Joint Conference on Pervasive and Ubiquitous Computing \& the 2023 ACM International Symposium on Wearable Computing}}. \bibinfo{pages}{366--373}.
\newblock


\end{thebibliography}

\newpage

\appendix


\begin{table}[htb!]
    \caption{Reviewed academic performance prediction work sorted by amount of \textbf{Data} needed for prediction. The prediction \textbf{Task} is either classifying students into groups  (such as below and above 3.2, in our case) or regression (continues GPA). All papers focus on end-of-term GPA except \cite{Nghe07}, which detects end-of-year GPA and \cite{wang2015smartgpa}, which detects cumulative GPA.  The data set used as \textbf{Input} for each paper includes logs of online learning system use, student academic records, behavioral data, and self reports; the \textbf{Metrics} for assessing the model varied significantly, making comparison difficult. Half of the prior work did not consider model \textbf{Explainability}. Most of prior work did not consider model \textbf{Generalizability} for their models. No prior work considered \textbf{Fairness} of their models to marginalized student groups. \textbf{Ref.} represents for reference.} 
   \renewcommand{\arraystretch}{1.1}
   \resizebox{\textwidth}{!}{
\begin{tabular}{L{0.05\textwidth} C{0.1\textwidth} L{0.15\textwidth} L{0.18\textwidth} L{0.16\textwidth} C{0.13\textwidth} C{0.13\textwidth} C{0.09\textwidth}}
    \hline\hline
    \multicolumn{1}{c}{\textbf{Ref.}} &
    \multicolumn{1}{c}{\textbf{Data}} & \multicolumn{1}{c}{\textbf{Task}} & \multicolumn{1}{c}{\textbf{Input}}  & \multicolumn{1}{l}{\textbf{Metrics}} &
    \multicolumn{1}{c}{\makecell[c]{\makecell[c]{\textbf{Model}\\\textbf{Explainability}}}}  &
    \multicolumn{1}{c}{\makecell[c]{\textbf{Model}\\ \textbf{Fairness}}}&
    \multicolumn{1}{c}{\makecell[c]{\textbf{Model}\\ \textbf{Generalizability}}} \\\hline
    \cite{Nghe07}  & A year & \makecell[c]{End-of-year GPA \\ (2-class, 3-class \\ \& 4-class)}  & 
    Academic records; admissions information & \makecell[l]{Accuracy$\approx$ \\ 72\% (4-class); \\80\% (3-class); \\93\% (2-class)} & $\times$ & $\times$ & $\times$ \\\hline
     \cite{bravo2021early}  & A term (14-17 wks)  & \makecell[c]{End-of-Year GPA \\ (continuous)} & Learning Management System Log data; academic records; demographics &  \makecell[l]{$R$ = 0.677 \\ $R^2$ = 0.458 }& \checkmark & $\times$  & $\times$ \\\hline
    \cite{yao2019predicting} &  A term (14-17 wks)  & \makecell[c]{Term GPA \\ (continuous)} & Campus smart card logs; academic records &  \makecell[l]{Avg $r$=0.43,\\ SD=0.01} & \checkmark & $\times$ & $\times$ \\\hline
    \cite{qu2019predicting} & 14 weeks & \makecell[c]{Term GPA\\(2-class)} & Learning Management System Log Data & \makecell[l]{Accuracy=93\%, \\ Recall =0.95} & $\times$ & $\times$ & $\times$ \\\hline
   \cite{sukhbaatar2019artificial}  & 12 wks  & \makecell[c]{Term GPA \\ (2-class)} & Learning Management System Log data; academic records &  \makecell[l]{Avg accuracy$\approx$92\%,\\ Avg sensitivity=65\%, \\Avg precision$\approx$75\%, \\Avg F1=66\% }& $\times$ & $\times$ & $\times$ \\\hline
   \cite{wang2015smartgpa}   & 10 wks & \makecell[c]{Cumulative GPA \\ (continuous)}  &  Behavioral data from sensor; self-reports & \makecell[l]{MAE=0.18, \\ $r$=0.81, \\ $R^2$=0.56} & \checkmark & $\times$ & $\times$ \\\hline
   \cite{lara2014system} &   10 wks  & \makecell[c]{Term GPA \\(2-class)} & Online Learning Log data  & \makecell[l]{Accuracy=94\%, \\Precision=0.82,\\ Recall=0.90, \\Specificity=0.95} & $\times$ & $\times$ & $\times$ \\\hline
   \cite{lu2018applying} & 6 wks &  \makecell[c]{Term GPA \\ (continuous)} & Learning Management System Log data; academic records & \makecell[l]{PMSE=159.71,\\ $R^2$=0.56} & \checkmark & $\times$ & $\times$ \\\hline
   \cite{waheed2023early} & 5 wks & \makecell[c]{Single-class GPA \\ (2-class)} & Online Learning Log data, demographics, assessment-relatd data & \makecell[l]{Accuracy=69\% \\ Precision=0.70 \\ Recall=0.70 \\ AUC=0.71} & \checkmark & $\times$ & $\times$ \\\hline
   \cite{yu2018improving} & 5  wks & \makecell[c]{Term GPA \\ (2-class)} & Academic records; self-evaluation comments & \makecell[l]{Accuracy=71\%,\\ F1=0.71} & $\times$ & $\times$ & $\times$ \\ \hline 
   \cite{sano2015recognizing} & 4 wks & \makecell[c]{Term GPA \\ (2-class)} & Behavioral data from sensor; self-reports & \makecell[l]{Accuracy=92\%} & \checkmark & $\times$ & $\times$ \\\hline
   \cite{chen2020utilizing} & 4 wks & \makecell[c]{Single-class GPA \\ (2-class)} & Learning Management System Log data  & \makecell[l]{AUC=0.75 (original) \\ AUC=0.63 (unseen data)} & $\times$ & $\times$ & \checkmark\\
   \hline
   \hline
   \end{tabular}}
   \label{tab:related_work}
\end{table}

\section{A Review of Algorithmic Bias Measures}\label{appx:algorithmic_bias}
There are several common definitions of subpopulation-based algorithmic fairness and corresponding evaluation metrics.  We review the three most applicable fairness measures below with respect to a binary classification setting. We use all three measures to assess the algorithmic fairness of our approaches.

\textbf{Demographic Parity }\cite{barocas2016big}. Also commonly referred to as \textit{Independence} and \textit{Statistical Parity}, it requires the prediction of positive outcome,  $\hat{Y} = 1$, to be the same regardless of whether the person is in a protected (\eg female, disabled, and underrepresented minority) group ($S = 1$). Note that one disadvantage of \textit{demographic parity} is that a fully accurate classifier may be seen as biased when the ratios of actual positive outcomes of the groups differ~\cite{pessach2022review}. Mathematically, it is computed as follows:
\begin{equation*}
\label{eqn:demographic_parity}
P[\hat{Y} = 1|S = 1] = P[\hat{Y} = 1|S \neq 1], 
\end{equation*}

\textbf{Equalized Odds}~\cite{zafar2017fairness}. This measure requires the protected and unprotected groups to have the same rates for true positives (TPRs) and false positives (FPRs)~\cite{mehrabi2021survey}. It was designed to overcome the disadvantage of \textit{demographic parity} described above~\cite{hardt2016equality}.  Mathematically, it is computed as follows:

\begin{equation*}
\label{eqn:equalized_odds}
    P[\hat{Y} = 1|S = 1, Y = 0] = P[\hat{Y} = 1|S \neq 1, Y=0], P[\hat{Y} = 1|S = 1, Y = 1] = P[\hat{Y} = 1|S \neq 1, Y=1], 
\end{equation*}

\textbf{Equal Opportunity}~\cite{hardt2016equality}. This is also commonly referred to as \textit{recall} or \textit{sensitivity}. It is less strict than \textit{equalized odds}, which only requires the protected and unprotected groups to have equal true positive rates (or false negative rates). Mathematically, it is computed as follows:

\begin{equation*}
\label{eqn:equal_opportunity}
    P[\hat{Y} = 1|S = 1, Y = 1] = P[\hat{Y} = 1|S \neq 1, Y=1]. 
\end{equation*}

To set fairness criteria, the definition of disparity that is expressed as a difference is often considered~\cite{ahmad2020fairness,kobayashi2021one}. For example, the \textit{demographic parity difference} is defined as the difference in the probability of prediction between the two groups. Similarly, one may calculate an \textit{equalized odds difference}, or the greater of two metrics, TPR difference and FPR difference between the two groups; and an \textit{equal opportunity difference}, which only compares the TPRs between the unprotected and protected groups. In each case, a difference of 0 indicates that the model is perfectly fair to the protected trait (it favors neither the protected nor the unprotected group). 

Another common fairness criteria is to compute a ratio between groups ~\cite{ahmad2020fairness}. For example, the \textit{demographic parity ratio} (also called disparate impact~\cite{us1978uniform}) is defined as the ratio between the probability of positive prediction for the unprotected group and the probability of positive prediction for the protected group. A ratio of 1 indicates that the model is fair relative to the protected trait (it favors neither the protected nor the unprotected group). In US law, a value of demographic parity ratio (or disparate impact) more than 0.8 indicates that there is not an unfair situation (80\% rule)~\cite{us1978uniform,kabacoff1997psychometric}. Similarly, \textit{Equalized odds ratio} is defined as the smaller value of two metrics, TPR ratio and FNR ratio~\cite{fairLearnPackage}, where TPR and FNR ratios are calculated as the rate of the unprotected group divided by the rate of the protected group. An equalized odds ratio of 1 means that all groups have the same true positive, true negative, false positive, and false negative rates, respectively~\cite{fairLearnPackage}.  \textit{Equal opportunity ratio} is calculated as the ratio of TPRs between the unprotected and protected groups. A value of 1 means that all groups have the same TPR, and that the model is within the ``fair'' range relative to the protected trait.

\begin{table}[htb!]
\caption{Data information. Statistics, dropout rate and data missingness for 2018 and 2019. Dropout count refers to the number of participants left out of the modeling due to not having enough of a particular data type. Note that providing GPA data was optional, it was not a requirement of the study.} 
\scalebox{0.85}{
    \centering 
    \begin{tabular}{c c c c c c c c} 
    \toprule[1pt]\hline
     \textbf{Year} & \textbf{Study Period} 
     & \makecell[c]{ \textbf{\mbox{GPA}}\\\textbf{\mbox{Provided}}}
     & \textbf{Data Type}
     & \makecell[c]{\textbf{Dropout}\\ \textbf{\mbox{Count (Rate)}}}
        & \makecell[c]{\textbf{Data}\\ \textbf{\mbox{Missingness}}}
         & \makecell[c]{\textbf{\mbox{Sample}} \\ \textbf{\mbox{ used for}}\\ \textbf{\mbox{Modeling}}} &\makecell[c]{\textbf{Retention}}
     \\
    \hline 
    \multirow{3}{*}{2018} & \multirow{3}{*}{\makecell[c]{March 26 \\ 
    | \\ June 03}} &
    \multirow{3}{*}{195} & 
    sensor & 7 (3\%)  & 33\%  & \multirow{3}{*}{188}
    & \multirow{3}{*}{96\%} \\
    & & & \textit{survey} & 0 (0\%) &  30\% \\
    && & EMA & 0 (0\%)  & 12.5\%  \\
    \hline
    \multirow{3}{*}{2019} & \multirow{3}{*}{\makecell[c]{April 1 \\
    | \\ June 07}}  &
  \multirow{3}{*}{201} & sensor & 5 (2\%)& 16\%  & \multirow{3}{*}{196} & \multirow{3}{*}{98\%} \\
  && & \textit{survey} & 0 (0\%) &  29\% \\
    && & EMA & 0 (0\%)  & 5.1\%  \\
  \hline
    \bottomrule[1pt]
    \end{tabular}}
    \label{tab:study_completion}
\end{table}

\begin{figure}[htb!]
    \centering
    \subfloat[2018 spring term GPA]{\includegraphics[scale=0.5]{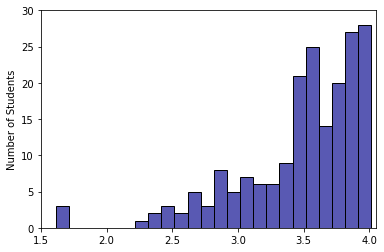}\label{fig:gpa_distribution_2018a}}
     \subfloat[2018 spring term GPA]{\includegraphics[scale=0.5]{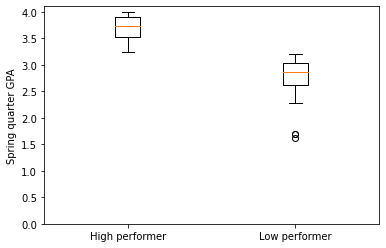}\label{fig:gpa_distribution_2018b}}
     
     \subfloat[2019 spring term GPA]{\includegraphics[scale=0.5]{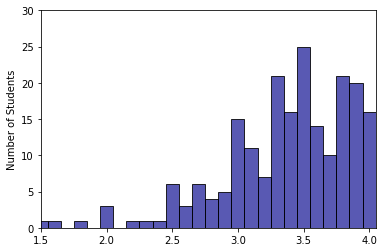}\label{fig:gpa_distribution_2019a}}
     \subfloat[2019 spring term GPA]{\includegraphics[scale=0.5]{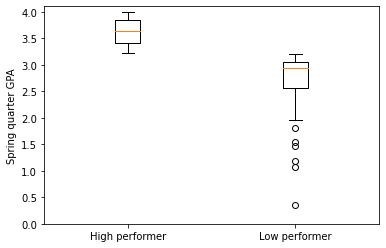}\label{fig:gpa_distribution_2019b}}
    \caption{Distributions of spring term GPA among all students and distribution of spring term GPA of high and low performers in 2018 and 2019. In 2018, 145 (77\%) out of 188 students are labeled as high performers ($M$ = 3.69, $std$ = 0.21), 43 (23\%) students are labeled as low performers ($M$ = 2.76, $std$ = 0.39). In 2019, 133 (68\%) out of 196 students are labeled as high performers ($M$ = 3.63, $std$ = 0.24), 66 (32\%) students are labeled as low performers ($M$ = 2.69, $std$ = 0.58). 
    } 
    \label{fig:group_gpa}
\end{figure}

\section{Behavioral Features}\label{appx:behavioral_features}

\begin{table}[htb!]
\caption{Passive-sensing data and extracted low-level behavior features.} 
\scalebox{0.85}{
    \begin{tabular}{c | c | c | c }
        \toprule[1pt]
        \hline
        \textbf{Source} & \textbf{Sensor} & \textbf{Sampling frequency} & \textbf{Low-level Behavior features}\\\hline
        \multirow{23}{*}{Smartphone} & Physical activity & Every minute&  \makecell[l]{Most common activity, number of activities} \\\cline{2-4}
        & Application usage & Event-based & \makecell[l]{Number of used apps, most commonly used app, \\  most common app category, apps used per minute}\\\cline{2-4}
        &Battery & Event-based & \makecell[l]{Number of charging sessions, total charging time}\\\cline{2-4}
        & Bluetooth & Every 3 minutes &  \makecell[l]{Number of scans, number of unique devices,\\number of scans of least frequent device,\\ number of scans of most frequent device, etc.}\\\cline{2-4}
        & Calls &Event-based & \makecell[l]{Number of incoming calls, number of outgoing calls,\\ number of missed calls, duration of incoming calls, \\ duration of outgoing calls, etc.}\\\cline{2-4}
        & Locations & Every minute & \makecell[l]{Total distance traveled, time spent at/near home,\\ average traveling speed, percentage of time moving,\\  time spent at top 3 location clusters, etc.}\\\cline{2-4}
        & Location map & Every minute & \makecell[l]{Time at exercise-labeled places, time at food-labeled places,\\ time at fraternity-labeled places, time at greens-labeled places,\\ time at living-labeled places, time at study-labeled places, etc.}\\\cline{2-4}
        & Screen &Event-based & \makecell[l]{Sum duration of phone interactions,\\  average duration of phone interactions, \\ standard deviation of interaction durations, \\ time of first unlock event, time of last unlock event, \\ number of unlocks per minute, etc.}\\\cline{2-4}
        & WiFi &Every 3 minutes & \makecell[l]{Number of unique access points, most frequent access point} \\\hline
        \multirow{5}{*}{Fitbit} & Sleep & Every minute & \makecell[l]{Time in bed, awake duration, asleep duration, \\ restless duration, sleep efficiency, etc.}\\\cline{2-4}
        & Step count & Every minute & \makecell[l]{Total step count, number of active bouts, \\ average duration of active bouts,  average steps per active bout,\\ start time of longest active bout, etc.} \\
        \bottomrule[1pt]
    \end{tabular}
    }
\label{tab:behavioral-features}
\end{table}

\subsection{Implementation of Low-level Behavior Features}\label{feature_implementation}
\textbf{Physical Activity.} For a given day/epoch, we counted the number of times a student's activity type changes (\eg from ``still'' to ``walking''), the number of unique activity types, and the most frequently logged activity type. 

\textbf{Application Usage.} We pre-processed the data to exclude system apps from our feature computation to focus mainly on user installed applications (UIA). For a given day/epoch, we calculated the number of unique apps used, and the most commonly used app and app category. We also calculated the average number of apps used per minute by the user.

\textbf{Battery.} We calculated the number of times users charge their phones and the total battery charging time to indicate how often and how long users charge their phones.

\textbf{Bluetooth.} We applied the K-means clustering algorithm to scanned Bluetooth addresses based on their frequency in the data set, and grouped the devices into 2 or 3 clusters depending on which can better separate the data points with more concentrated clusters, to differentiate the person’s own devices (labeled as ``self'') and other people’s devices (labeled as ``others'') \cite{doryab2018extraction}. We then calculated statistical features for each group of devices, such as the number of scans of most/least frequent device of self/others, the number of unique devices of self/others, total and average number of scans of all devices of self/others, \textit{etc}.

\textbf{Calls.} The call logs provide session information for incoming, outgoing and missed calls. We computed the total number as well as the total duration of calls belonging to each call type.

\textbf{Locations.} We extracted location variance, radius of gyration, total distance traveled and circadian movement features described in \cite{doryab2018extraction}. We used DBSCAN \cite{ester1996density} to group static location samples into clusters, and calculated the statistical features (\eg sum, mean, standard deviation, maximum, and minimum) on the duration of stay at each cluster. In addition, we calculated the entropy of the duration of stay at each cluster to evaluate how students distributed their time. We inferred students’ home locations by clustering their location data at night (12am to 6am). We considered a potential cluster to be a home location if the student stays there for more than 3 days in a row, and the dwelling time at the cluster is at least 80\% of each night. We then calculated the total time spent at home (within 10 meters from home) and near home (within 100 meters from home) accordingly based on their home locations. 

\textbf{Location Map.} To better map GPS location data to meaningful places, we hand labeled the boundaries of places of interest (\ie  exercise, food, frat house, greens, dorms/living and study) on campus to create a location map. For each location sample, we assigned a map label to it by comparing it against the location map. We then grouped consecutive samples of the same map label into bouts and calculated statistical features on the durations of the bouts.

\textbf{Screen.} We used screen data to define a phone interaction session as a time series with a screen status of ``on'' at the beginning and a screen status of ``off'' or ``locked'' at the end of the session. Similarly, we defined a screen unlock session to be a time series with a screen status of ``unlocked'' at the beginning and a screen status of ``locked'' at the end. We then computed statistical features (\eg sum, max, min, average, standard deviation) on the duration of interaction and unlock sessions. In addition, we extracted the time information of the first and last occurrence of different types of screen events (\ie on, off, unlock and lock), and calculated the average number of unlocks per minute to indicate the frequency of a user initiating a phone interaction. 

\textbf{WiFi.} We counted the number of unique WiFi access points sensed by the phone and identified the most frequently detected access point.

\textbf{Sleep.} We obtained students’ sleep logs through the Fitbit API (v1.0), which contain per-minute data of sleep status (\ie asleep, restless, and awake) throughout each sleep episode. We grouped consecutive sleep samples of the same status into sleep bouts and calculate statistical features on the asleep, restless and awake bouts such as the total number of awake bouts, the start time, end time, max, min and average duration of asleep bouts, restless bouts, \textit{etc}. We also considered Fitbit summary data \cite{fitbitsleep} as part of our daily features, including duration and efficiency of sleep, time in bed, \textit{etc}.

\textbf{Step Count.} We computed step features from the minute-by-minute data returned by the Fitbit API. Epidemiological studies report a mean daily cadence of 7.7 steps per minute at the population level \cite{tudor2018fast}. We used 12 steps per minute as a threshold to determine if a person is active or not in that minute. We grouped consecutive active or inactive samples into active or sedentary bouts, and calculated statistical features on the duration and step count of the bouts. We also extracted the start and end time of the active bout with the longest duration and the bout with the most steps.

\subsection{Implementation of High-level Behavior Features}\label{high-level_feature_implementation}
\textbf{Activity Duration.} We implemented this feature by grouping consecutive activity data samples with non-stationary labels (\ie  on foot, walking, running, and on bicycle) into activity bouts, and then computing the total activity duration of the student by summing up the duration of the bouts .

\textbf{Study duration and study focus.} We included any dwelling time of 20 minutes or greater at study labeled locations (\eg libraries, teaching buildings, and cafes) in the estimation of a student's study duration. We considered students being stationary at the study locations to be more focused on studying. By fusing location data and activity data, we calculated study focus as the percentage of the dwelling time with stationary activity labels (\ie  still and tilting) with respect to the total study duration.

\textbf{Dorm duration.} We computed this feature as the total amount of time a student spend at places labeled as ``dorm'' or ``living''.

\textbf{Party duration.} We considered students staying at the fraternity houses on campus any time from 6pm to 12pm the next day with a dwelling time of 30 minutes or above to be partying and calculate this feature by summing up the dwelling time. We excluded the students who live at the fraternity houses from the calculation.

\textbf{Indoor and outdoor mobility.} Similar to~\cite{wang2015smartgpa}, we fused location and activity data and calculate indoor mobility as the total amount of time when a student is walking or running indoors. We calculated outdoor mobility as the total distance traveled by the student when he/she is outdoors.

\textbf{Class Attendance.} We computed class attendance related features using both students’ class schedules and location data. Similar to location map features, we hand labeled the locations of all teaching buildings on campus. For each class period a student was scheduled to attend, we compared the student’s location during the class time against the teaching building of the scheduled class. We calculated the amount of time a student was at the correct building as a percentage of the total class duration, and considered the student attending the class only if the percentage is more than 50\%.

\textbf{Behavioral Change.} We divided each academic term into individual weeks and capture a student’s overall behavioral changes within each week. We followed a similar approach to~\cite{wang2015smartgpa} and computed slopes and breakpoints on a weekly basis for all the above-mentioned behavior features. We defined Thursday as the midpoint of each week (starting on Monday), and fit linear regression models to the data of the first half (midpoint excluded), second half (midpoint included) and the entire week, respectively. We designated the slopes of the above three linear regression models as first-half slope, second-half slope, and slope all. Note that, slope captures 1) the direction of behavioral change (\ie increases or decreases in sleep duration) and 2) magnitude of the behavioral change (\ie steep or gradual changes in sleep duration) within the first week, as well as the first half (Monday to Wednesday) and second half (Thursday to Sunday) of the week. Separate from the midpoint, we also computed breakpoints that capture the specific day in the first week when a student’s behavioral pattern shows a directional change (\ie the day when their sleep duration increases or decreases).

\section{Data Preprocessing}

\subsection{Common Data Cleaning}\label{appx:common_data_cleaning}

Before modeling, we assigned each participant a unique participant ID to ensure privacy, with all analyses conducted using anonymized data. Missing values primarily arose due to data collection challenges, such as app crashes, phones running out of battery, or participants failing to comply with study protocols, like not wearing their Fitbit or skipping questionnaires. Features that were 100\% missing were removed from the dataset. We handled numeric outliers by capping them based on the interquartile range (IQR) calculated for each student individually. Categorical features were transformed using one-hot encoding to prepare the data for model training.

\subsection{Customized Data Preprocessing for The LR Approach}\label{appx:data_prep_lr}
\subsubsection{Missing Value Handling} During data preprocessing, features with 100\% missingness across the dataset were removed. For remaining features with missing values, we tested two imputation methods: (1) imputing missing values with a default value (999), and (2) imputing values with the mean of the training set. Based on model performance in 2018, we selected the second method. During leave-one-subject-out cross-validation (LOSO-CV), if a feature in the training set was entirely missing, the default value of 999 was used.

\subsubsection{Class Imbalance Handling} Our data from 2018 and 2019 is imbalanced, with only 23\% and 32\% of students having lower GPAs, respectively. To address this, we experimented with SMOTE and ADASYN for oversampling the minority class in the training set to balance the classes. SMOTE was chosen based on the 2018 model performance~\cite{chawla2002smote}.

\subsubsection{Collinearity Removal} To avoid issues of collinearity that could distort model estimation, we removed features from the training and test sets that were highly correlated ($|r| >$ 0.7) based on training set data~\cite{dormann2013collinearity}.

\subsubsection{Feature Selection} We employed correlation-based feature selection (CFS~\cite{hall1999correlation}) to identify features significantly correlated with end-of-term GPA ($p < 0.05$). For each round of LOSO-CV, we performed a grid search to determine an optimal correlation threshold $r$, selecting the $r$ value that maximized the performance advantage ($a_{diff} = a_{test} - a_{train}$). We note that while the use of test data in determining $r$ introduces some leakage, this was only during feature inclusion, and no leakage occurred when applied to the 2019 data.

\subsection{Customized Data Preprocessing for The 1D-CNN Approach}\label{appx:data_prep_1d-cnn}

\subsubsection{Missing Value Handling}\label{subsub:data_imputation}
Since the data used in the deep learning model is time series data, we employed forward filling to impute missing values initially, followed by backward filling for any remaining gaps. This is a standard technique for handling missing data in time series~\cite{che2018recurrent}. Unlike the LR pipeline, which used mean imputation, this approach ensures that the imputed values reflect the temporal sequence of the data, as aggregating by week (as done in LR) does not apply to continuous time series data.

\subsubsection{Class Imbalance Handling}\label{subsub:class_imbalanced}
To address class imbalance, we adopted a simple oversampling approach by randomly duplicating samples from the minority class (i.e., low performers) to equalize the ratio between the two classes (1:1) in the training set.

\subsubsection{Data Standardization and Transformation}\label{subsub:data_trans}
We standardized all features and transformed the data into a three-dimensional time series format, suitable for deep learning models, structured as [number of participants, number of days, number of features].

\subsubsection{Architecture of 1D-CNN Model}\label{appx:1d-cnn_arch}

The architecture of the 1D-CNN model includes a single 1D convolutional layer (1D-CNN) followed by a rectified linear unit (ReLU) activation function. To prevent overfitting, we applied a dropout layer immediately after the 1D-CNN layer, masking 85\% of its output~\cite{gal2016theoretically}. This is followed by a max pooling layer, which reduces the spatial size by applying a max filter to non-overlapping subregions of the dropout layer’s output. The pooled output is then flattened into a single vector via a flattening layer. Finally, the model contains two dense (fully connected) layers: the first uses a ReLU activation function, while the output dense layer employs a softmax function to return class probabilities for the binary classification task. The model was optimized using the Adam optimizer, with categorical cross-entropy as the loss function. The training process used 150 epochs with a batch size of 6. Hyperparameters, including a learning rate of 0.0001, were selected using grid search. Additionally, early stopping was employed, halting training after 10 steps without improvement.

\section{Evaluations}

\subsection{Fairness Evaluation Results}\label{appx:fairness_eval}
\renewcommand{\arraystretch}{1.2}
\begin{table}[htb!]
\caption{Fairness evaluation of three approaches (LR, 1D-CNN, and MTL-1D-CNN) on four protected traits using the difference and ratio of demographic parity, equalized odds, and equal opportunity. Results that are considered reasonably fair (based on the literature) are in \textbf{bold}.}
\renewcommand{\arraystretch}{1.1}
\footnotesize
   \resizebox{\textwidth}{!}{
    \begin{tabular}{c|lcccccccccc}\hline\hline
    \multirow{2}{*}{\textbf{Approach}} && \multirow{2}{*}{\textbf{Protected Trait}}&&\multicolumn{2}{c}{\textbf{Demographic Parity}}&&\multicolumn{2}{c}{\textbf{Equalized Odds}}&&\multicolumn{2}{c}{\textbf{Equal Opportunity}} \\
    \cline{5-6}\cline{8-9}\cline{11-12}
    && && difference  & ratio && difference  &  ratio && difference & ratio \\\hline
    \multirow{4}{*}{LR} && \makecell[l]{Race} &&  \textbf{0.033} & \textbf{0.955}   &&  0.324 & 0.353   && \textbf{0.050}& \textbf{0.978}     \\\cline{2-12}
    && \makecell[l]{First-generation} &&  0.147 & \textbf{0.808}   &&   \textbf{0.023} & \textbf{0.949}  &&   \textbf{0.023} & \textbf{0.976}    \\\cline{2-12}
    && \makecell[l]{Gender} &&  \textbf{0.050} & \textbf{0.933}   &&  0.107 & 0.625   &&   \textbf{0.062} &\textbf{0.937}      \\\cline{2-12}
    && \makecell[l]{Sexual Orientation} &&  \textbf{0.095} & \textbf{0.882}   &&  \textbf{0.051} & \textbf{0.829} &&  \textbf{0.051} & \textbf{1.054}    \\\hline
    \multirow{4}{*}{1D-CNN} && \makecell[l]{Race} &&  \textbf{0.027} &  \textbf{0.963} && \textbf{0.085}& \textbf{0.882} &&\textbf{0.085} &  \textbf{0.882}  \\\cline{2-12}
    && \makecell[l]{First-generation} && 0.105  & \textbf{0.868} &&  0.170 &  0.798 && 0.170  &  1.253  \\\cline{2-12}
    && \makecell[l]{Gender} && \textbf{0.062} &  \textbf{0.917}  &&  0.221 &  0.733  && \textbf{0.013} & \textbf{0.982} \\\cline{2-12}
    && \makecell[l]{Sexual Orientation} && \textbf{0.006} &  \textbf{0.992} &&  \textbf{0.076} & \textbf{0.905} &&  \textbf{0.030} &  \textbf{0.958}    \\\hline
    
    \multirow{4}{*}{MTL-1D-CNN} && \makecell[l]{Race} &&  \textbf{0.068} & \textbf{0.900}   &&  0.353 & 0.471   &&   0.282 & 0.649       \\\cline{2-12}
  && \makecell[l]{First-generation} && 0.141 & \textbf{0.801} && 0.154 & \textbf{0.817} &&   0.154 & \textbf{0.817 }      \\\cline{2-12}
  && \makecell[l]{Gender} &&  \textbf{0.024} & \textbf{0.965}   &&  \textbf{0.051} & \textbf{0.893}  &&    \textbf{0.051} & \textbf{1.070}     \\\cline{2-12}
  && \makecell[l]{Sexual Orientation} &&  0.158 & \textbf{0.805}   &&   0.238 & 0.603  &&   \textbf{0.080}& \textbf{1.100}     \\\hline
    \hline                       
    \end{tabular}} 
\label{tab:fairness_evaluation}
\end{table}
\renewcommand{\arraystretch}{1.2}

\section{Academic-related Patterns and Factors}\label{appx:behavioral_patterns}
Below, we summarize behavioral patterns and factors and discuss their implications for early intervention strategies. These results are derived from Tables~\ref{tab:top30_features_2018} and \ref{tab:top30_features_2019}, where features suggesting similar patterns have been grouped together. 


\paragraph{Weekday vs Weekend Behaviors} One interesting observation is that many of the identified behavioral shifts (breakpoints in daily routines) during the first week of the Spring term, for both years, occur on Thursdays (\eg 2018-R7 to 2018-R10, 2018-R18, 2019-R2, 2019-R3). This suggests that students' weekend behaviors may begin on Fridays rather than Saturdays for a substantial portion of the population. This distinction could offer valuable insights for targeted interventions, as shifts in behavioral patterns earlier in the week may indicate opportunities for academic support or engagement efforts before the weekend. 

\paragraph{Class Attendance}
Not surprisingly, average class attendance during week one is positively associated with end-of-term GPA (2018-R14). This finding aligns with existing literature, which shows a strong relationship between class attendance and both individual course grades and overall GPA~\cite{crede2010class,boumi2021quantifying}. This consistency reinforces that the features identified in our study as predictors of academic performance are meaningful and worth exploring further. It also suggests that educators should take note of students' attendance early in the term and proactively check in with those who are not attending to understand potential barriers. Since flexibility in attendance is critical for addressing accessibility needs~\cite{zhang2022impact}, such outreach should avoid mandating physical presence, as this could place additional stress on students with disabilities or those experiencing mental health challenges.

\paragraph{Phone Usage} An increase in phone usage is negatively associated with students' end-of-term GPA (2018-R7 and 2019-R19), a finding supported by prior research showing a negative correlation between smartphone usage and academic performance~\cite{lepp2015relationship,giunchiglia2018mobile,yeboah2014impact}. Our results suggest that this effect is particularly pronounced on \textit{weekdays}, adding nuance to the general understanding of this relationship. Studies show that in-class phone use can significantly hinder student performance~\cite{sumuer2021effect}, with in-class usage being nearly double that of outside-classroom use~\cite{felisoni2018cell}. Whether phone use is a cause or consequence of struggling academically---or perhaps a related factor such as stress---is unclear, but these patterns suggest that students who are distracted by their phones during weekdays may not be setting themselves up for academic success. 

Phone usage is also linked to stress~\cite{vivsnjic2018relationship}, and excessive use can act as a negative coping strategy~\cite{augner2012associations}, which is further supported by our finding that feeling helpless in difficult situations (2018-R12) is negatively associated with GPA. This highlights an opportunity for student wellness programs or new student orientation initiatives to address the role of stress, coping strategies, and phone use in academic success. Interestingly, our findings also suggest that phone usage during weekends may help students unwind, as more frequent use in the evenings and nights later in the weekend is positively associated with academic outcomes (2018-R1 and 2018-R15), indicating that weekend phone use might serve as a way to relax after a week of hard work.

\paragraph{Time Spent at Different Locations}  An increase in time spent at living places, such as home or dorms, during evenings or at night on weekdays is positively associated with end-of-term GPA (2019-R3 and 2019-R16). This may reflect the value of students engaging in campus life during the day and then spending time with roommates or dorm mates in the evenings, possibly studying or socializing. This type of evening engagement could alleviate feelings of isolation or a lack of belonging, as students who reported knowing less about school than their peers (2018-R13) exhibited a negative association with GPA. However, longer durations spent at living places during the day or in the morning on both weekdays and weekends were associated with lower end-of-term GPA (2018-R9, 2018-R25, 2018-R29, 2019-R6, 2019-R11, and 2019-R21). This suggests that while evening time at living places may be beneficial for academic success, excessive time spent indoors during the day or morning, especially on weekends, could detract from opportunities to engage with the broader campus environment, which might support students' academic and social integration.

Additionally, an increase in time spent at exercise locations throughout the day on weekdays is positively associated with academic performance (2018-R22), suggesting that physical activity during the week may contribute to better academic outcomes. However, spending more time at exercise places during the evening on weekends is associated with worse end-of-term GPA (2019-R4 and 2019-R5), which may indicate a potential disruption to academic focus or preparation for the upcoming week. Similarly, time spent in green spaces during the afternoon and evening is positively associated with academic performance (2019-R2 and 2019-R17). This could reflect the value of engaging with the campus environment, benefiting from outdoor activity and social interaction, especially during times that support mental well-being without conflicting with academic obligations like class attendance. These patterns highlight the importance of balanced engagement in physical and outdoor activities during the week, while suggesting that weekend activities might need to be managed to avoid negatively affecting academic outcomes.

Furthermore, an increase in time spent at food places in the evening on weekends is positively associated with student academic performance (2019-R12 and 2019-R24), suggesting that social or leisurely activities in such settings may serve as a beneficial break for students. Interestingly, while party duration at night during the first week is negatively associated with academic performance (2019-R22), time spent at Greek houses on weekend nights---regardless of whether students live in them---is positively associated with end-of-term GPA (2019-R15). Although previous research has suggested that Greek membership may negatively impact academic performance~\cite{grubb2006does}, our findings indicate that moderate socializing at these locations may not be inherently harmful. This suggests that relaxing at a party or gathering during weekends can be a healthy activity for students. Further research could explore whether students who are confident in their academic performance are more likely to attend social events and examine what types of social behaviors, including those in Greek life, are most supportive for students dealing with academic or personal stressors.

\paragraph{Sleep} Longer periods of restless sleep are negatively associated with student academic performance (2018-R18), as are extended periods of being awake, such as getting up early and staying awake late or pulling all-nighters (2019-R18). This is consistent with previous research that highlights the detrimental impact of poor sleep on academic outcomes~\cite{okano2019sleep,gomes2011sleep}. Interestingly, the longer the shortest duration of staying awake in a 24-hour period (2018-R6)---which could indicate a nap or a period of insomnia---is positively associated with higher end-of-term GPA. This finding warrants further investigation, as it contrasts with existing literature that shows sufficient sleep, good sleep quality, and greater sleep consistency are positively linked to academic performance~\cite{gomes2011sleep}. Understanding the nuanced relationship between short wakefulness periods and academic outcomes could provide deeper insights into student sleep patterns and their impact on academic success.

In addition to the above behavioral patterns that are relatively easy to interpret, we note that some location patterns are harder to explain. For example, an increase in time spent at the third-ranked location cluster (2018-R11), time spent at second-ranked location cluster (2018-R28, 2018-R30, 2019-R7, 2019-R8, and 2019-R30) at any time in a day was negatively associated with end-of-term GPA.

\paragraph{Self-reported Stressors} 
We also observed several serious stressors from students’ self-reports that were, unsurprisingly, strongly negatively associated with GPA. These included issues with romantic partners (2018-R13), health concerns (2018-R21, 2019-R25), and traumatic experiences (2019-R29). Additionally, academic-related stressors such as receiving lower grades than expected in a prior term (2018-R3) and obtaining a lower GPA than anticipated (2018-R16) were also negatively associated with end-of-term GPA. These findings align with prior research that highlights the negative impact of stressors on academic outcomes~\cite{de2016relationship,pereira2018types}. The strong association between stressors and academic performance suggests that early intervention strategies focusing on mental health support and academic counseling could be highly beneficial. Proactively addressing these issues at the beginning of the term might help students better manage their stress and improve their academic resilience.

Interestingly, we find that type of phone service provider is also associated with students' academic performance. To better understand this feature, we compare the proportion of each service provider use between students with higher and lower GPAs. We found that high performers used AT\&T, Cricket, and Sprint more, while low performers were prone to use Virgin Mobile and other providers. A chi-square test of independence was performed to examine the relation between high/low performers and the service provider they used. The relation between these variables was significant, $\chi^2$(6, $N$ = 188) = 17.1, $p$ < .01.  This certainly means that there is some other variable that is not being captured in our feature set that connects phone service provider and performance, perhaps something related to income or childhood home locale or other context, and highlights the importance of comparing features to behavioral science knowledge. 

\end{document}